\documentclass[prd,reprint,superscriptaddress,altaffillsymbol,amsmath,nofootinbib,longbibliography]{revtex4-1}
\pdfoutput=1

\usepackage{ifthen}
\usepackage{epsfig}
\usepackage{booktabs} 
\usepackage{natbib}
\usepackage{wasysym}
\usepackage{slashed} 
\usepackage{bbm}
\usepackage{mathrsfs}
\usepackage{amssymb}
\usepackage{dsfont}
\usepackage[export]{adjustbox}

\usepackage{color}

\definecolor{ColorXenon1}{rgb}{0.25,0.46,0.89}
\definecolor{ColorXenon2}{rgb}{0.24,0.33,0.83}
\definecolor{ColorArgon}{rgb}{0.62,0.2,0.7}

\usepackage[
colorlinks=true, 
linkcolor=ColorXenon2,
citecolor=ColorArgon,
urlcolor=ColorXenon1,
]{hyperref}

\allowdisplaybreaks

\newcommand{\erf}{\mathop{\mathrm{erf}}}
\newcommand{\beqra}{\begin{eqnarray}}
\newcommand{\eeqra}{\end{eqnarray}}
\newcommand{\beq}{\begin{equation}}
\newcommand{\eeq}{\end{equation}}
\newcommand{\dd}{\mathrm{d}}

\renewcommand{\epsilon}{\varepsilon}

\renewcommand{\vec}[1]{\mathbf{#1}}
\renewcommand{\bar}{\overline}

\newcommand{\ket}[1]{\left| #1 \right\rangle}
\newcommand{\bra}[1]{\left\langle #1 \right|}

\newcommand{\Kummer}[3]{{}_1F_{1}\left(#1,#2,#3 \right)}
\newcommand{\Hypergeometric}[4]{{}_2F_{1}\left(#1,#2,#3,#4\right)}

\newcommand{\wignerj}[6]{\ensuremath{\begin{pmatrix} #1 & #2 & #3 \\ #4 & #5 & #6\end{pmatrix}}}

\newcommand{\vPerpEl}{\mathbf{v}_{\rm el}^\perp}

\begin{document}

\title{Atomic responses to general dark matter-electron interactions}

\author{Riccardo Catena}
\email{catena@chalmers.se}
\affiliation{Chalmers University of Technology, Department of Physics, SE-412 96 G\"oteborg, Sweden}

\author{Timon Emken}
\email{emken@chalmers.se}
\affiliation{Chalmers University of Technology, Department of Physics, SE-412 96 G\"oteborg, Sweden}

\author{Nicola A. Spaldin}
\email{nicola.spaldin@mat.ethz.ch}
\affiliation{Department of Materials, ETH Z\"urich, CH-8093 Z\"urich, Switzerland}

\author{Walter Tarantino}
\email{walter.tarantino@dsf.unica.it}
\affiliation{Department of Materials, ETH Z\"urich, CH-8093 Z\"urich, Switzerland}

\begin{abstract}
In the leading paradigm of modern cosmology, about 80\% of our Universe's matter content is in the form of hypothetical, as yet undetected particles.~These do not emit or absorb radiation at any observable wavelengths, and therefore constitute the so-called dark matter (DM) component of the Universe.~Detecting the particles forming the Milky Way DM component is one of the main challenges for astroparticle physics and basic science in general.~One promising way to achieve this goal is to search for rare DM-electron interactions in low-background deep underground detectors.~Key to the interpretation of this search is the response of detectors' materials to elementary DM-electron interactions defined in terms of electron wave functions' overlap integrals.~In this work, we compute the response of atomic argon and xenon targets used in operating DM search experiments to general, so far unexplored DM-electron interactions.~We find that the rate at which atoms can be ionized via DM-electron scattering can in general be expressed in terms of four independent atomic responses, three of which we identify here for the first time.~We find our new atomic responses to be numerically important in a variety of cases, which we identify and investigate thoroughly using effective theory methods.~We then use our atomic responses to set 90\% confidence level~(C.L.) exclusion limits on the strength of a wide range of DM-electron interactions from the null result of DM search experiments using argon and xenon targets.

\end{abstract}

\maketitle

\section{Introduction}
\label{sec:introduction}
One of the major unsolved mysteries in modern physics is the elusive nature of dark matter (DM)~\cite{Bertone:2016nfn}.~Revealed via the gravitational pull it exerts on stars, galaxies, light and the Universe at large, DM is widely believed to be made of hypothetical particles which have so far escaped detection~\cite{Bertone:2004pz}.~Detecting the particles forming the Milky Way DM component and understanding their properties in terms of particle physics models is a top priority in astroparticle physics~\cite{appec}.~The experimental technique known as direct detection will play a major role in elucidating the nature of DM in the coming years~\cite{Undagoitia:2015gya}.~It searches for signals of DM interactions in low background deep underground detectors~\cite{Drukier:1983gj,Goodman:1984dc}.~Such interactions could produce observable nuclear recoils via DM-nucleus scattering or induce electronic transitions via DM-electron scattering in the detector material~\cite{Bernabei:2007gr,Kopp:2009et}.~DM~particles that are gravitationally bound to our galaxy do not have enough kinetic energy to produce an observable nuclear recoil if their mass is approximately below 1~GeV/c$^2$.~On the other end, they have enough kinetic energy to ionize an atom in a target material or induce the transition from the valence to the conduction band in a semiconductor crystal if their mass is above about 1~MeV/$c^2$~\cite{Essig:2011nj}.~Motivated by recent experimental efforts in the field of astroparticle physics~\cite{Alexander:2016aln}, here we focus on DM~particles of mass in the 1 -- 1000 MeV/$c^2$ range and on their interactions with the electrons in materials used in direct detection experiments.~In the literature, this framework is referred to as the ``sub-GeV'' or ``light DM'' scenario~\cite{Battaglieri:2017aum}.

Materials used in direct detection experiments which have set limits on the strength of DM-electron interactions include dual-phase argon~\cite{Agnes:2018oej} and xenon~\cite{Essig:2012yx,Essig:2017kqs,Aprile:2019xxb} targets, as well as germanium and silicon semiconductors~\cite{Essig:2011nj,Graham:2012su,Lee:2015qva,Essig:2015cda,Crisler:2018gci,Agnese:2018col,Abramoff:2019dfb,Aguilar-Arevalo:2019wdi}, and sodium iodide crystals~\cite{Roberts:2016xfw}.~Furthermore, graphene~\cite{Hochberg:2016ntt,Geilhufe:2018gry}, 3D Dirac materials~\cite{Hochberg:2017wce,Geilhufe:2019ndy}, polar crystals~\cite{Knapen:2017ekk}, scintillators~\cite{Derenzo:2016fse,Blanco:2019lrf}, as well as superconductors~\cite{Hochberg:2015pha,Hochberg:2015fth,Hochberg:2019cyy} have also been proposed recently as target materials to probe DM-electron interactions.

In order to interpret the results of direct detection experiments searching for signals of DM-electron interactions, it is crucial to understand quantitatively how detector materials respond to general DM-electron interactions.~Materials' responses to DM-electron interactions can be quantified in terms of the overlap between initial (before scattering) and final (after scattering) electron wave functions.~The stronger the material response, the larger the expected rate of DM-electron interactions in the target.

In the pioneering works~\cite{Kopp:2009et,Essig:2011nj,Essig:2015cda}, detector materials' responses to DM-electron interactions have been computed under the assumption that the amplitude for DM scattering by free electrons only depends on the initial and final state particle momenta through the momentum transferred in the scattering\footnote{For more recent works on the theoretical framework of so-called spin-independent DM-electron interactions and a comparison of targets, we also refer to~\cite{Trickle:2019nya} and~\cite{Griffin:2019mvc} respectively.}.~While this restriction significantly simplifies the calculation of the predicted DM-electron interaction rate in a given detector, it prevents computations of physical observables whenever the scattering amplitude exhibits a more general momentum dependence.~A more general dependence on the particle momenta occurs in a variety of models for DM~\cite{DelNobile:2018dfg}, including those where DM interacts with electrons via an anapole or magnetic dipole coupling~\cite{Kavanagh:2018xeh}.

In this paper, we calculate the response of materials used in operating direct detection experiments to a general class of nonrelativistic DM-electron interactions.~We build this class of nonrelativistic interactions by using effective theory methods developed in the context of DM-nucleus scattering, e.g. Refs.~\cite{Fan:2010gt,Fitzpatrick:2012ix,DelNobile:2018dfg,Catena:2019hzw}.~In terms of dependence on the incoming and outgoing particle momenta, this class of interactions generates the most general amplitude for the nonrelativistic scattering of DM~particles by free electrons.~The predicted amplitude is compatible with momentum and energy conservation, and is invariant under Galilean transformations (i.e.~constant shifts of particle velocities) and three-dimensional rotations.~At the same time, it depends on the incoming and outgoing particle momenta through more than only the momentum transferred.~It can explicitly depend on a second, independent combination of particle momenta, as well as on the DM~particle and electron spin operators.~We present our results focusing on DM scattering from electrons bound in isolated argon and xenon atoms, since these are the targets used in leading experiments such as XENON1T~\cite{Aprile:2019xxb} and DarkSide-50~\cite{Agnes:2018oej}.
~However, most of the expressions derived in this work also apply to other target materials, such as semiconductors.
~They do not apply to scenarios where incoming DM~particles induce collective excitations, such as magnons or plasmons~\cite{Trickle:2019ovy,Kurinsky:2020dpb}.

Within this general theoretical framework, we find that the rate of DM-induced electronic transitions in a given target material can be expressed in terms of four, independent material response functions.~They depend on scalar and vectorial combinations of electron momenta and wave functions.~We numerically evaluate these response functions for argon and xenon targets, and use them to set 90\% C.L. exclusion limits on the strength of DM-electron interactions from the current null result of the XENON10~\cite{Angle:2011th}, XENON1T~\cite{Aprile:2019xxb} and DarkSide-50~\cite{Agnes:2018oej} experiments.~We also find that the contribution of each material response function to the rate of DM-induced electronic transitions is weighted by a corresponding combination of particle momenta which is independent of the initial and final electron wave functions.~We refer to these weights as ``DM response functions'', in analogy to previous studies in the context of DM-nucleus scattering~\cite{Fitzpatrick:2012ix}.

Previous works could only identify one of the four material response functions found here because they assumed that the amplitude for DM scattering by free electrons solely depends on the momentum transfer, e.g. Ref.~\cite{Essig:2015cda}.~This previously found response function is not only generated by DM-electron scattering.~It also arises when materials are probed with known particles, such as photons or electrons which interact via the electromagnetic force.~In contrast, the three novel response functions found in this work can only arise from a general treatment of DM-electron interactions.~We will show this using an effective theory approach to DM-electron interactions.~Specific examples of scenarios where the new response functions appear include models where DM interacts via an anapole or magnetic dipole coupling to photons.

These general considerations lead us to the intriguing conclusion that DM particles may represent a new type of probe for the electronic structure of a sample material.~At the moment, the information we can get about the actual electronic structure of a material is limited by the probes we can use.~If alongside standard probes (electrons, photons, alpha particles, etc\dots) we could use a particle that interacts in a totally different way with electrons, we might be able to get as yet ``hidden'' features of the electronic structures that would be missed with standard probes.~While this would not change how a material responds to standard particles, which ultimately is what is currently relevant for applications in the real world, it would help us to know more about the material, and the more we know, the easier is to design and produce new materials with desired properties of technological relevance.~Consequently, the detection of DM~particles at direct detection experiments might not only solve one of the most pressing problems in astroparticle physics, but also open up a new window on the exploration of materials' responses to external probes.
~From this perspective, the results of this work constitute the first step within a far-reaching program which has the potential to open an entirely new field of research.

While the notion of DM as a probe for the electronic structure of a sample material would require good knowledge about the local properties of the galactic DM halo, the novel material responses identified here only depend on the materials' electronic structure and are not degenerate (i.e.~cannot be confused) with variations in the assumed local DM velocity distribution.

\begin{figure*}
\centering
\includegraphics[width=0.32\textwidth]{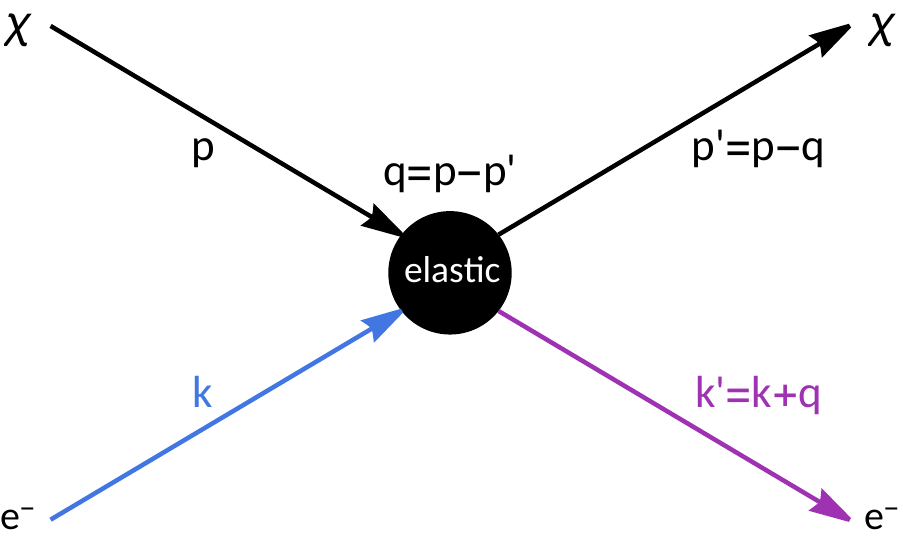}
\includegraphics[width=0.32\textwidth]{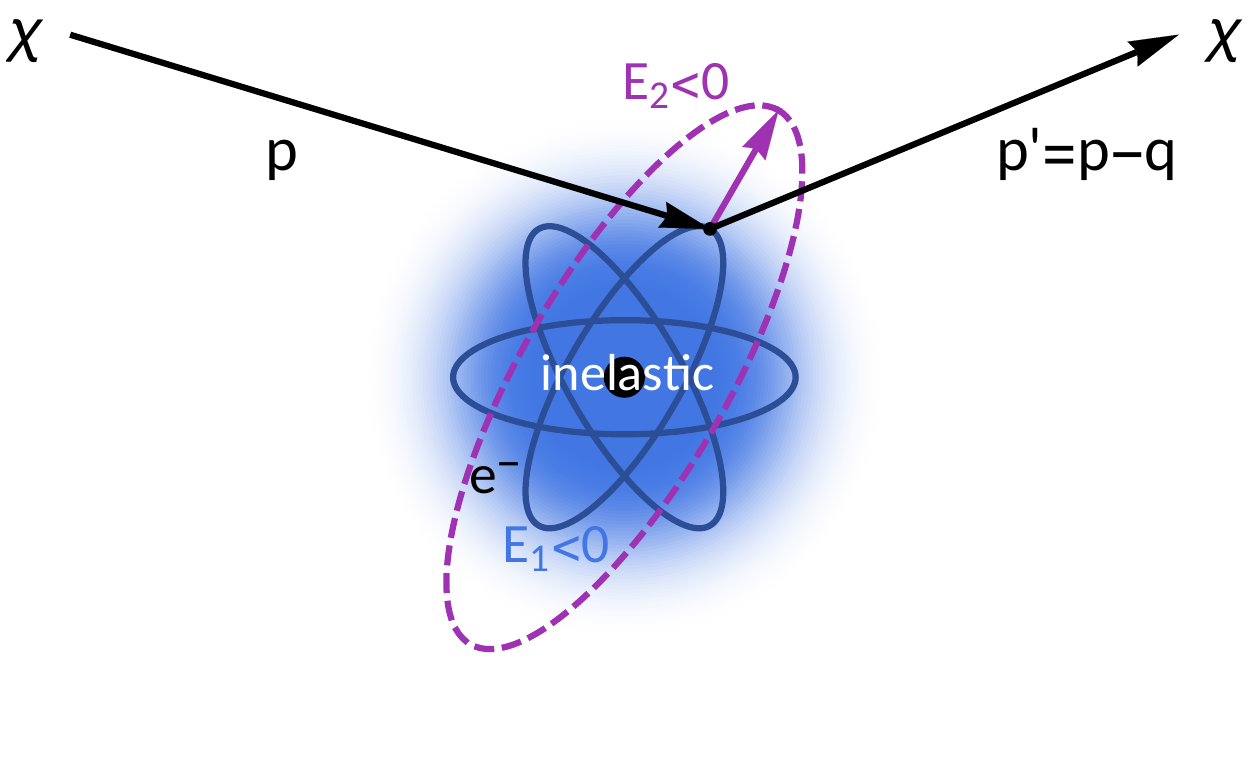}
\includegraphics[width=0.3225\textwidth]{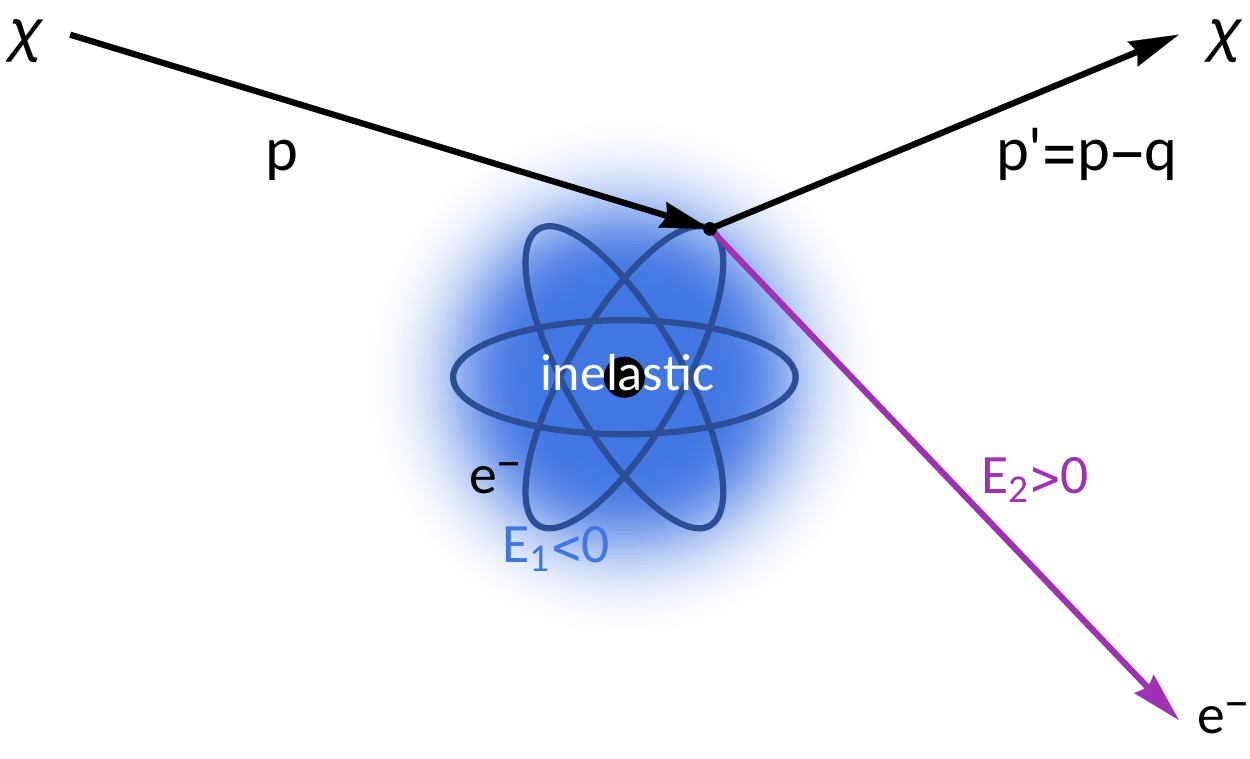}
\caption{Illustration of elastic~(left) and inelastic~(middle, right) DM-electron scatterings. In the inelastic case, the electron gets excited by~DM, transitioning from an atomic bound state with negative energy~$E_1$ to a final state with higher energy~$E_2$, which is either still negative for excitation~(middle) or positive for ionization~(right).}
\label{fig: kinematics}
\end{figure*}

This work is organized as follows.~In Sec.~\ref{sec:scattering} we derive a general expression for the rate of DM-induced electronic transitions in target materials which we then specialize to the case of general, nonrelativistic DM-electron interactions, focusing on argon and xenon as target materials.~In Sec.~\ref{sec:FF}, we express the rate of DM-induced electronic transitions in terms of DM and atomic response functions, while in Sec.~\ref{sec:results} we apply our results to compute physical observables for argon and xenon targets and set limits on the strength of DM-electron interactions from the current null results of the XENON10, XENON1T and DarkSide-50 experiments.~We present our conclusions in Sec.~\ref{sec:conclusions} and provide the details of our calculations in the appendices. In addition, we provide the numerical code used to evaluate the atomic response functions~\cite{Emken2019}.\footnote{An archived version can be found under~\href{https://doi.org/10.5281/zenodo.3581334}{[DOI:10.5281/zenodo.3581334]}.}\\[1em]

Throughout this paper, we employ natural units, i.e.,~$c=\hbar=1$. 

\section{Dark matter-induced electronic transitions}
\label{sec:scattering}
In this section we derive a general expression for the rate of electronic transitions induced by the scattering of Milky Way DM~particles in target materials (Sec.~\ref{sec:rate}).~This result applies to arbitrary DM-electron interactions (including relativistic interactions) and target materials.~We then specialize our general expression to the case of {\it nonrelativistic} DM-electron interactions (Sec.~\ref{sec:interactions}) and to argon and xenon as target materials (Sec.~\ref{sec:targets}).~We use these results to model the response of atoms to general DM-electron interactions in Sec.~\ref{sec:FF} and set limits on the strength of such interactions from the null results of operating direct detection experiments in Sec.~\ref{sec:results}.

\subsection{Electron transition rate}
\label{sec:rate}
We start by calculating the rate of transitions from an initial state $|i\rangle\equiv|\mathbf{e}_1,\mathbf{p}\rangle=|\mathbf{e}_1\rangle \otimes |\mathbf{p}\rangle$ to a final state $|f\rangle\equiv|\mathbf{e}_2,\mathbf{p}'\rangle=|\mathbf{e}_2\rangle \otimes |\mathbf{p}'\rangle$, where initial and final DM~particle states are labeled by the corresponding tridimensional momenta, $\mathbf{p}$ and $\mathbf{p}'$, respectively, while $|\mathbf{e}_1\rangle$ ($|\mathbf{e}_2\rangle$) denotes the initial (final) electron state.~Here $|\mathbf{e}_1\rangle$ and $|\mathbf{e}_2\rangle$ are eigenstates of the electron's energy with eigenvalue~$E_1$ and~$E_2$, but not of momentum. This is illustrated in the middle and right panel of Fig.~\ref{fig: kinematics}.~Following~\cite{Essig:2015cda}, single-particle states are normalized as
\begin{equation}
\label{eq:norm}
\langle\mathbf{p}|\mathbf{p}\rangle = (2\pi)^3 \delta^{(3)}(0) = \int {\rm d}^3x \equiv V\,,
\end{equation}
and, similarly, e.g., $\langle\mathbf{e}_1|\mathbf{e}_1\rangle=V$.~The divergent volume factor, $V$, will not appear in the expressions for physical observables.

The first non-trivial term in the Dyson expansion of the S-matrix element corresponding to the $|i\rangle \rightarrow |f\rangle$ transition is
\begin{widetext}
\begin{eqnarray}
\label{eq:Sfi}
S_{fi}&=&-i \langle f| \int {\rm d}^4 x \, \mathscr{H}_I(x) |i\rangle \nonumber \\
&=&-i \int {\rm d}^4 x \, \langle \mathbf{p}', \mathbf{e}_2 | e^{i H_0 t} \mathscr{H}_S(x) e^{-i H_0 t} |\mathbf{e}_1,\mathbf{p}\rangle \nonumber\\ 
&=&-i \int {\rm d}^3 x \, \langle \mathbf{p}', \mathbf{e}_2 | \mathscr{H}_S(x) |\mathbf{e}_1,\mathbf{p}\rangle \int {\rm d}t \,e^{i (E_f - E_i) t} \nonumber\\ 
&=&-i (2\pi) \delta(E_f-E_i) \int {\rm d}^3 x \, \langle \mathbf{p}', \mathbf{e}_2 | \mathscr{H}_S(x) |\mathbf{e}_1,\mathbf{p}\rangle \nonumber\\ 
&=&-i (2\pi) \delta(E_f-E_i) \int {\rm d}^3 x   \int  \frac{{\rm d}^3 k}{(2 \pi)^3}\int  \frac{{\rm d}^3 k'}{(2 \pi)^3} \, \langle \mathbf{e}_2 |\mathbf{k}'\rangle \langle \mathbf{k}' , \mathbf{p}' | \mathscr{H}_S(x) | \mathbf{p},\mathbf{k}\rangle \langle \mathbf{k} | \mathbf{e}_1\rangle 
\end{eqnarray}
\end{widetext}
where $\mathscr{H}_I(x)$ ($\mathscr{H}_S(x)$) is the interaction Hamiltonian density in the interaction (Schr\"odinger) picture at the spacetime point $x$, and $H_0$ is the Hamiltonian of the DM-electron system with $\mathscr{H}_I(x)$ set to zero which obeys $H_0|i\rangle = E_i |i\rangle$ and $H_0|f\rangle = E_f |f\rangle$.~We also used the identity 
\begin{equation}
\label{eq:1}
\int \frac{{\rm d}^3k}{(2 \pi)^3} \, |\mathbf{k}\rangle \langle\mathbf{k}| = \mathds{1}\,,
\end{equation}
where $|\mathbf{k}\rangle$ are eigenstates of the free-electron Hamiltonian.~The asymptotic states $|i\rangle$ and $|f\rangle$ are eigenstates of $H_0$ because $\mathscr{H}_I$ is assumed to be different from zero within a finite time window only (the so-called adiabatic hypothesis).~In the nonrelativistic limit, the corresponding eigenvalues, denoted here by $E_i$ and $E_f$, respectively, are given by
\begin{align}
    E_i &= m_\chi + m_e + \frac{m_\chi}{2}v^2 + E_1\, , \label{eq: energy initial}\\
    E_f &= m_\chi + m_e + \frac{|m_\chi\vec{v}-\mathbf{q}|^2}{2m_\chi} + E_2\,, \label{eq: energy final}
\end{align}
where $\mathbf{q}=\mathbf{p}-\mathbf{p}^\prime$ is the momentum transferred in the scattering, $\mathbf{v}$ ($v$) is the initial DM~particle velocity (speed), $m_\chi$ and $m_e$ are the DM and electron mass, respectively, and $E_1$ ($E_2$) is the energy of the initial (final) electron bound state.~Defining $\Delta E_{1\rightarrow 2}\equiv E_2-E_1$, the energy difference $E_f-E_i$ reads
\begin{align}
E_f-E_i &= \Delta E_{1\rightarrow 2} + \frac{q^2}{2m_\chi}-q v \cos \theta_{qv}\,,
\end{align}
where $\theta_{qv}$ is the angle between the vectors $\mathbf{q}$ and $\mathbf{v}$, while $q=|\mathbf{q}|$.~Notice that $S_{fi}$ has dimension [energy]$^{-6}$, because of our choice of single-particle state normalization, Eq.~(\ref{eq:norm}).

Let us now consider the process $|\mathbf{k},\mathbf{p} \rangle \rightarrow |\mathbf{k}',\mathbf{p}' \rangle$, where a DM~particle of initial (final) momentum  $\mathbf{p}$ ($\mathbf{p}'$) scatters elastically off a free electron of initial (final) momentum $\mathbf{k}$ ($\mathbf{k}'$) as shown in the left panel of Fig.~\ref{fig: kinematics}.~The first non-trivial term in the Dyson expansion of the S-matrix element corresponding to this process is 
\begin{equation}
\label{eq:Sfree1}
S_{fi}^{\rm free}=-i (2\pi) \delta(\tilde{E}_f-\tilde{E}_i) \int {\rm d}^3 x \, \langle \mathbf{p}', \mathbf{k}' | \mathscr{H}_S(x) |\mathbf{k},\mathbf{p}\rangle \,,
\end{equation}
where we defined $\tilde{E}_f\equiv E_{\mathbf{k}'}+E_{\mathbf{p}'}$ and $\tilde{E}_i\equiv E_{\mathbf{k}}+E_{\mathbf{p}}$.~The above S-matrix element can equivalently be expressed as follows
\begin{eqnarray}
\label{eq:Sfree2}
S_{fi}^{\rm free}&=&i (2\pi)^4\delta(\tilde{E}_f-\tilde{E}_i) \delta^{(3)}(\mathbf{p}'+\mathbf{k}'-\mathbf{p}-\mathbf{k}) \nonumber\\
&\times&\frac{1}{\sqrt{2 E_{\mathbf{p}'} 2 E_{\mathbf{k}'} 2 E_{\mathbf{k}} 2 E_{\mathbf{p}}}} \, 
\mathcal{M}(\mathbf{k},\mathbf{p},\mathbf{k}',\mathbf{p}')\,,
\end{eqnarray}
where $\mathcal{M}$ is the amplitude for DM scattering by a free electron.~Compared to Eq.~(4.73) of~\cite{Peskin:1995ev}, Eq.~(\ref{eq:Sfree2}) differs by a factor of $1/\sqrt{2 E_{\mathbf{p}'} 2 E_{\mathbf{k}'} 2 E_{\mathbf{k}} 2 E_{\mathbf{p}}}$ because of our different normalization of single-particle states (see Eq.~(\ref{eq:norm})).~By comparing Eq.~(\ref{eq:Sfree1}) with Eq.~(\ref{eq:Sfree2}), we obtain the following identity
\begin{eqnarray}
\label{eq:identity}
  \int {\rm d}^3 x \,\langle \mathbf{p}', \mathbf{k}' | \mathscr{H}_S(x) |\mathbf{k},\mathbf{p}\rangle &=& - (2\pi)^3\delta^{(3)}(\mathbf{p}'+\mathbf{k}'-\mathbf{p}-\mathbf{k}) \nonumber\\
  &\times& \frac{\mathcal{M}(\mathbf{k},\mathbf{p},\mathbf{k}',\mathbf{p}')}{\sqrt{2 E_{\mathbf{p}'} 2 E_{\mathbf{k}'} 2 E_{\mathbf{k}} 2 E_{\mathbf{p}}}} \,,
\end{eqnarray}
where in the nonrelativistic limit $2 E_{\mathbf{p}'} 2 E_{\mathbf{k}'} 2 E_{\mathbf{k}} 2 E_{\mathbf{p}} = 16 m_\chi^2 m_e^2$.~By substituting the free scattering result of Eq.~(\ref{eq:identity}) back into the general S-matrix element in Eq.~(\ref{eq:Sfi}), we obtain  
\begin{eqnarray}
\label{eq:Sfi1}
S_{fi}&=&(2\pi) \delta(E_f-E_i) \frac{1}{4 m_{\chi} m_e}  \int  \frac{{\rm d}^3 k}{(2 \pi)^3} \langle \mathbf{e}_2 |\mathbf{k}+\mathbf{q}\rangle  \langle \mathbf{k} | \mathbf{e}_1\rangle  \nonumber\\
&\times& i\mathcal{M}(\mathbf{k},\mathbf{p},\mathbf{k}+\mathbf{q},\mathbf{p}-\mathbf{q}) \,,
\end{eqnarray}
where we used the nonrelativistic expression for $\sqrt{2 E_{\mathbf{p}'} 2 E_{\mathbf{k}'} 2 E_{\mathbf{k}} 2 E_{\mathbf{p}}}$.~We now introduce the electron wave functions $\psi_1(\mathbf{k})=\langle \mathbf{k} | \mathbf{e}_1\rangle/\sqrt{V}$ and $\psi_2^*(\mathbf{k}+\mathbf{q})=\langle \mathbf{e}_2 |\mathbf{k}+\mathbf{q}\rangle/\sqrt{V}$.~Using Eq.~(\ref{eq:norm}) and Eq.~(\ref{eq:1}), a simple calculation shows that the wave functions $\psi_1$ and $\psi_2$ are normalized to one.~In terms of $\psi_1$ and $\psi_2$, Eq.~(\ref{eq:Sfi1}) can be written as follows,
\begin{eqnarray}
\label{eq:Sfi2}
S_{fi}&=&(2\pi) \delta(E_f-E_i)  \frac{V}{4 m_\chi m_e} \int  \frac{{\rm d}^3 k}{(2 \pi)^3} \psi_2^*(\mathbf{k}+\mathbf{q}) \psi_1(\mathbf{k}) \nonumber\\
&\times&i\mathcal{M}(\mathbf{k},\mathbf{p},\mathbf{k}+\mathbf{q},\mathbf{p}-\mathbf{q}) \,.
\end{eqnarray}
The probability for the transition $|\mathbf{e}_1,\mathbf{p}\rangle \rightarrow |\mathbf{e}_2,\mathbf{p}'\rangle$ is then given by $|S_{fi}|^2/V^4$, where we divide by $V^4$ in order to obtain unit-normalized single-particle states from Eq.~(\ref{eq:norm}) (remember that $S_{fi}$ has dimension [energy]$^{-6}$).~Consistently, the probability, $\mathscr{P}(\mathbf{p})$, that a DM~particle with initial momentum $\mathbf{p}$ and a bound electron in the $| e_1\rangle$ state scatter to a group of final states with DM momenta in the infinitesimal interval ($\mathbf{p}'$, $\mathbf{p}'+{\rm d}\mathbf{p}'$) and the electron in the $| e_2\rangle$ state is given by $|S_{fi}|^2/V^4$ times the number of states in the ($\mathbf{p}'+{\rm d}\mathbf{p}'$) interval, namely
\begin{eqnarray}
\label{eq:Pp}
\mathscr{P}(\mathbf{p})&=&\frac{|S_{fi}|^2}{V^4} \frac{V{\rm d}^3 p'}{(2 \pi)^3}\nonumber\\
&=&\frac{|S_{fi}|^2}{V^4} \frac{V {\rm d}^3 q}{(2 \pi)^3} \nonumber\\
&=& (2\pi) \delta(E_f-E_i)  \frac{T {\rm d}^3 q}{(2 \pi)^3 V} \frac{1}{16 m^2_{\chi} m^2_e} \nonumber\\
&\times&\left|\int \frac{{\rm d}^3 k}{(2 \pi)^3 } \, \psi_2^*(\mathbf{k}+\mathbf{q})\mathcal{M}(\mathbf{k},\mathbf{p},\mathbf{q})
\psi_1(\mathbf{k}) \right|^2 
\end{eqnarray}
where the divergent factor $T=\int {\rm d}t$ arises when squaring the one-dimensional Dirac delta in Eq.~(\ref{eq:Sfi2}).~While $T$ enters in Eq.~(\ref{eq:Pp}) explicitly, physical observables will not depend on $T$.~In order to simplify the notation, in Eq.~(\ref{eq:Pp}) we replaced $\mathcal{M}(\mathbf{k},\mathbf{p},\mathbf{k}+\mathbf{q},\mathbf{p}-\mathbf{q})$ with $\mathcal{M}(\mathbf{k},\mathbf{p},\mathbf{q})$.~The rate per unit DM number density for the transition $|\mathbf{e}_1,\mathbf{p}\rangle \rightarrow |\mathbf{e}_2,\mathbf{p}'\rangle$ with $\mathbf{p}'$ within ($\mathbf{p}'$, $\mathbf{p}'+{\rm d}\mathbf{p}'$) is therefore given by $\mathscr{P}(\mathbf{p})V/T$.~Indeed, there is just one DM~particle with initial momentum  $\mathbf{p}$ within ($\mathbf{p}'$, $\mathbf{p}'+{\rm d}\mathbf{p}'$) in a phase-space element of three-dimensional volume $V$~\cite{Mandl:1985bg}.

Finally, integrating $\mathscr{P}(\mathbf{p})V/T$ over the initial DM~particle velocity distribution $f_{\chi}$ and the transferred momentum $\mathbf{q}$, and multiplying by the local DM number density, $n_{\chi}$\footnote{The details of the DM~density and velocity distributions are summarized in Appendix~\ref{app: experiments}.}, for the total rate of DM-induced $|\mathbf{e}_1\rangle \rightarrow |\mathbf{e}_2\rangle$ transitions we find
\begin{align}
\mathscr{R}_{1\rightarrow 2}&=\frac{n_{\chi}}{16 m^2_{\chi} m^2_e}\nonumber\\
& \times \int \frac{{\rm d}^3 q}{(2 \pi)^3} \int {\rm d}^3 v f_{\chi}(\mathbf{v}) (2\pi) \delta(E_f-E_i) \overline{\left| \mathcal{M}_{1\rightarrow 2}\right|^2}\, . \label{eq:transition rate}
\end{align}
Here, we defined the {\it squared electron transition amplitude}, $\overline{\left| \mathcal{M}_{1\rightarrow 2}\right|^2}$, in terms of $\psi_1$, $\psi_2$ and the free scattering amplitude,
\begin{align}
    \overline{\left| \mathcal{M}_{1\rightarrow 2}\right|^2}\equiv \overline{\left|\int  \frac{{\rm d}^3 k}{(2 \pi)^3} \, \psi_2^*(\mathbf{k}+\mathbf{q})  
\mathcal{M}(\mathbf{k},\mathbf{p},\mathbf{q})
\psi_1(\mathbf{k}) \right|^2}\, , \label{eq:transition amplitude}
\end{align}
where a bar denotes an average (sum) over initial (final) spin states.

In the following two sections, we motivate and explore the implications of our assumptions for $\mathcal{M}$, $\psi_1$ and $\psi_2$.~Here, we model $\mathcal{M}$ within a nonrelativistic effective theory of DM-electron interactions (Sec.~\ref{sec:interactions}) and use initial and final electron wave functions derived for isolated atoms in argon and xenon targets~\cite{Bunge:1993jsz} for $\psi_1$ and $\psi_2$ (Sec.~\ref{sec:targets}).

\subsection{Non-relativistic dark matter-electron interactions}
\label{sec:interactions}
In this section, we investigate the nonrelativistic limit of DM-electron interactions.~Our goal is to show that the standard assumption in the context of DM direct detection, $\mathcal{M}=\mathcal{M}(\mathbf{q})$, is highly restrictive.~Rather than basing our argument on a specific model, we build a general nonrelativistic effective theory where the active degrees of freedom are electrons and DM~particles.~The basic symmetries governing the DM-electron scattering, together with the assumption that both the DM~particle and the electron are nonrelativistic, will determine the allowed interactions between the active degrees of freedom in our effective theory, and allow us to show that $\mathcal{M}=\mathcal{M}(\mathbf{q})$ is in general a too restrictive assumption.~The symmetries that govern the nonrelativistic DM-electron scattering are the invariance under time and space translations (leading to energy and momentum conservation), the invariance under Galilean transformations (namely, constant shifts of particle velocities) and the invariance under three-dimensional rotations.~Galilean invariance replaces the invariance under Lorentz boosts of relativistic theories.~Let us now explore the constraints set by these symmetries on the amplitude for nonrelativistic DM-electron scattering.

While the nonrelativistic scattering of Milky Way DM~particles by free electrons is characterized by the three-dimensional momenta $\mathbf{k}$ ($\mathbf{k}'$) and $\mathbf{p}$ ($\mathbf{p}'$) of the incoming (outgoing) electron and DM~particle, respectively, momentum conservation and Galilean invariance imply that only two out of the four three-dimensional vectors are actually independent.~A convenient choice of independent momenta is $\mathbf{q}=\mathbf{p}-\mathbf{p}'$, the momentum transferred in the scattering, and 
\begin{align}
\mathbf{v}_{\rm el}^\perp &= \frac{\left( \mathbf{p} + \mathbf{p}' \right)}{2 m_{\chi}} - \frac{\left( \mathbf{k} + \mathbf{k}' \right)}{2 m_e}  \\
&=\mathbf{v} - \frac{\mathbf{q}}{2\mu_e} - \frac{\mathbf{k}}{m_e}\, ,
\end{align}
where~$\mu_e$ denotes the reduced mass of the DM~particle and the electron, and~$\mathbf{v}\equiv \mathbf{p}/m_\chi$ is the incoming DM~velocity. In the case of elastic DM-electron scattering, $\mathbf{v}_{\rm el}^\perp\cdot \mathbf{q}=0$ because of energy conservation.~In contrast, when DM scatters from electrons inelastically (and initial and final electrons populate different energy levels), $\mathbf{v}_{\rm el}^\perp\cdot \mathbf{q}\neq0$.~In this case, one could define
\begin{align}
    \mathbf{v}_{\rm inel}^\perp \equiv\mathbf{v} -\frac{\mathbf{q}}{2m_\chi}-\frac{\Delta E_{1\rightarrow 2}}{q^2}\mathbf{q}\, ,
\end{align}
such that $\mathbf{v}_{\rm inel}^\perp\cdot \mathbf{q}=0$ by construction.~While $\mathcal{M}$ can equivalently be expressed in terms of $\mathbf{v}_{\rm inel}^\perp$ or $\mathbf{v}_{\rm el}^\perp$~\cite{DelNobile:2018dfg}, we find that it is more convenient to use $\mathbf{v}_{\rm el}^\perp$ as a basic variable when matching explicit models to the nonrelativistic effective theory built in this section (compare for example Eqs.~(\ref{eq:M}) and~(\ref{eq:M3}) below).

To summarize, within our effective theory the free amplitude for nonrelativistic DM-electron scattering reads
\begin{equation}
\label{eq:M}
\mathcal{M}(\mathbf{k},\mathbf{k}',\mathbf{p},\mathbf{p}') 
= \mathcal{M}(\mathbf{q},\mathbf{v}_{\rm el}^\perp)\,,
\end{equation}
where we emphasized the dependence of the amplitude $\mathcal{M}$ on the vectors $\mathbf{q}$ and $\mathbf{v}_{\rm el}^\perp$.~In addition to $\mathbf{q}$ and $\mathbf{v}_{\rm el}^\perp$, $\mathcal{M}$ can also depend on matrix elements of the electron and DM~particle spin operators, denoted here by $\mathbf{S}_\chi$ and $\mathbf{S}_e$, respectively.~Within the nonrelativistic effective theory of DM-electron interactions developed here, the amplitude $\mathcal{M}$ does not explicitly depend on the properties that characterize the particle mediating the interactions between DM and electrons, such as the mediator mass, $m_{\rm med}$.~There are two scenarios in which our effective theory approach can be applied.~In the first one, $m_{\rm med}$ is much larger than the momentum transfer in DM-electron scattering (contact interaction).~In the second one, the mediator is effectively mass-less (long-range interaction).~These are the two limiting cases that we investigate here.~For $m_{\rm med} \sim|\mathbf{q}|$, $\mathcal{M}$ is model dependent, and some of the considerations made here do not apply.

\begin{table}[t]
    \centering
    \begin{tabular*}{\columnwidth}{@{\extracolsep{\fill}}ll@{}}
    \toprule
        $\mathcal{O}_1 = \mathds{1}_{\chi}\mathds{1}_e$  & $\mathcal{O}_{11} = i\vec{S}_\chi\cdot\frac{\vec{q}}{m_e}\mathds{1}_e$    \\
        $\mathcal{O}_3 = i\vec{S}_e\cdot\left(\frac{\vec{q}}{m_e}\times\vPerpEl\right)\mathds{1}_\chi$ & $\mathcal{O}_{12} = \vec{S}_\chi\cdot \left(\vec{S}_e \times\vPerpEl \right)$  \\
        $\mathcal{O}_4 = \vec{S}_\chi\cdot \vec{S}_e$ & $\mathcal{O}_{13} =i \left(\vec{S}_\chi\cdot \vPerpEl\right)\left(\vec{S}_e\cdot \frac{\vec{q}}{m_e}\right)$ \\
        $\mathcal{O}_5 = i\vec{S}_\chi\cdot\left(\frac{\vec{q}}{m_e}\times\vPerpEl\right)\mathds{1}_e$ &  $\mathcal{O}_{14} = i\left(\vec{S}_\chi\cdot \frac{\vec{q}}{m_e}\right)\left(\vec{S}_e\cdot \vPerpEl\right)$ \\
        $\mathcal{O}_6 = \left(\vec{S}_\chi\cdot\frac{\vec{q}}{m_e}\right) \left(\vec{S}_e\cdot\frac{\vec{q}}{m_e}\right)$ &$\mathcal{O}_{15} = 
        i\mathcal{O}_{11}
        \left[ \left(\vec{S}_e\times \vPerpEl \right) \cdot \frac{\vec{q}}{m_e}\right] $  \\
        $\mathcal{O}_7 = \vec{S}_e\cdot \vPerpEl\mathds{1}_\chi$ & $\mathcal{O}_{17}=i \frac{\vec{q}}{m_e} \cdot \boldsymbol{\mathcal{S}} \cdot \vPerpEl \mathds{1}_e$  \\
        $\mathcal{O}_8 = \vec{S}_\chi\cdot \vPerpEl\mathds{1}_e$ & $\mathcal{O}_{18}=i \frac{\vec{q}}{m_e} \cdot \boldsymbol{\mathcal{S}}  \cdot \vec{S}_e$  \\
        $\mathcal{O}_9 = i\vec{S}_\chi\cdot\left(\vec{S}_e\times\frac{\vec{q}}{m_e}\right)$ & $\mathcal{O}_{19} = \frac{\vec{q}}{m_e} \cdot \boldsymbol{\mathcal{S}} \cdot \frac{\vec{q}}{m_e}$  \\
       $\mathcal{O}_{10} = i\vec{S}_e\cdot\frac{\vec{q}}{m_e}\mathds{1}_\chi$   & $\mathcal{O}_{20} = \left(\vec{S}_e \times \frac{\vec{q}}{m_e} \right) \cdot \boldsymbol{\mathcal{S}} \cdot \frac{\vec{q}}{m_e}$  \\
    \bottomrule
    \end{tabular*}
    \caption{At second order in $\mathbf{q}$ and at linear order in $\mathbf{v}_{\rm el}^\perp$, $\mathbf{S}_\chi$ and $\mathbf{S}_e$, the first fourteen operators in this table are the only three-dimensional scalars that one can build while preserving Galilean invariance and momentum conservation when DM has spin 1/2.~Here,~$\mathds{1}_{\chi}$ and $\mathds{1}_e$ are $2\times 2$ identity matrices in the DM~particle and electron spin space, respectively.~The operators $\mathcal{O}_{17}$, $\mathcal{O}_{18}$, $\mathcal{O}_{19}$ and $\mathcal{O}_{20}$ were found to arise from the nonrelativistic reduction of models for spin 1 DM with a vector mediator.~We follow the numbering used for nonrelativistic DM-nucleon interactions~\cite{Catena:2019hzw}, and neglect $\mathcal{O}_{2}$, which is quadratic in $\mathbf{v}_{\rm el}^\perp$, and $\mathcal{O}_{16}$, which can be expressed as a linear combination of $\mathcal{O}_{12}$ and $\mathcal{O}_{15}$.~For further details, see, for example, Ref.~\cite{DelNobile:2018dfg}.}
\label{tab:operators}
\end{table}

In Eq.~(\ref{eq:M}) we are assuming that both DM~particle and electron are nonrelativistic.~Indeed, Milky Way DM~particles have a typical speed of the order of $10^{-3}$ in natural units, and the typical speed of electrons bound in atoms is $\alpha Z_{\rm eff}$, where $\alpha=1/137$ and $Z_{\rm eff}$ is 1 for electrons in outer shells (the most interesting ones from a detection point of view) and larger for inner shells.~Notice also that for $m_\chi \ge 1$~MeV/$c^2$ the electron is the fastest and lightest particle in the nonrelativistic DM-electron scattering, which implies that the typical momentum transferred in the scattering is of the order of $Z_{\rm eff} \alpha m_e$~\cite{Essig:2015cda}.~Equation~(\ref{eq:M}) can then be expanded in powers of $|\mathbf{q}|/m_e$ and $|\mathbf{v}_{\rm el}^\perp|$.~Each term in this expansion of $\mathcal{M}$ must be both Galilean invariant and scalar under three-dimensional rotations.~At linear order in $\mathbf{v}_{\rm el}^\perp$ and at second order in $\mathbf{q}$, there are fourteen combinations of $\mathbf{q}$, $\mathbf{v}_{\rm el}^\perp$, $\mathbf{S}_\chi$ and $\mathbf{S}_e$ fulfilling the above requirements when DM has spin 1/2~\cite{Dobrescu:2006au,Fan:2010gt,Fitzpatrick:2012ix}.~Specifically, these three-dimensional scalar combinations are operators in the DM/electron spin space.~They correspond to the first fourteen entries in Table~\ref{tab:operators}.~The remaining entries were found to arise from the nonrelativistic reduction of models for spin 1 DM coupling to nucleons via a vector mediator~\cite{Catena:2019hzw}.~We include them in Table~\ref{tab:operators} after replacing nucleon with electron operators.~Based on the above considerations, the amplitude $\mathcal{M}$ can in general be expressed as the linear combination~\footnote{The standard treatment of DM-electron interactions corresponds to $\mathcal{M}(\mathbf{q}) = c_1 F_{\rm DM}(q^2) \,\langle \mathcal{O}_1 \rangle  $, where the so-called DM form factor $F_{\rm DM}$ can be either 1 (contact interaction) or $1/q^2$ (long-range interaction).}

\begin{equation}
 \label{eq:Mnr}
\mathcal{M}(\mathbf{q},\mathbf{v}_{\rm el}^\perp) = \sum_i \left(c_i^s +c^\ell_i \frac{q_{\rm ref}^2}{|\mathbf{q}|^2} \right) \,\langle \mathcal{O}_i \rangle  \,,
\end{equation}
where the reference momentum, $q_{\rm ref}$, is given by $q_{\rm ref}\equiv \alpha m_e$, the interaction operators $\mathcal{O}_i$ are defined in Table~\ref{tab:operators} (with $i$ running on subsets of Table~\ref{tab:operators}, depending on the DM~particle spin), and the coefficients $c_i^s$ ($c_i^\ell$) refer to contact (long-range) interactions.~The case $m^2_{\rm med} \gg |\mathbf{q}|^2$ corresponds to the limit of contact interaction, $c_i^\ell=0$.~In contrast, for $m_{\rm med}^2\ll|\mathbf{q}|^2$ the interaction between DM and electrons is long-range, i.e.~$c_i^s=0$.~Angle brackets in Eq.~(\ref{eq:Mnr}) denote matrix elements between the two-dimensional spinors $\xi^\lambda$ and $\xi^s$ ($\xi^{\lambda'}$ and $\xi^{s'}$) associated with the initial (final) state electron and DM~particle, respectively.~For example, $\langle \mathcal{O}_1 \rangle = \xi^{s'} \xi^{s} \xi^{\lambda'} \xi^\lambda$.~An analogous expansion for $\mathcal{M}$ was found previously in studies of nonrelativistic DM-nucleon interactions (see e.g.~\cite{DelNobile:2018dfg} for a review).

It is important to mention that the distinction between contact and long-range interactions in Eq.~\eqref{eq:Mnr} is not this straight-forward in general.~While t-channel scattering processes mediated by a scalar or vector boson are associated with either long- or short-range interactions depending on the mediator mass, mixed cases are also frequent.~For example, as we shall see in Sec.~\ref{sec:ame}, magnetic dipole interactions, mediated by the massless photon, correspond to a combination of two ``contact'' couplings~$c_i^s$ and two ``long-range'' couplings~$c_i^\ell$.

As an illustrative example, let us compute $\mathcal{M}$ for DM-electron interactions arising from the DM anapole moment and match the result of this calculation to the nonrelativistic expansion in Eq.~(\ref{eq:Mnr}).~The DM anapole moment is potentially important being the leading coupling between DM and photons if DM is made of Majorana fermions.~It generates DM-electron interactions via the electromagnetic current.~The Lagrangian of the model reads
\begin{equation}
    \mathscr{L} = \frac{1}{2}\frac{g}{\Lambda^2} \bar{\chi} \gamma^\mu \gamma^5 \chi\, \partial^\nu F_{\mu\nu} \,,
\end{equation}
where $\chi$ is a Majorana fermion describing the DM~particle, $g$ is a dimensionless coupling constant, $\Lambda$ a mass scale and $F_{\mu\nu}$ the electromagnetic field strength tensor.~The corresponding nonrelativistic scattering amplitude is given by
\begin{eqnarray}
\label{eq:M3}
    \mathcal{M} &=& \frac{4 e g}{\Lambda^2} m_\chi m_e \Bigg\{
    2 \left(\mathbf{v}_{\rm el}^\perp \cdot \xi^{\dagger s'} \mathbf{S}_\chi \xi^s \right) \delta^{\lambda' \lambda}
    \nonumber\\
    &+&
    g_e \left( \xi^{\dagger s'} \mathbf{S}_\chi \xi^s \right) \cdot \left( i\frac{\mathbf{q}}{m_e} \times \xi^{\dagger \lambda'} \mathbf{S}_e \xi^\lambda  \right) 
    \Bigg\}\,,
\end{eqnarray}
where $g_e$ is the electron $g$-factor, while $\xi^\lambda$ and $\xi^s$ are two-component spinors for the electron and DM~particle, respectively.~The first line in Eq.~(\ref{eq:M3}) depends on $\mathbf{v}_{\rm el}^\perp$ and is proportional to $\langle\mathcal{O}_8\rangle$, whereas the second line is linear in $\mathbf{q}$ and proportional to $\langle \mathcal{O}_9 \rangle$.~This example shows that the assumption commonly made in the interpretation of direct detection experiments searching for DM-electron interactions, $\mathcal{M}(\mathbf{q},\mathbf{v}_{\rm el}^\perp)=\mathcal{M}(\mathbf{q})$, is likely restrictive, and contributions to $\mathcal{M}$ depending on $\mathbf{v}_{\rm el}^\perp$ are in general expected.

Let us now calculate the squared transition amplitude assuming that the expansion in Eq.~(\ref{eq:Mnr}) applies.~This will lead us to introduce two {\it atomic form factors} depending on $\psi_1$, $\psi_2$ and $\mathcal{M}$.~We start by noticing that the free electron scattering amplitude in Eq.~(\ref{eq:Mnr}) can also be written as
\begin{align}
    \label{eq:Mnr2}
    \mathcal{M}(\mathbf{q},\mathbf{v}_{\rm el}^\perp) &= \mathcal{M}(\mathbf{q},\mathbf{v}_{\rm el}^\perp)_{\mathbf{k}=0} \nonumber\\ &+ \left(\frac{\mathbf{k}}{m_e}\right)\cdot m_e\nabla_{\mathbf{k}} \mathcal{M}(\mathbf{q},\mathbf{v}_{\rm el}^\perp)_{\mathbf{k}=0}\,,
\end{align}
where we used $\mathbf{v}_{\rm el}^\perp=\mathbf{v}-\mathbf{k}/m_e-\mathbf{q}/(2 \mu_{\chi e})$.~This is not an approximation because Eq.~(\ref{eq:Mnr}) is at most linear in $\vPerpEl$.~Substituting Eq.~(\ref{eq:Mnr2}) into the squared transition amplitude in Eq.~\eqref{eq:transition amplitude}, we find
\begin{widetext}
\begin{align}
    \overline{\left| \mathcal{M}_{1\rightarrow 2}\right|^2}&=\Bigg\{ \overline{| \mathcal{M}(\mathbf{q},\mathbf{v}_{\rm el}^\perp) |^2}   | f_{1\rightarrow 2}(\mathbf{q}) |^2 
+  2m_e\overline{ \Re \left[ \mathcal{M}(\mathbf{q},\mathbf{v}_{\rm el}^\perp) f_{1\rightarrow 2}(\mathbf{q}) \nabla_{\mathbf{k}} \mathcal{M}^*(\mathbf{q},\mathbf{v}_{\rm el}^\perp) \cdot \mathbf{f}^{\,*}_{1\rightarrow 2}(\mathbf{q}) \right]} \nonumber\\ &
+m^2_e\overline{|\nabla_{\mathbf{k}} \mathcal{M}(\mathbf{q},\mathbf{v}_{\rm el}^\perp) \cdot \mathbf{f}_{1\rightarrow 2}(\mathbf{q}) |^2}\Bigg\}_{\mathbf{k}=0}\, . \label{eq:transition amplitude 2}
\end{align}
\end{widetext}
While the first term in Eq.~(\ref{eq:transition amplitude 2}) has already been considered in previous studies, the second and third terms are new, and arise from our general modeling of the free electron scattering amplitude.~Indeed, in the analysis of direct detection experiments searching for DM-electron interaction signals, it is standard to assume that the DM-electron free scattering amplitude $\mathcal{M}$ depends solely on the momentum transfer~$\mathbf{q}$ and not on the velocity $\mathbf{v}_{\rm el}^\perp$ and, therefore, on the electron's initial momentum~$\mathbf{k}$.~In this case, the scattering amplitude $\mathcal{M}(\mathbf{q})$ can be taken out of the integral in Eq.~\eqref{eq:transition rate}, and only the first term in Eq.~(\ref{eq:transition amplitude 2}) contributes to the squared transition amplitude~$\overline{\left| \mathcal{M}_{1\rightarrow 2}\right|^2}$.~The advantage of this standard assumption is that $\mathcal{M}=\mathcal{M}(\mathbf{q})$ implies a convenient factorization of atomic physics and DM physics.~Following the notation of Ref.~\cite{Essig:2015cda}, for $\mathcal{M}=\mathcal{M}(\mathbf{q})$ one can define a {\it scalar atomic form factor},
\begin{align}
	&f_{1\rightarrow 2}(\mathbf{q}) = \int\frac{\dd^3 k}{(2\pi)^3}\psi^*_{2}(\mathbf{k}+\mathbf{q})\psi_{1}(\mathbf{k})\, , \label{eq:standard atomic form factor}
	\end{align}
and obtain the transition amplitude, $\mathcal{M}_{1\rightarrow 2}(\mathbf{q})$, by multiplying the free scattering amplitude by the form factor $f_{1\rightarrow 2}(\mathbf{q})$,
\begin{align}
&\mathcal{M}_{1\rightarrow 2}(\mathbf{q}) = \mathcal{M}(\mathbf{q})\times f_{1\rightarrow 2}(\mathbf{q})\, .
\end{align}
Since the momentum space wave functions $\psi_1$ and $\psi_2$ have dimension [energy]$^{-3/2}$,~$f_{1\rightarrow 2}(\mathbf{q})$ is a dimensionless quantity.

Besides the well-known scalar atomic form factor of Eq.~\eqref{eq:standard atomic form factor}, a new, {\it vectorial atomic form factor} appears in Eq.~(\ref{eq:transition amplitude 2}),
\begin{align}
\label{eq:new atomic form factor}
	\mathbf{f}_{1\rightarrow 2}(\mathbf{q})&= \int \frac{\dd^3 k}{(2\pi)^3}\psi^*_2(\mathbf{k}+\mathbf{q})\,\left(\frac{\mathbf{k}}{m_e}\right)\,\psi_1(\mathbf{k})\, .
\end{align}
The details of the evaluation of the scalar, $f_{1\rightarrow 2}$, and vectorial, $\mathbf{f}_{1\rightarrow 2}$, atomic form factors are presented in Appendix~\ref{app: atomic form factors}.

\subsection{Ionization of isolated atoms in argon and xenon targets}
\label{sec:targets}
Our treatment of DM-electron scattering in target materials has so far been general in terms of initial and final state electron wave functions, $\psi_1$ and $\psi_2$, respectively.~From now onward, however, we specialize our results to the case of DM-induced ionization of isolated atoms.~In this case the initial state (formerly simply denoted by ``1'') is a bound state characterized by the principal, angular and magnetic quantum numbers~$(n,\ell,m)$.~The final state (``2'') is a free electron state at large distance from the atom, but still affected by the remaining ion's presence at low distance.~It is defined by the quantum numbers $(k^\prime,\ell^\prime,m^\prime)$, where~$k^\prime$ is the asymptotic momentum of the electron and $\ell^\prime,m^\prime$ are its angular and magnetic quantum numbers.~We obtain the total ionization rate~$\mathscr{R}_{\rm ion}^{n\ell}$ of a full atomic orbital~$(n,\ell)$ by summing the transition rate $\mathscr{R}_{1\rightarrow 2}$ over all occupied initial electron states and integrating over the allowed final electron states. 

For the initial bound electrons, the rate has to be summed over all values of the magnetic quantum number~$m$ and multiplied by~2 to account for the spin degeneracy.~In order to account for all the allowed electron final states, we have to act on $\mathscr{R}_{1\rightarrow 2}$ with the integral operator~\cite{Essig:2015cda}
\begin{align}
\frac{V}{2}\sum_{\ell^\prime=0}^\infty\sum_{m^\prime=-\ell^\prime}^{\ell^\prime}\int \frac{k^{\prime 3} \dd \ln E_e}{(2\pi)^3}\,.\label{eq: final state phase space}
\end{align}
Here, $E_e = k^{\prime 2}/(2m_e)$ is the ionized electron's final energy, and the number of final states with asymptotic momentum between  $\mathbf{k}'$ and $\mathbf{k}'+{\rm d}\mathbf{k}'$ is $V{\rm d}^3k'/(2\pi)^3$.~Summarizing, the total ionization rate for the~$(n,\ell)$ orbital is given by
\begin{align}
    \mathscr{R}_{\rm ion}^{n\ell} &= \sum_{m=-\ell}^\ell \sum_{\ell^\prime=0}^\infty\sum_{m^\prime=-\ell^\prime}^{\ell^\prime}\int\dd \ln E_e \frac{V k^{\prime 3}}{(2\pi)^3} \mathscr{R}_{1\rightarrow 2}\, ,
\end{align}
and the corresponding final state electron ionization energy spectrum by
\begin{align}
    \frac{\dd\mathscr{R}_{\rm ion}^{n\ell}}{\dd\ln E_e} &=\sum_{m=-\ell}^\ell \sum_{\ell^\prime=0}^\infty\sum_{m^\prime=-\ell^\prime}^{\ell^\prime} \frac{V k^{\prime 3}}{(2\pi)^3} \mathscr{R}_{1\rightarrow 2}\, . \label{eq:ionization spectrum R1->2}
\end{align}
The divergent factor $V$ in Eq.~(\ref{eq:ionization spectrum R1->2}) cancels with the $1/V$ factor arising from the normalization of $\psi_{k'm'\ell'}$ (see Appendix~\ref{app: wave functions}).~Substituting Eq.~\eqref{eq:transition rate}, we can rewrite the spectrum as
\begin{align}
    \frac{\dd\mathscr{R}_{\rm ion}^{n\ell}}{\dd\ln E_e}&= \frac{n_{\chi}}{128 \pi \,m^2_{\chi} m^2_e}\nonumber\\
&\times \int \dd q \; q \int \frac{{\rm d}^3 v}{v} \,f_{\chi}(\mathbf{v}) \Theta(v-v_{\rm min}) \overline{\left| \mathcal{M}^{n\ell}_{\rm ion}\right|^2}\, , \label{eq:ionization spectrum}
\end{align}
where, following~\cite{Essig:2015cda}, we integrated over $\cos\theta_{qv}$ while assuming that $f_\chi(\mathbf{v})=f_\chi(v)$, and then replaced $f_\chi(v)$ with $f_\chi(\mathbf{v})$ in the final expression\footnote{This approach significantly simplifies the evaluation of Eq.~(\ref{eq:transition rate}) and is justified by the fact that we are not interested in a directional analysis of the predicted signal.~If Eq.~(\ref{eq:transition rate}) is to be used in directional analyses of energy spectra, this simplified treatment of the velocity integral should be refined along the lines discussed in Ref.~\cite{Geilhufe:2019ndy}.}.

\begin{figure*}
    \centering
    \includegraphics[width=0.48\textwidth]{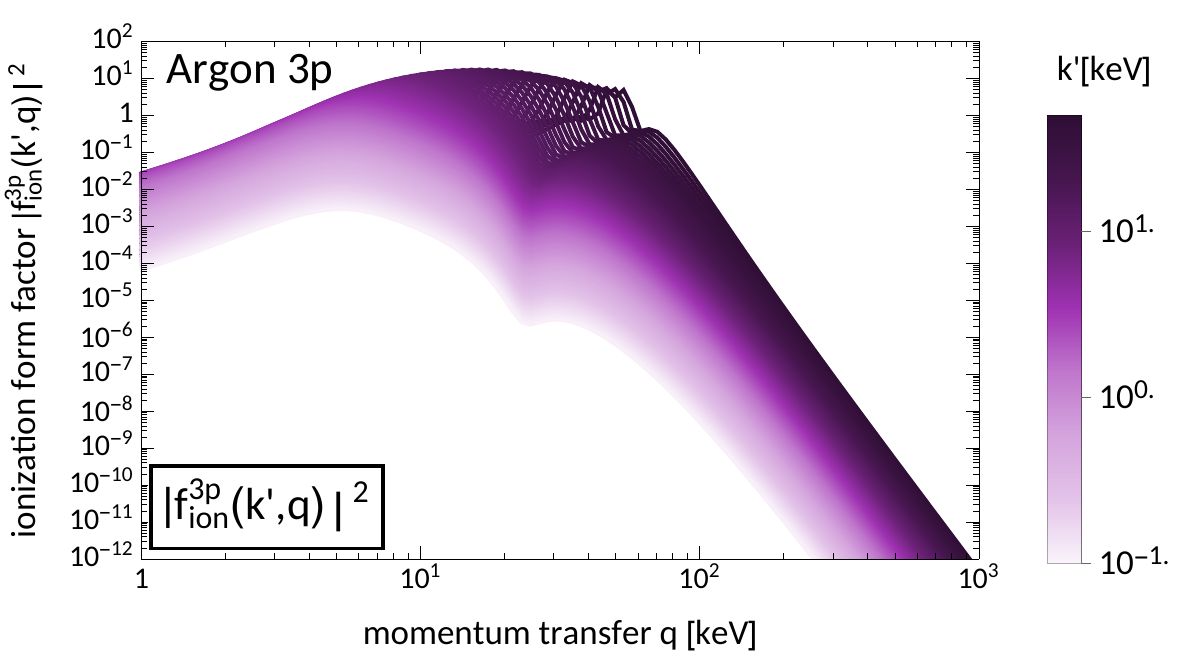}
    \includegraphics[width=0.48\textwidth]{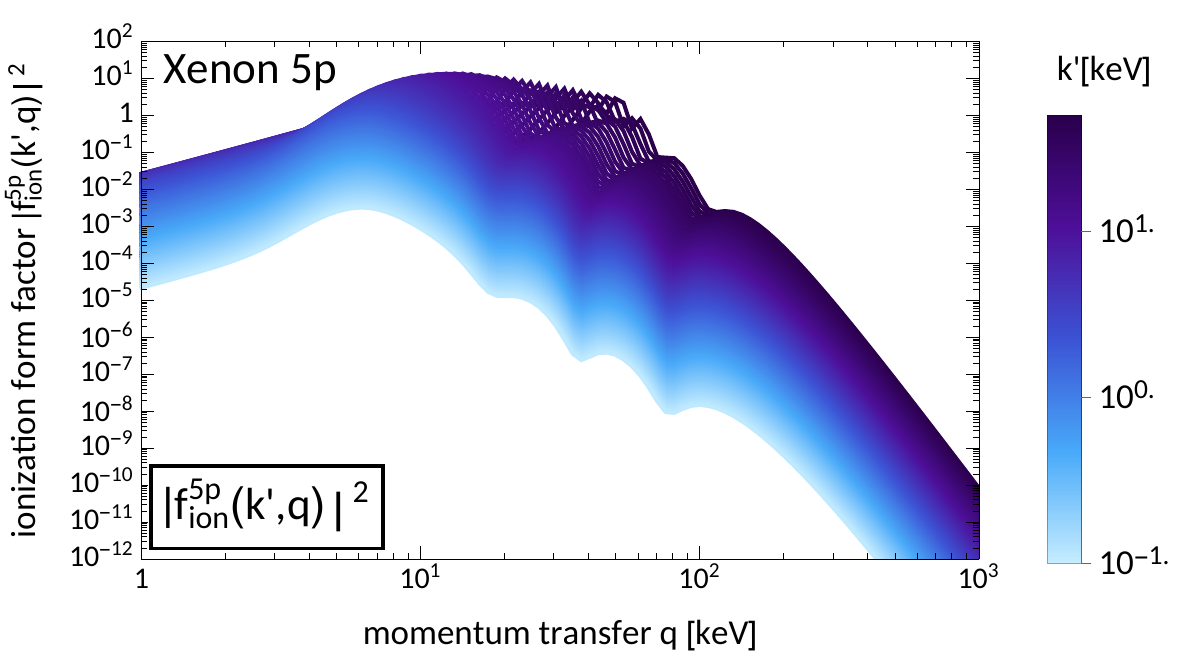}
    \caption{The ionization form factor~$\left|f^{n\ell}_{\rm ion}(k^\prime,q)\right|^2$ defined in Eq.~\eqref{eq:standard ionization form factor} for the outer electron orbital of argon (left) and xenon (right).}
    \label{fig: standard ionization form factor}
\end{figure*}

In order to emphasise the connection to previous work~\cite{Essig:2011nj,Essig:2012yx,Essig:2017kqs}, we can extract a differential ionization cross section from Eq.~\eqref{eq:ionization spectrum}, i.e., the number of events per unity of flux,
\footnote{In the derivation of Eq.~(\ref{eq:transition rate}) we refrained from introducing a scattering cross section.
 Properly defining a scattering cross section between an incoming DM~particle and a bound electron, not being in a momentum eigenstate, is hampered by the fact that the relative velocity is not well defined. Instead one would have to treat the whole atom as a multi-particle target, whose center of mass is indeed in a total momentum eigenstate. }
\begin{align}
    \frac{\dd\sigma_{\rm ion}^{n\ell}}{\dd\ln E_e}&= \frac{1}{128\pi m_\chi^2 m_e^2 v^2} \int \dd q\; q\, \overline{\left| \mathcal{M}^{n\ell}_{\rm ion}\right|^2}\, ,
    \end{align}
such that
\begin{align}
   \frac{\dd\mathscr{R}_{\rm ion}^{n\ell}}{\dd\ln E_e}&=  n_{\chi} \int \dd^3v \,v\,f_{\chi}(\mathbf{v})   \Theta(v-v_{\rm min})\frac{\dd\sigma_{\rm ion}^{n\ell}}{\dd\ln E_e}\, .
\end{align}

The {\it squared ionization amplitude}, $\overline{\left| \mathcal{M}^{n\ell}_{\rm ion}\right|^2}$, appearing in Eq.~(\ref{eq:ionization spectrum}) is defined as follows 
\begin{align}
    \overline{\left| \mathcal{M}^{n\ell}_{\rm ion}\right|^2} &\equiv V \frac{4k^{\prime 3}}{(2\pi)^3} \sum_{m = -\ell}^\ell \sum_{\ell^\prime=0}^\infty \sum_{m^\prime = -\ell^\prime}^{\ell^\prime} \overline{\left| \mathcal{M}_{1\rightarrow 2}\right|^2}\, , \label{eq:ionization amplitude}
\end{align}
and can explicitly be expressed in terms of the amplitude $\mathcal{M}(\mathbf{q},\mathbf{v}_{\rm el}^\perp)$,
\begin{widetext}
\begin{align}
    \overline{\left| \mathcal{M}^{n\ell}_{\rm ion}\right|^2} &= \Bigg\{ \overline{| \mathcal{M}(\mathbf{q},\mathbf{v}_{\rm el}^\perp) |^2} \left|f_{\rm ion}^{n\ell}(k^\prime,q)\right|^2
    +V \frac{4k^{\prime 3}}{(2\pi)^3} \sum_{m = -\ell}^\ell \sum_{\ell^\prime=0}^\infty \sum_{m^\prime = -\ell^\prime}^{\ell^\prime}\Bigg[ 2m_e\overline{ \Re \left[ \mathcal{M}(\mathbf{q},\mathbf{v}_{\rm el}^\perp) f_{1\rightarrow 2}(\mathbf{q})  \nabla_{\mathbf{k}}  \mathcal{M}^*(\mathbf{q},\mathbf{v}_{\rm el}^\perp) \cdot \mathbf{f}^*_{1\rightarrow 2}(\mathbf{q})  \right]} \nonumber\\
&+m_e^2 \overline{|\nabla_{\mathbf{k}} \mathcal{M}(\mathbf{q},\mathbf{v}_{\rm el}^\perp) \cdot \mathbf{f}_{1\rightarrow 2}(\mathbf{q}) |^2}\Bigg]\Bigg\}_{\mathbf{k}=0;\, \mathbf{v}\cdot \mathbf{q}/(qv)=\xi}\, .\label{eq:ionization amplitude 2}
\end{align}
\end{widetext}
Here, we defined $\xi=\Delta E_{1\rightarrow 2}/(qv)+q/(2m_\chi v)$.~To further clarify the connection to the results and notation of previous works, in Eq.~(\ref{eq:ionization amplitude 2}) we introduced the dimensionless {\it ionization form factor},
\begin{align}
    \left|f_{\rm ion}^{n\ell}(k^\prime,q)\right|^2 = V\frac{4k^{\prime 3}}{(2\pi)^3} \sum_{\ell^\prime=0}^\infty \sum_{m = -\ell}^\ell \sum_{m^\prime = -\ell^\prime}^{\ell^\prime} \left|f_{1\rightarrow 2}(q)\right|^2 \label{eq:standard ionization form factor}\, .
\end{align}
This ionization form factor is depicted in Fig.~\ref{fig: standard ionization form factor} for the outer atomic orbitals of argon and xenon. 

In Eq.~(\ref{eq:ionization spectrum}), the minimum DM speed, $v_{\rm min}$, required to deposit an energy of~$\Delta E_{1\rightarrow 2}$ given a momentum transfer~$q$ is
\begin{align}
    v_{\rm min} = \frac{\Delta E_{1\rightarrow 2}}{q} + \frac{q}{2 m_\chi}\,. \label{eq: vmin}
\end{align}
Finally, the wave functions of the initial and final state electrons can be expanded in spherical harmonics,
\begin{align}
    \psi_{n\ell m}(\mathbf{x}) &= R_{n\ell}(r)\,Y_{\ell}^m(\theta,\phi)\, ,\label{eq: wave function general}\\
    \psi_{k^\prime\ell^\prime m^\prime}(\mathbf{x}) &= R_{k^\prime\ell^\prime}(r)\,Y_{\ell^\prime}^{m^\prime}(\theta,\phi)\label{eq: wave function general final}\, .
\end{align}
Their radial parts, $R_{n\ell}(r)$ and $R_{k'\ell'}(r)$, are given in Appendix~\ref{app: wave functions} for the case of isolated argon and xenon atoms~\cite{Bunge:1993jsz}.~While these wave functions have also been used in previous works on DM direct detection~\cite{Agnes:2018oej}, we note that they are not fully applicable to dense liquid argon and xenon systems~\cite{Sorensen:2011bd}.~Our neglect of the difference in electronic structure between the isolated atoms and the liquid state makes our results conservative since the electron binding energies in liquid nobles are smaller than those of the isolated atoms.~Furthermore, this treatment of the wave functions can be improved by including relativistic corrections~\cite{Roberts:2015lga,Roberts:2016xfw} and many-body effects~\cite{Pandey:2018esq}.~However, both improvements, in particular the relativistic corrections, are expected to have limited impact on theoretical predictions for sub-GeV DM. 

\section{General dark matter and atomic response functions}
\label{sec:FF}

\begin{figure*}
\centering
\includegraphics[width=0.48\textwidth]{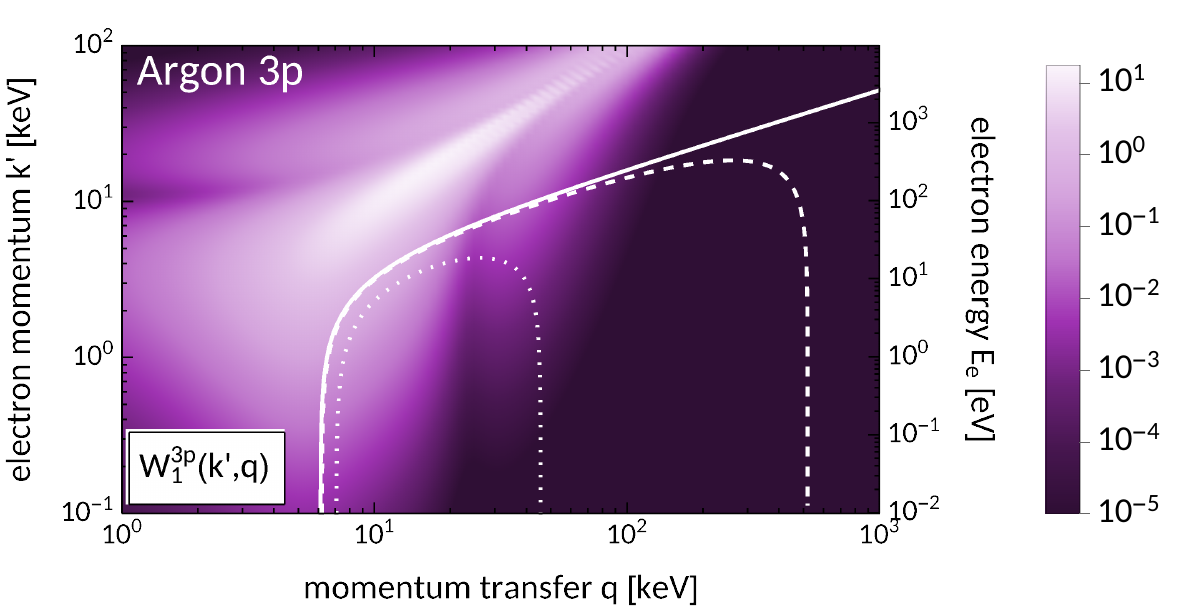}
\includegraphics[width=0.48\textwidth]{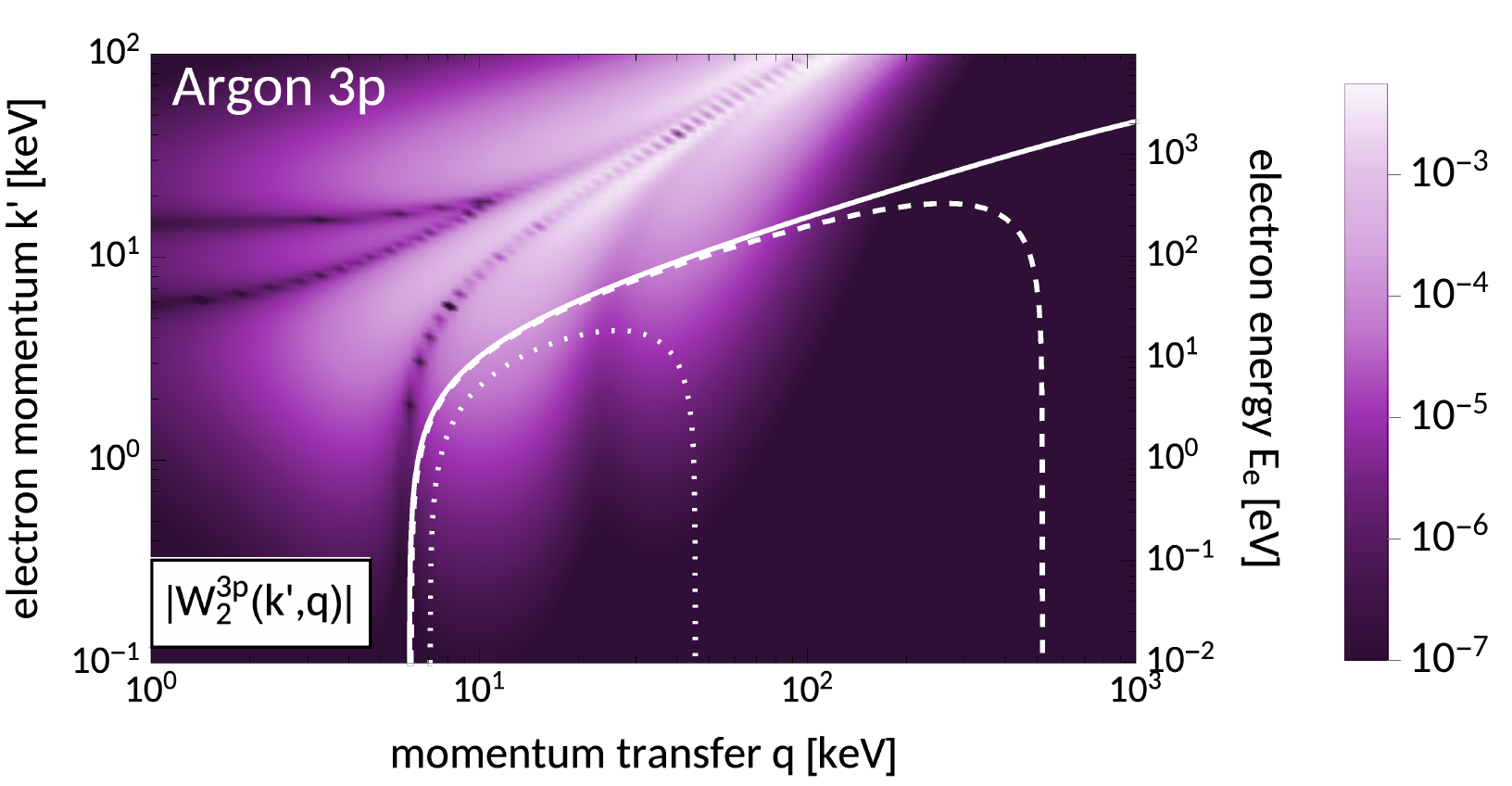}
\includegraphics[width=0.48\textwidth]{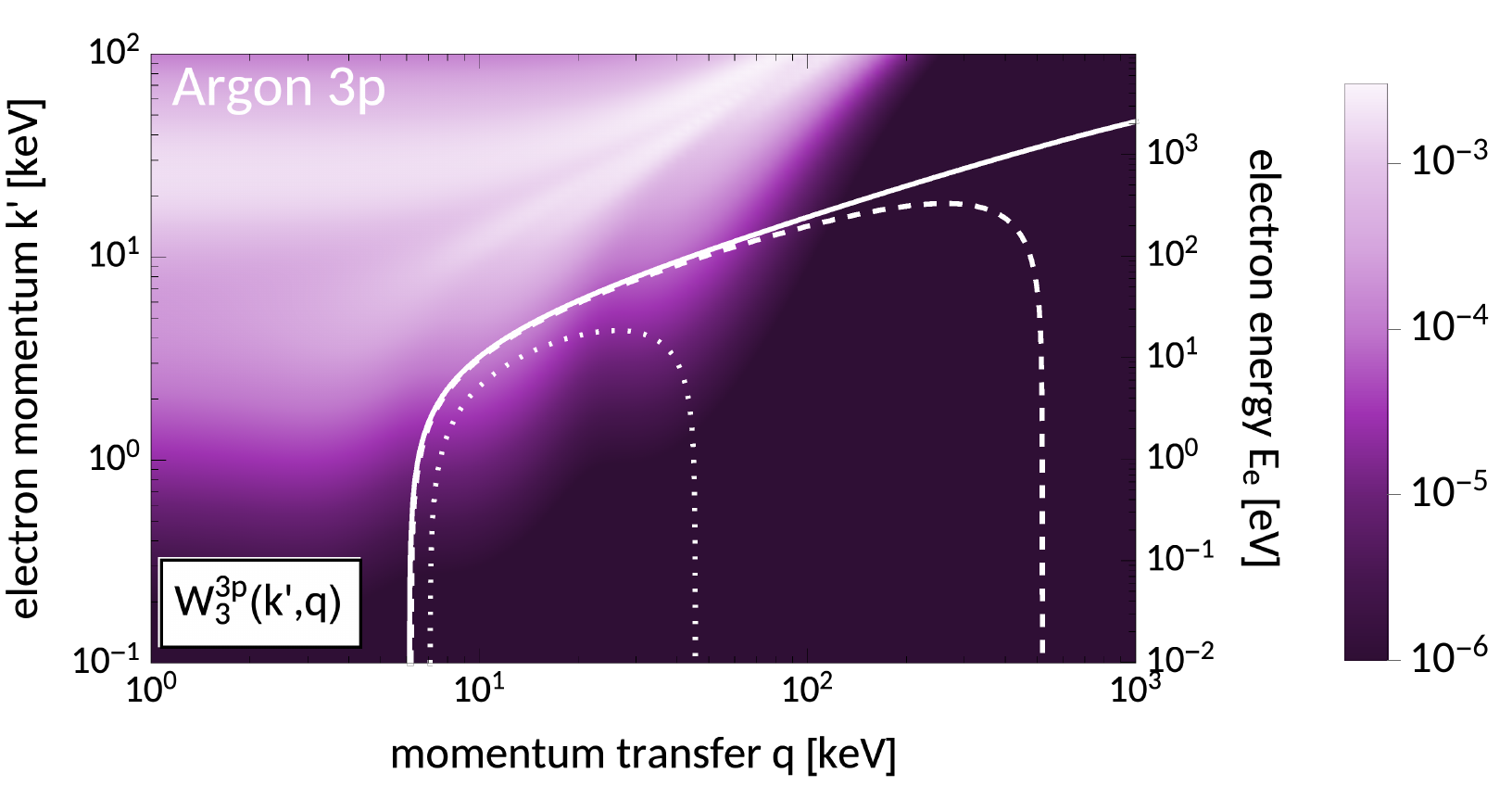}
\includegraphics[width=0.48\textwidth]{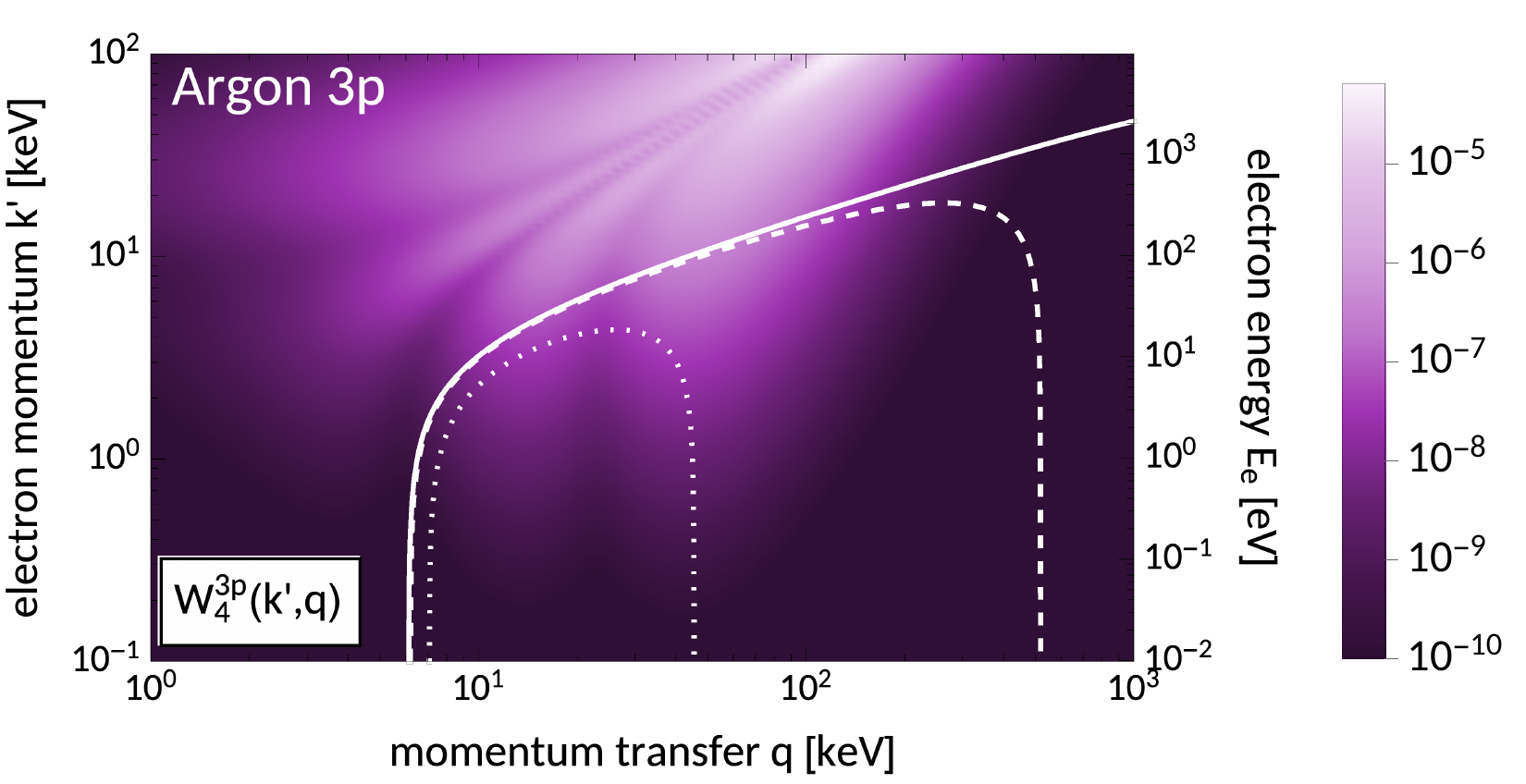}
\caption{The four atomic responses for the outer atomic orbital of argon. Above the white dotted/dashed/solid line on the left plots, the minimum speed~$v_{\rm min}$ in Eq.~\eqref{eq: vmin} exceeds the maximum speed~$v_{\rm max}=v_\oplus + v_{\rm esc}$ for a DM mass of 10~MeV/100~MeV/$\rightarrow\infty$, respectively (see Appendix~\ref{app: kinematics} and~\ref{app: experiments}). The top left panel shows the same function as the left panel of Fig.~\ref{fig: standard ionization form factor}, but now with the final state electron asymptotic momentum $k'$ on the $y$-axis. Note the different color bar scales in the four panels.}
\label{fig: atomic responses argon}
\end{figure*}

In this section, we reformulate the expressions found in Sec.~\ref{sec:targets} in order to obtain compact equations which are suitable for numerical applications.~By doing so, we also investigate and characterize the atomic response to DM-electron interactions of the general form considered here.~We start by observing that the squared ionization amplitude, $\overline{| \mathcal{M}_{\rm ion}^{n\ell}|^2}$, can in general be rewritten as follows:
\begin{align}
\label{eq:RW}
   \overline{| \mathcal{M}_{\rm ion}^{n\ell}|^2}= \sum_{i=1}^4    R^{n\ell}_i\left(\vPerpEl,\frac{\mathbf{q}}{m_e}\right) W_{i}^{n\ell}(k^\prime,\mathbf{q})\, .
\end{align}
Here, we define a set of DM~response functions~$R^{n\ell}_i\left(\vPerpEl,\frac{\mathbf{q}}{m_e}\right)$, which are functions of the couplings in Eq.~\eqref{eq:Mnr}, and atomic response functions~$W_{i}^{n\ell}(k^\prime,\mathbf{q})$, which encapsulate the information on the electron's initial and final state wave functions.\footnote{This factorization is analogous to that introduced previously, e.g.,~in Ref.~\cite{Fitzpatrick:2012ix}, in the treatment of nonrelativistic DM-nucleus scattering~\cite{Fitzpatrick:2012ix}.}~By means of Eq.~(\ref{eq:RW}), we are again able to disentangle the particle physics input from that of the atomic physics.~However, our general treatment of DM-electron interactions predicts not one but four atomic response functions.
At this point, we summarize the final expressions for the DM and atomic response functions.~Their detailed derivations can be found in Appendixes~\ref{app: response derivation} and~\ref{app: atomic responses}.

The four DM~response functions, $R^{n\ell}_i\left(\vPerpEl,\frac{\mathbf{q}}{m_e}\right)$, are given by
\begin{widetext}
\begin{subequations}
\label{eq: DM response functions}
\begin{align}
   R^{n\ell}_1\left(\vPerpEl,\frac{\mathbf{q}}{m_e}\right) &\equiv c_1^2 +\frac{c_3^2}{4}\left( \frac{\mathbf{q}}{m_e}\right)^2 (\vPerpEl)^2 -\frac{c_3^2}{4} \left(\frac{\mathbf{q}}{m_e}\cdot \vPerpEl\right)^2 +\frac{c_7^2}{4}(\vPerpEl)^2 + \frac{c_{10}^2}{4}\left(\frac{\mathbf{q}}{m_e}\right)^2   \nonumber\\
   &+ \frac{j_\chi(j_\chi+1)}{12}\Bigg\{ 3c_4^2 +c_6^2 \left(\frac{\mathbf{q}}{m_e}\right)^4 +(4c_8^2+2c_{12}^2)(\vPerpEl)^2 +(2c_9^2+4c_{11}^2+2c_4c_6)\left(\frac{\mathbf{q}}{m_e}\right)^2\nonumber \\
   &+\left(4c_5^2+c_{13}^2+c_{14}^2-2c_{12}c_{15}\right)\left(\frac{\mathbf{q}}{m_e}\right)^2(\vPerpEl)^2+c_{15}^2\left(\frac{\mathbf{q}}{m_e}\right)^4 \left(\vPerpEl\right)^2\nonumber\\
   &-c_{15}^2 \left(\frac{\mathbf{q}}{m_e}\right)^2 \left(\vPerpEl\cdot \frac{\mathbf{q}}{m_e}\right)^2+\left(-4c_5^2+2c_{13}c_{14}+2c_{12}c_{15}\right)\left(\vPerpEl\cdot\frac{\mathbf{q}}{m_e}\right)^2\Bigg\}\, ,\label{eq: DM response 1}\\
R^{n\ell}_2\left(\vPerpEl,\frac{\mathbf{q}}{m_e}\right) &\equiv\left(\frac{\mathbf{q}}{m_e}\cdot \vPerpEl\right)\left[-\frac{c_7^2}{2}\left(\frac{\mathbf{q}}{m_e}\right)^{-2} - \frac{j_\chi(j_\chi+1)}{6}\left\{(4c_8^2+2c_{12}^2)\left(\frac{\mathbf{q}}{m_e}\right)^{-2}+(c_{13}+c_{14})^2\right\}\right]\, ,\label{eq: DM response 2}\\
   R^{n\ell}_3\left(\vPerpEl,\frac{\mathbf{q}}{m_e}\right) &\equiv\frac{c_3^2}{4}\left(\frac{\mathbf{q}}{m_e}\right)^{2}+\frac{c_7^2}{4}+\frac{j_\chi(j_\chi+1)}{12}\left\{4c_8^2+2c_{12}^2+\left(4c_5^2+c_{13}^2+c_{14}^2-2c_{12}c_{15}\right)\left(\frac{\mathbf{q}}{m_e}\right)^{2}+c_{15}^2\left(\frac{\mathbf{q}}{m_e}\right)^{4}\right\} \, ,\label{eq: DM response 3}\\
   R^{n\ell}_4\left(\vPerpEl,\frac{\mathbf{q}}{m_e}\right) &\equiv -\frac{c_3^2}{4}+\frac{j_\chi(j_\chi+1)}{12}\left\{-4c_5^2-c_{15}^2\left(\frac{\mathbf{q}}{m_e}\right)^{2}+2c_{12}c_{15}+2c_{13}c_{14}\right\} \label{eq: DM response 4}\, ,
\end{align}
\end{subequations}
\end{widetext}
where $c_i\equiv c_i^s + (q_{\rm ref}^2/ |\mathbf{q}|^2)c_i^\ell$.~They indirectly depend on the $(n,\ell)$ quantum numbers through the variable $\xi=\Delta E_{1\rightarrow 2}/(qv)+q/(2m_\chi v)$, see also Appendix~\ref{app: kinematics}.~In the above expressions, ~$\vPerpEl$ is evaluated at $\mathbf{k}=\mathbf{0}$ due to the nonrelativistic expansion of $\mathcal{M}$.~The four atomic response functions $W_{i}^{n\ell}(k^\prime,\mathbf{q})$, which we define next in Eq.~(\ref{eq: atomic responses}), are one of the main results of this work: 
\begin{subequations}
\label{eq: atomic responses}
\begin{align}
    W_{1}^{n\ell}(k^\prime,\mathbf{q}) &\equiv V\frac{4k^{\prime 3}}{(2\pi)^3} \sum_{m = -\ell}^\ell \sum_{\ell^\prime=0}^\infty \sum_{m^\prime = -\ell^\prime}^{\ell^\prime} \left|f_{1\rightarrow 2}(q)\right|^2 \, ,\label{eq: atomic response 1}\\
    W_{2}^{n\ell}(k^\prime,\mathbf{q}) &\equiv V\frac{4k^{\prime 3}}{(2\pi)^3} \nonumber\\
    &\times\sum_{m = -\ell}^\ell \sum_{\ell^\prime=0}^\infty \sum_{m^\prime = -\ell^\prime}^{\ell^\prime}\frac{\mathbf{q}}{m_e}\cdot f_{1\rightarrow 2}(\mathbf{q})\mathbf{f}^{\,*}_{1\rightarrow 2}(\mathbf{q}) \, ,\label{eq: atomic response 2}\\
    W_{3}^{n\ell}(k^\prime,\mathbf{q}) &\equiv V\frac{4k^{\prime 3}}{(2\pi)^3} \sum_{m = -\ell}^\ell \sum_{\ell^\prime=0}^\infty \sum_{m^\prime = -\ell^\prime}^{\ell^\prime} |\mathbf{f}_{1\rightarrow 2}(\mathbf{q})|^2\, ,\label{eq: atomic response 3}\\
    W_{4}^{n\ell}(k^\prime,\mathbf{q}) &\equiv V\frac{4k^{\prime 3}}{(2\pi)^3}\nonumber\\
    &\times\sum_{m = -\ell}^\ell \sum_{\ell  ^\prime=0}^\infty \sum_{m^\prime = -\ell^\prime}^{\ell^\prime} \left|\frac{\mathbf{q}}{m_e}\cdot\mathbf{f}_{1\rightarrow 2}(\mathbf{q})\right|^2\, .\label{eq: atomic response 4}
\end{align}
\end{subequations}

\begin{figure*}
\centering
\includegraphics[width=0.48\textwidth]{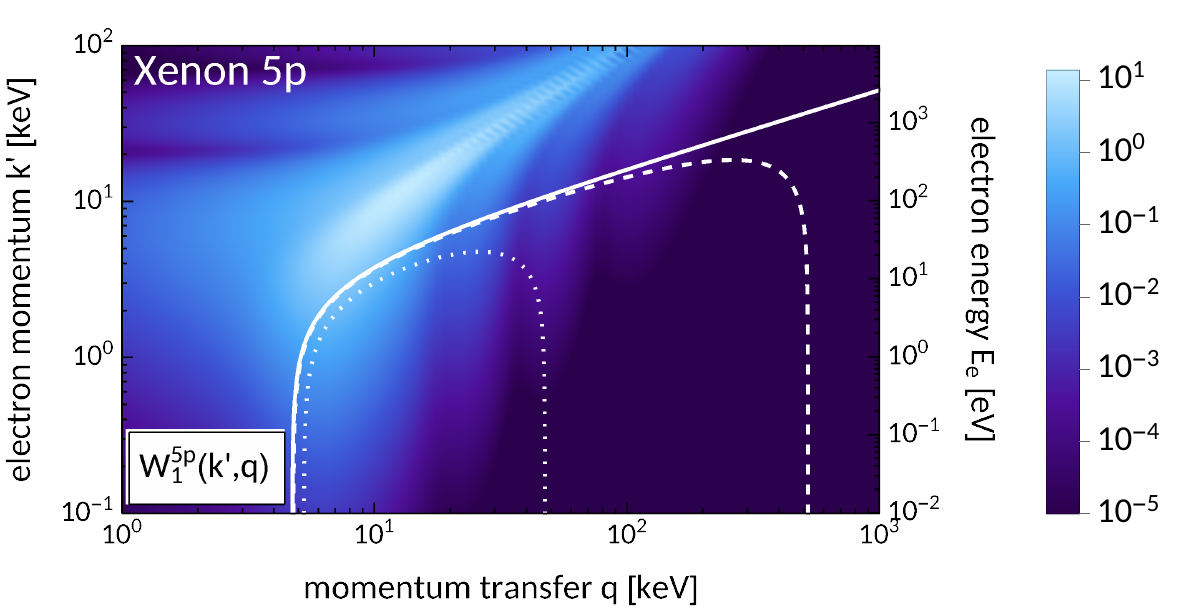}
\includegraphics[width=0.48\textwidth]{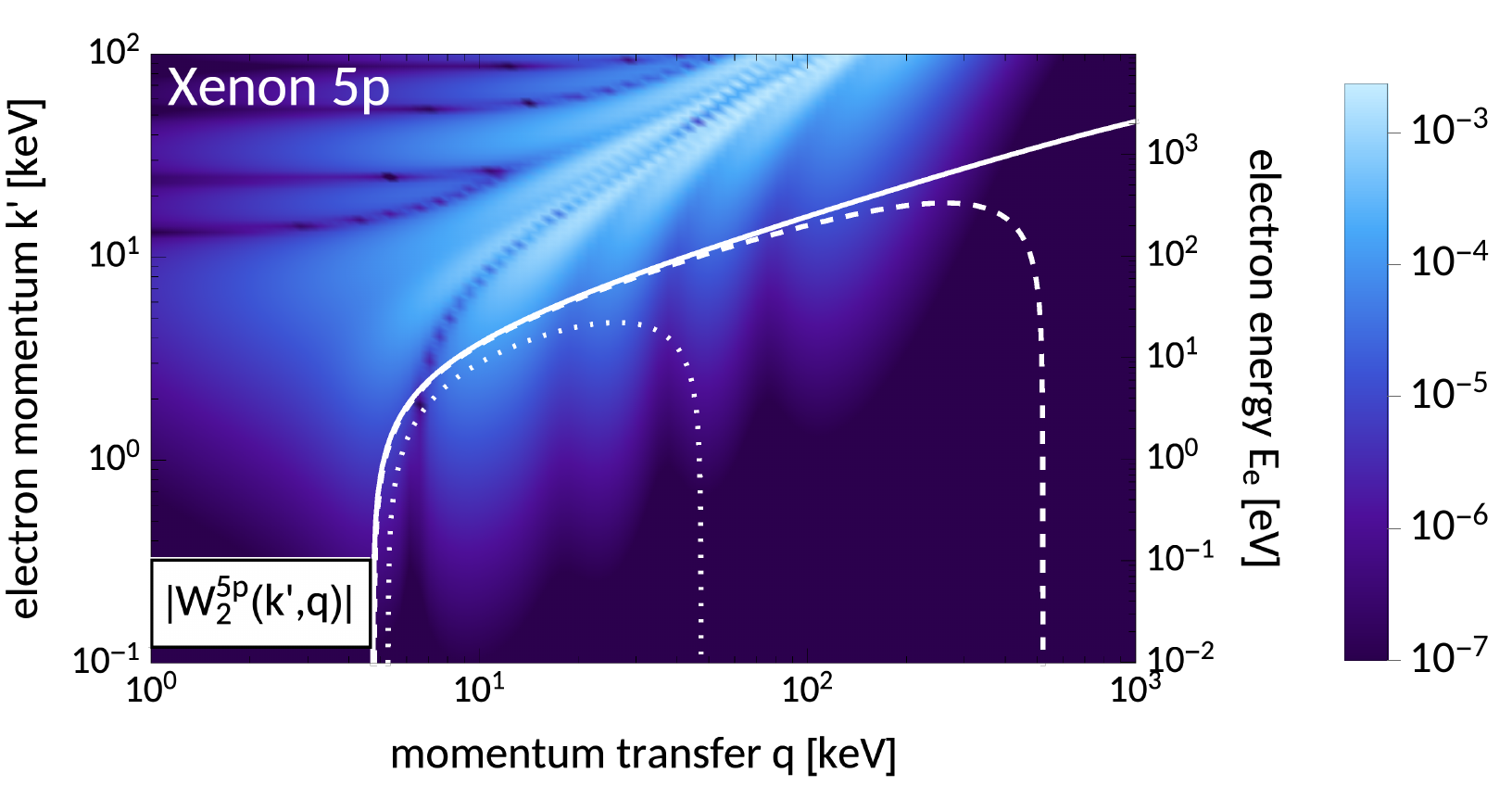}
\includegraphics[width=0.48\textwidth]{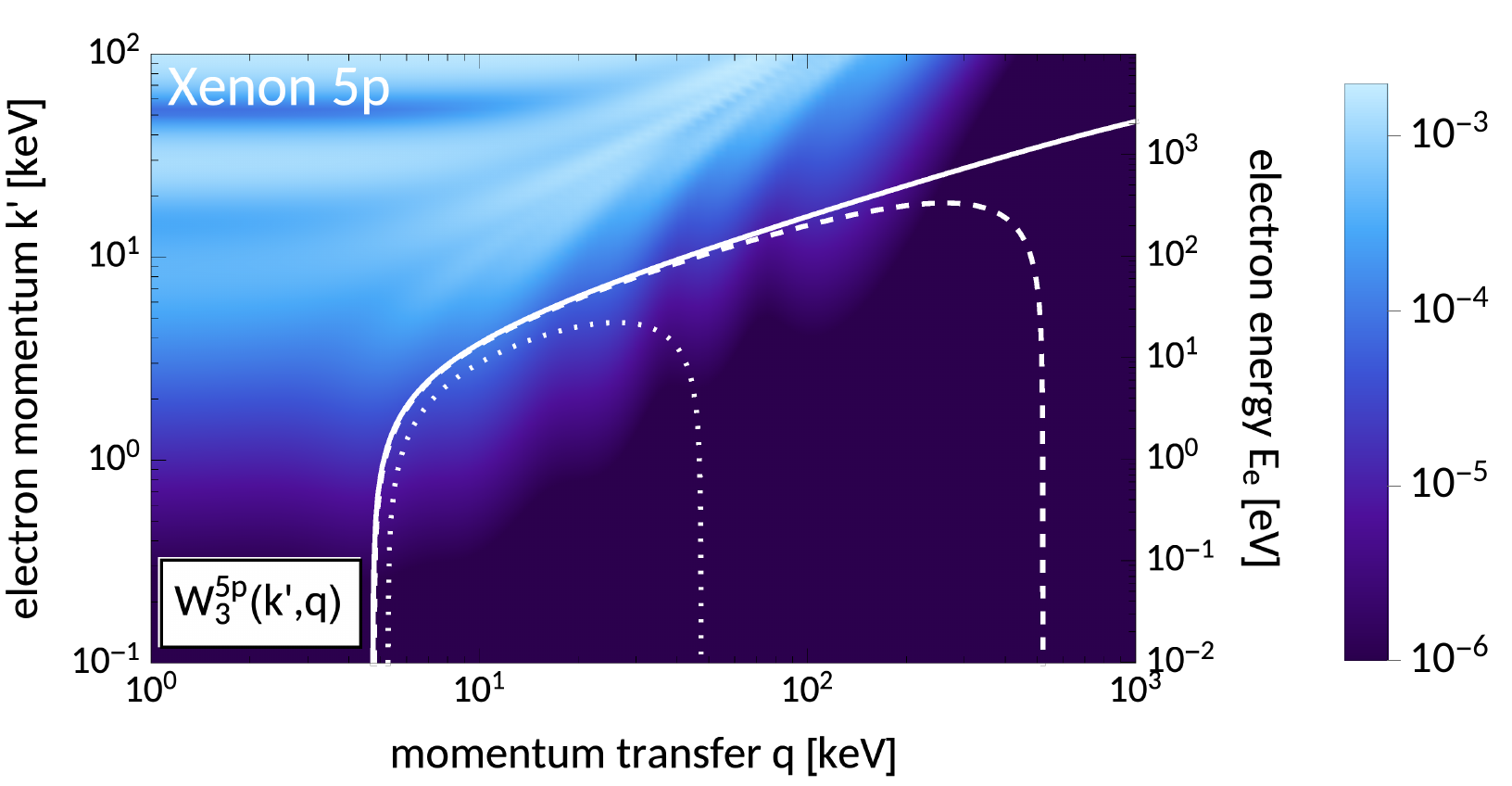}
\includegraphics[width=0.48\textwidth]{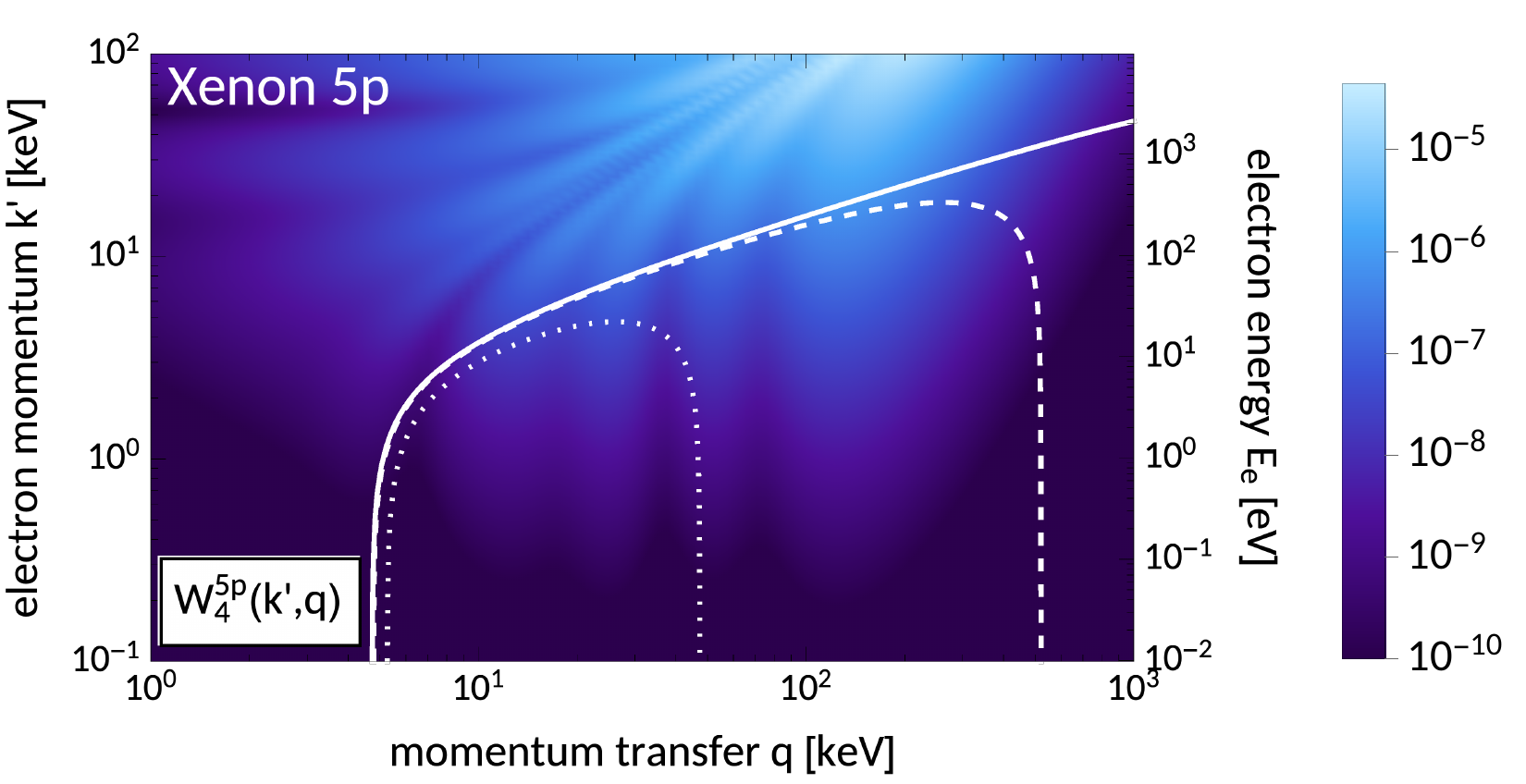}
\caption{The four atomic responses for the outer atomic orbital of xenon. The white lines have the same meaning as in Fig.~\ref{fig: atomic responses argon}. Again, the function in the top left panel was shown before in the right panel of Fig.~\ref{fig: standard ionization form factor}.~However, here the final state electron asymptotic momentum $k'$ is on the $y$-axis.}
\label{fig: atomic responses xenon}
\end{figure*}

The first atomic response function~$W_{1}^{n\ell}(k^\prime,\mathbf{q})$ can be identified with the ionization form factor $\left|f^{n\ell}_{\rm ion}(k^\prime,q)\right|^2$ commonly used in the sub-GeV DM detection literature~\cite{Essig:2015cda}.~The atomic response functions $W_{j}^{n\ell}(k^\prime,\mathbf{q})$, $j=2,3,4$ were not considered in previous studies and can only arise from a general treatment of DM-electron interactions.~While  $W_{4}^{n\ell}/W_{1}^{n\ell}$ roughly scales as particle momenta to the power of 4, and $W_{2,3}^{n\ell}/W_{1}^{n\ell}$ as particle momenta to the power of 2, numerically we find that the products $R_{j}^{n\ell}W_{j}^{n\ell}$ are often of comparable magnitude for several DM-interaction types (see below for further details).

In order to clarify the physical meaning of the new atomic responses, let us evaluate them in the plane-wave limit, where $\psi_1(\mathbf{x})=\exp(i\mathbf{k}\cdot \mathbf{x})/\sqrt{V}$ and $\psi_2(\mathbf{x})=\exp(i\mathbf{k}'\cdot \mathbf{x})/\sqrt{V}$ are eigenstates of the electron momentum operator with eigenvalues $\mathbf{k}$ and $\mathbf{k}'$, respectively.~In this limit, initial and final state electron are not bound to an atom and propagate as free particles.~Substituting $\psi_1(\mathbf{x})$ and $\psi_2(\mathbf{x})$ in Eqs.~(\ref{eq: atomic response 1})-(\ref{eq: atomic response 4}), we find
\begin{align}
|f_{1\rightarrow 2}(\mathbf{q})|^2 &= \frac{(2\pi)^3}{V}\delta^{(3)}(\mathbf{q}+\mathbf{k}-\mathbf{k}') \nonumber\\
\frac{\mathbf{q}}{m_e}\cdot f_{1\rightarrow2}(\mathbf{q})\mathbf{f}_{1\rightarrow2}(\mathbf{q})  &= -\left(\frac{\mathbf{k}}{m_e}\cdot\frac{\mathbf{q}}{m_e}\right)\nonumber\\
&\times\frac{(2\pi)^3}{V}\delta^{(3)}(\mathbf{q}+\mathbf{k}-\mathbf{k}') \nonumber\\
|\mathbf{f}_{1\rightarrow 2}(\mathbf{q})|^2 &= \left|\frac{\mathbf{k}}{m_e}\right|^2\frac{(2\pi)^3}{V}\delta^{(3)}(\mathbf{q}+\mathbf{k}-\mathbf{k}') \nonumber\\ 
\left|\frac{\mathbf{q}}{m_e}\cdot\mathbf{f}_{1\rightarrow 2}(\mathbf{q})\right|^2 &= \left|\frac{\mathbf{k}}{m_e}\cdot\frac{\mathbf{q}}{m_e}\right|^2\nonumber\\
&\times\frac{(2\pi)^3}{V}\delta^{(3)}(\mathbf{q}+\mathbf{k}-\mathbf{k}') \,.
\end{align}
In the plane-wave limit, we can define a laboratory frame in which the initial electron is at rest and, therefore, $\mathbf{k}=0$.~In this frame, $W_{j}^{n\ell}(k^\prime,\mathbf{q})=0$, $j=2,3,4$ and Eq.~(\ref{eq:RW}) reduces to the expression for the modulus squared of the amplitude for DM scattering by a point-like proton found in~\cite{Anand:2013yka} (with the proton mass replaced by the electron mass).

Consequently, the new atomic responses describe distortions in the ionization spectrum induced by the fact that the initial state electron obeys a momentum distribution with a finite dispersion, being the electron bound to an atom.   

Figures~\ref{fig: atomic responses argon} and~\ref{fig: atomic responses xenon} show the four atomic response functions, $W_j^{n\ell}$, $j=1,\dots,4$, for the 3p and 5p atomic orbitals of argon and xenon, respectively.

\begin{figure*}
\centering
\includegraphics[width=0.45\textwidth]{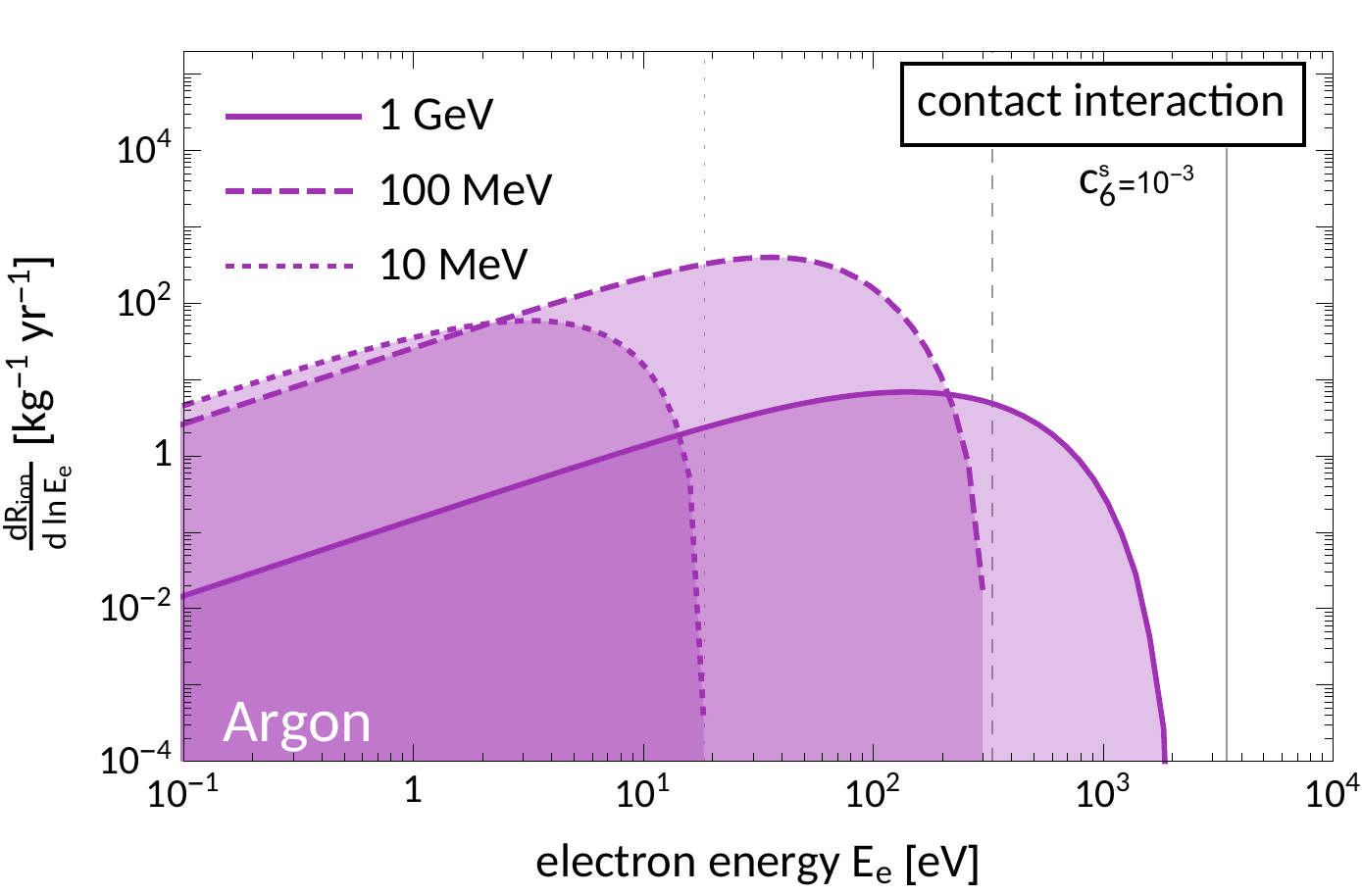}
\includegraphics[width=0.45\textwidth]{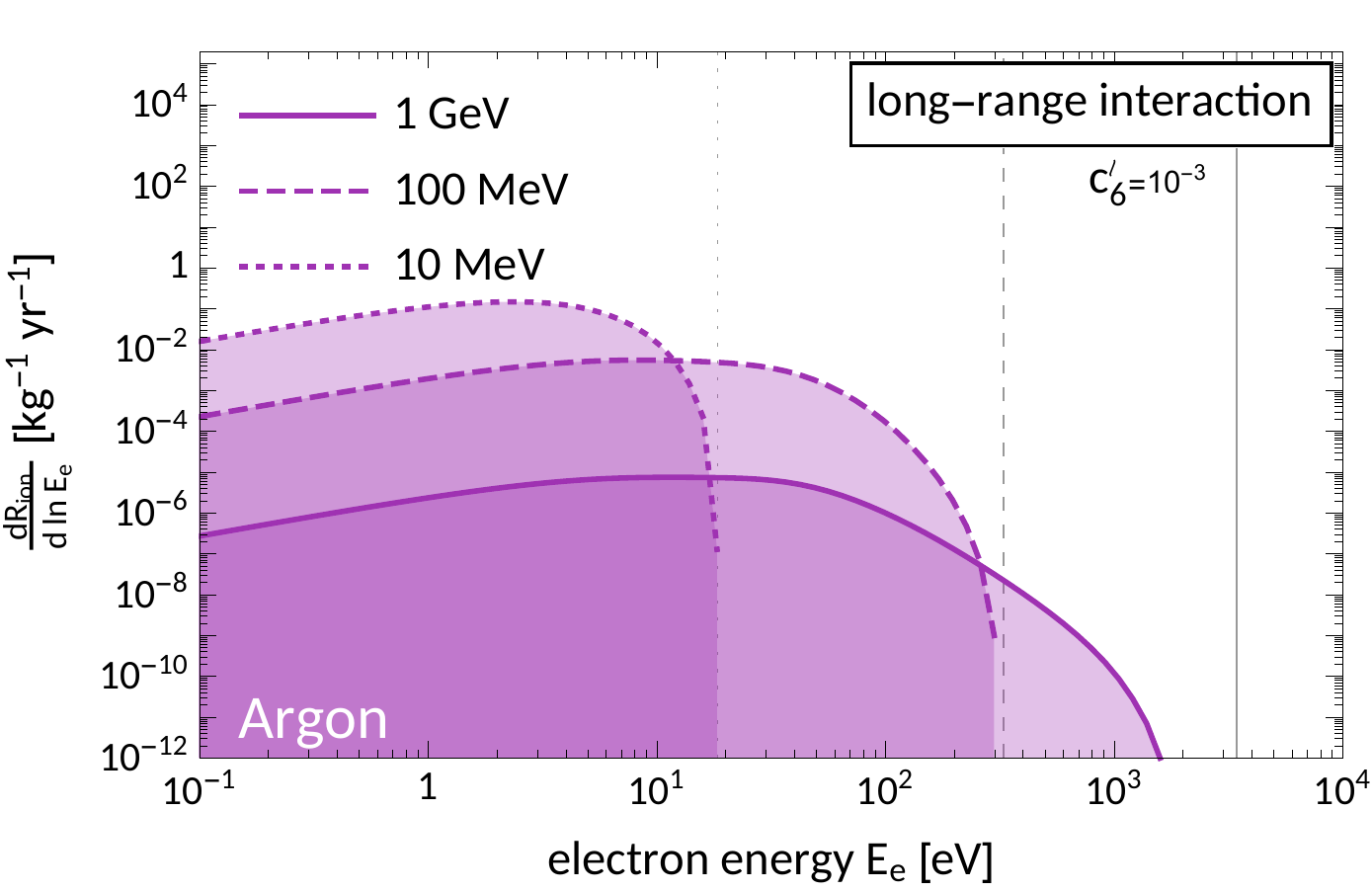}
\includegraphics[width=0.45\textwidth]{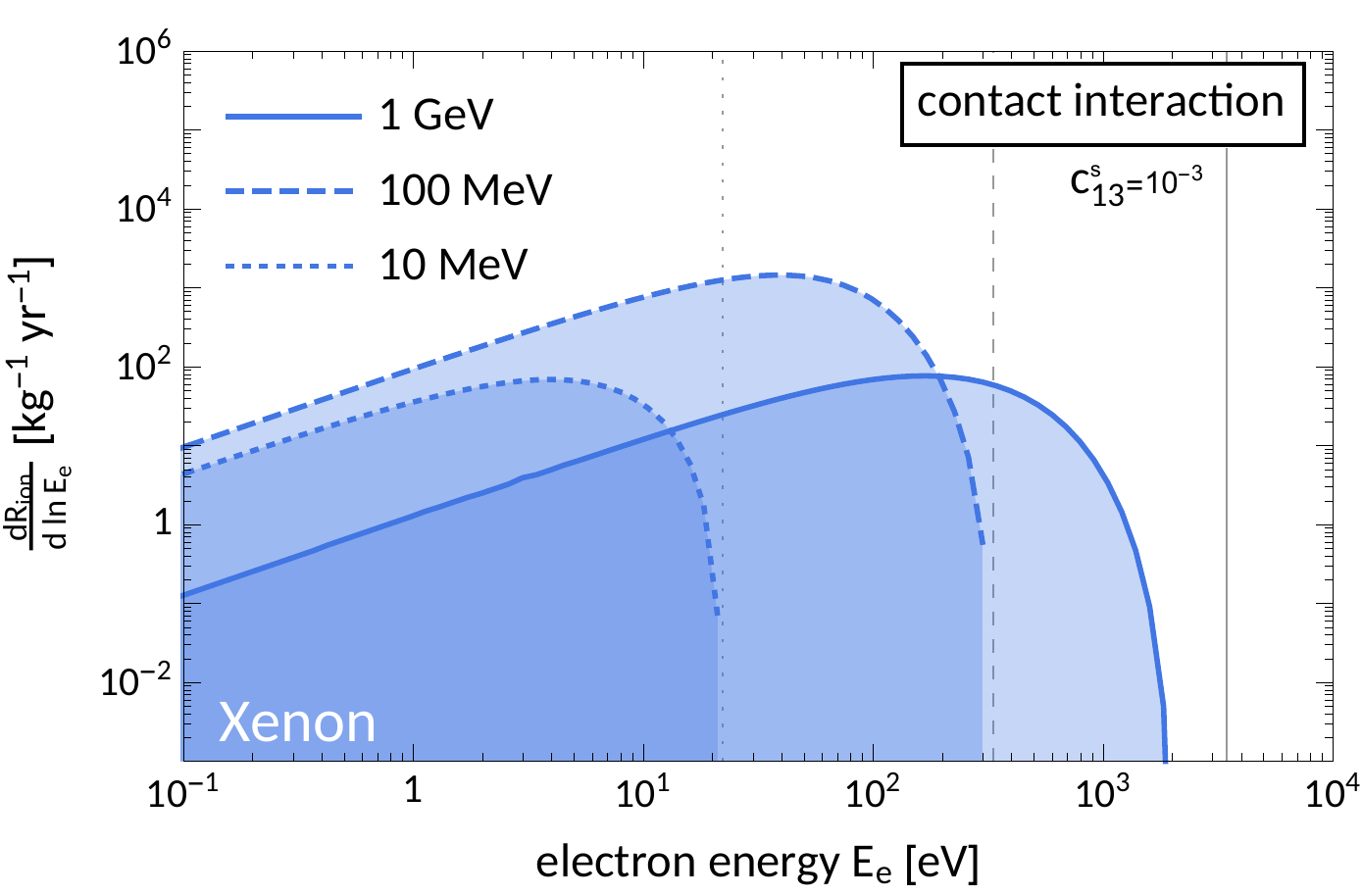}
\includegraphics[width=0.45\textwidth]{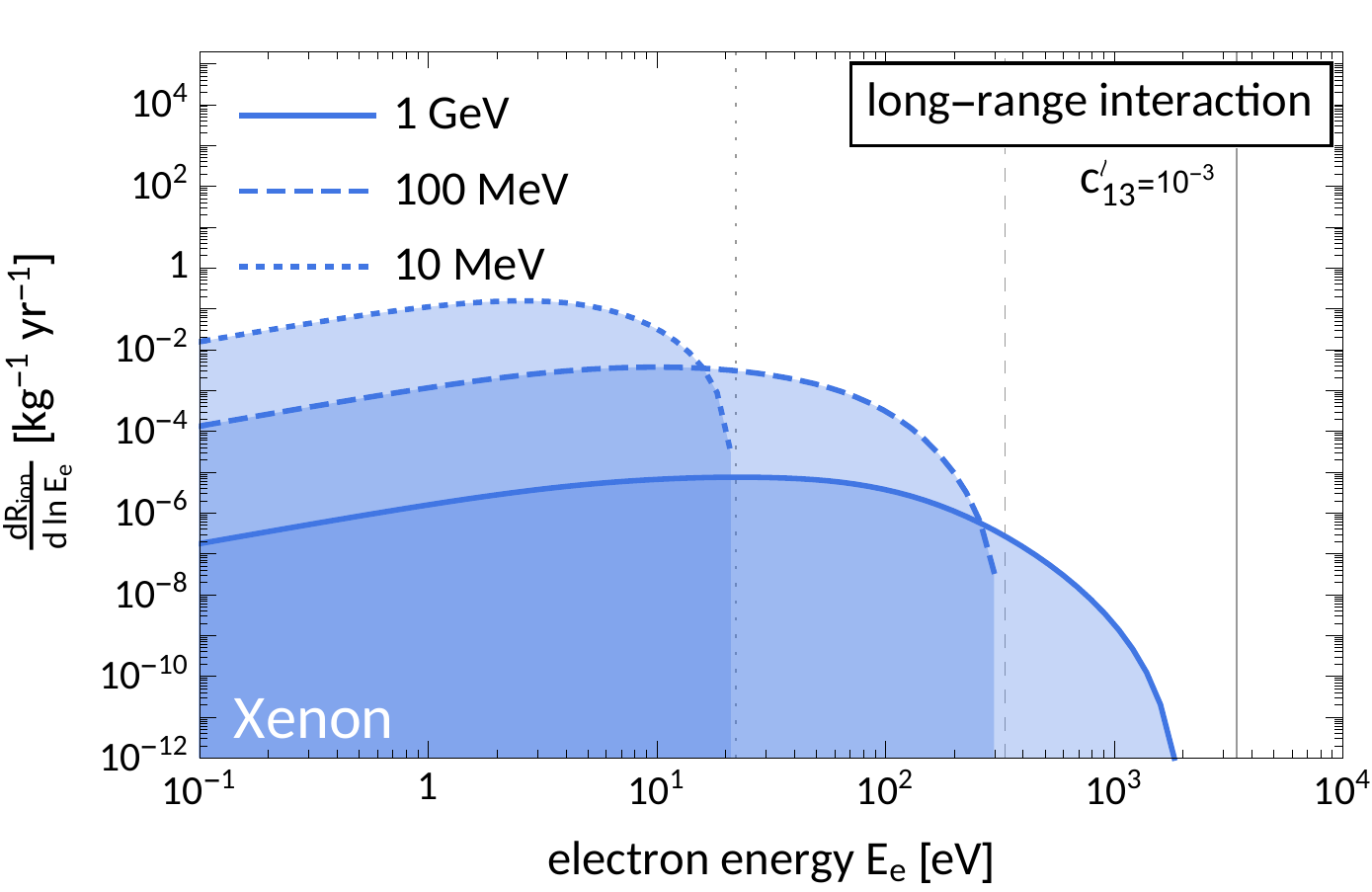}
\caption{Selected ionization spectra for argon for contact and long-range interactions with coupling~$c_6$ (top), and for xenon for contact and long-range interactions with coupling~$c_{13}$ (bottom), for three different DM~masses. The faint vertical lines indicate the kinematic cut-off of the spectra.}
\label{fig: spectra masses}
\end{figure*}

\section{First application to direct detection}
\label{sec:results}
The nonrelativistic effective theory of DM-electron interactions presented in the previous sections culminated in a general expression for the electron ionization energy spectrum of isolated atoms due to generic DM-electron interactions given by the Eqs.~\eqref{eq:ionization spectrum} and~\eqref{eq:RW}. This main result allows us to make predictions for direct searches of sub-GeV DM~particles by using an almost model-independent framework and a generic expression for the scattering amplitude in Eq.~\eqref{eq:Mnr} in terms of effective operators~$\mathcal{O}_i$.

In the following sections, we apply this effective theory to evaluate ionization spectra.
There are a number of large-scale dual-phase time-projection-chamber~(TPC) detectors with xenon and argon targets, which is why we focus on these two elements.~By re-interpreting the observational data by XENON10~\cite{Angle:2011th,Essig:2012yx}, XENON1T~\cite{Aprile:2019xxb}, and DarkSide-50~\cite{Agnes:2018oej}, we can set exclusion limits on either the effective coupling constants~$c^s_i$ and $c_i^\ell$ of Eq.~\eqref{eq:Mnr} in front of individual operators, or on the parameters of specific DM models generating combinations of interaction operators, as in the case of DM with anapole and magnetic dipole couplings.
In the latter case, one has to compute the elastic scattering amplitude between a DM~particle and an electron, take the nonrelativistic limit, and identify the different contributions with one or several of the effective operators in Table~\ref{tab:operators}.
We have already shown the example of anapole interactions, which corresponded to a combination of~$\mathcal{O}_8$ and $\mathcal{O}_9$, in Sec.~\ref{sec:interactions}.

It is also worth pointing out how the most-used interaction model in the sub-GeV~DM literature, i.e.~the ``dark photon'' model, emerges in this framework.
The dark photon model extends the Standard Model of particle physics by an additional U(1)~gauge group under which the DM~particle is charged.~Interactions between the DM~particle and electrically charged particles in the Standard Model arise from a ``kinetic mixing'' term in the interaction Lagrangian between the photon and the gauge boson~$A^\prime$ associated with the new U(1)~gauge group~\cite{Holdom:1985ag,Essig:2011nj}, i.e.~$\epsilon F_{\mu\nu}F^{\prime \mu\nu}$.~Here, $\epsilon$ is a coupling constant and $F^{\prime}_{ \mu\nu}$ the $A^\prime$ field strength tensor.  
This model can arguably be considered as the ``standard model'' of sub-GeV~DM detection.
The new massive gauge boson~$A^\prime$ is called the dark photon and acts as the interaction's mediator.
Its mass is usually assumed to be either much larger than the typical momentum transfer~$q_{\rm ref}$ in the scattering process (contact interaction), or much lower (long-range interaction).
By considering the nonrelativistic DM-electron scattering amplitude, this interaction can be identified with~$\mathcal{O}_1$, and the connection between the effective couplings in Eq.~(\ref{eq:Mnr}) and fundamental parameters is given by
\begin{align}
    c_1^{s}&=\frac{4 m_\chi m_e g_D e \epsilon}{m_{A^\prime}^2}\, ,& c_1^\ell &= 0\, ,
    \intertext{or}
    c_1^\ell&=\frac{4 m_\chi m_e g_D e \epsilon}{q_{\rm ref}^2}\, , &c_1^{s}&= 0\, ,
\end{align}
for contact and long-range interactions, respectively. Here, $g_D$ is the gauge coupling of the additional ``dark'' U(1)~gauge group, while $e$ denotes the charge of the electron.

\begin{figure*}
\centering
\includegraphics[width=0.9\textwidth]{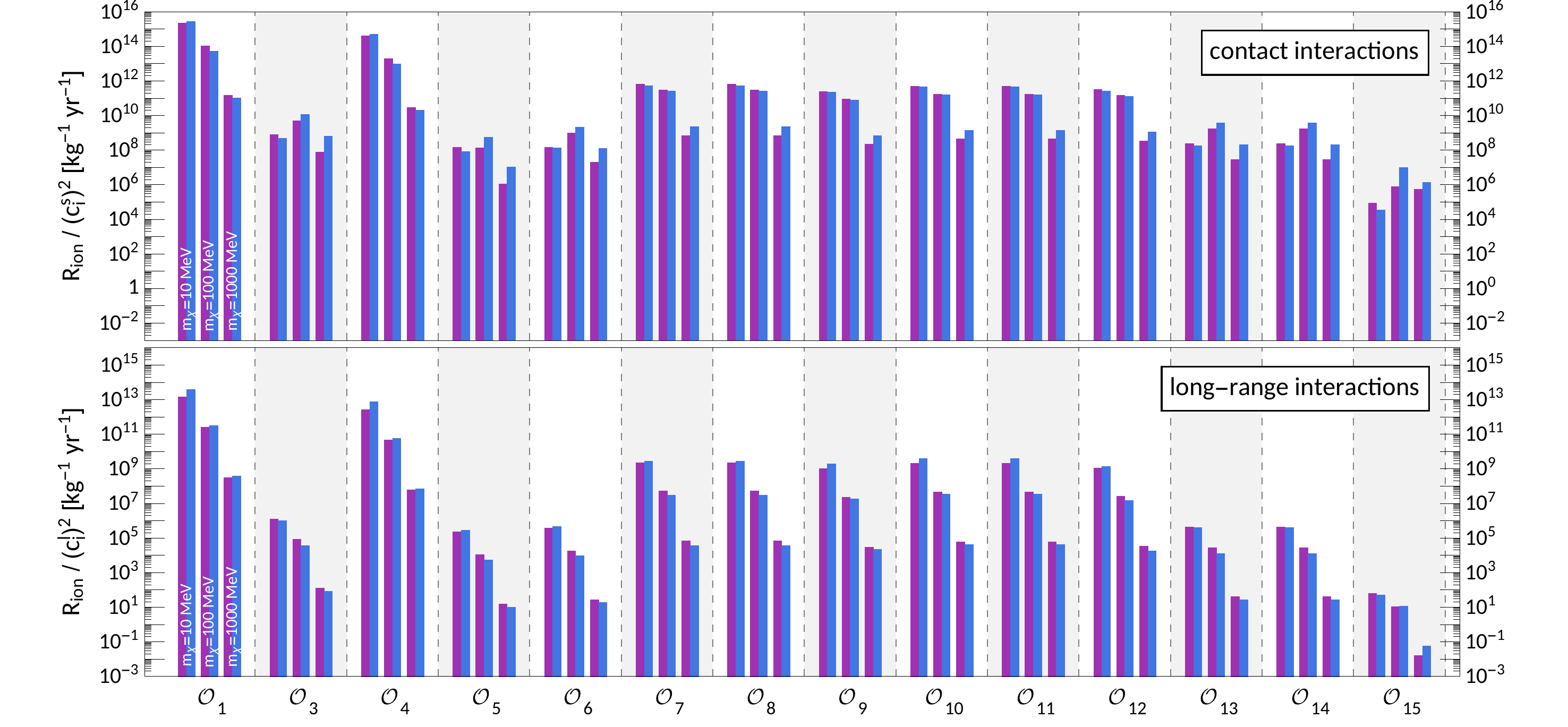}
\caption{Integrated ionization spectra for individual DM-electron interaction operators divided by the corresponding coupling constant squared.~We present results for argon (purple) and xenon (blue), as well as for contact interactions (top) and long-range interactions (bottom). A comparison of the different histograms shows the operators' relative strength when coupling constants are set to the same value. As indicated for the case of the ~$\mathcal{O}_1$ operator, we assume a DM~mass of 10/100/1000~MeV for the left/middle/right histograms.}
\label{fig: operator comparison}
\end{figure*}

As anticipated, the scattering amplitude arising from a given DM~model can correspond to either one or several of the effective operators in Table~\ref{tab:operators}. This is why we start by focusing on single operators, compute and study the corresponding ionization spectra, and set exclusion limits on the constants $c_i^s$ and $c_i^\ell$ for individual $\mathcal{O}_i$.
To illustrate the case of amplitudes being linear combinations of two or more of the effective operators, we study three relevant exemplary cases of DM-electron interactions, the anapole, magnetic dipole, and electric dipole interaction.
These arise as leading and next-to-leading terms in an effective theory expansion of the DM coupling to the ordinary photon~\cite{Kavanagh:2018xeh}.~We briefly review these models in Appendix~\ref{app:DMphoton}.

\subsection{Individual effective operators}
Let us consider the case where the free electron scattering amplitude, $\mathcal{M}$, consists of only one of the~$\mathcal{O}_i$~operators, or, in other words, only one of the couplings~$c^{s,\ell}_i$ in Eq.~\eqref{eq:Mnr} is different from zero.
On the one hand, this is instructive and allows us to study the impact of single operators in isolation.
On the other hand, this situation might also arise from specific models for DM interactions, with the dark-photon model being the most obvious example.

From Table~\ref{tab:operators}, we can identify the operators that generate
the three new atomic responses, i.e. $W_{j}^{n\ell}(k^\prime,\mathbf{q})$, $j=2,3,4$, by their dependence on~$\vPerpEl$.
They are $\mathcal{O}_{3}$, $\mathcal{O}_{5}$, $\mathcal{O}_{7}$, $\mathcal{O}_{8}$, $\mathcal{O}_{12}$, $\mathcal{O}_{13}$, $\mathcal{O}_{14}$, and $\mathcal{O}_{15}$.
Figure~\ref{fig: spectra masses} shows the {\it total ionization spectrum}, $\dd R_{\rm ion}/\dd \ln E_e$, for two example operators, $\mathcal{O}_{6}$ (top panels) and~$\mathcal{O}_{13}$ (bottom panels).~In the case of the $\mathcal{O}_{6}$ operator, we compute the ionization spectrum for argon.~For the $\mathcal{O}_{13}$ operator, we focus on xenon.~Furthermore, the left panels in Fig.~\ref{fig: spectra masses} refer to contact interactions, whereas the right panels refer to long-range interactions.~Finally, in all panels solid, long-dashed and short-dashed lines refer to different DM~particle masses.~We compute the total ionization spectrum by summing the ionization spectrum for each atomic orbital~$(n,\ell)$, Eq.~(\ref{eq:ionization spectrum}), over the five outermost occupied orbitals for xenon~(4s,4p,4d,5s, and~5p), and all occupied orbitals for argon~(1s, 2s, 2p, 3s, and~3p).~From the overall scale of the spectra, we can draw two conclusions.~Firstly, we find that ionization spectra are smaller in the case of long-range interactions than for contact interactions due to the additional $q_{\rm ref}^2/q^2$~factor in the scattering amplitude.~This suppression is more severe for large DM masses than for $m_\chi$ around 1 MeV/$c^2$.
Secondly, we find that the magnitude of the ionization spectrum associated with $\mathcal{O}_6$ (and scattering on argon) is comparable with that of $\mathcal{O}_{13}$ (and scattering on xenon).~This is expected, because $\mathcal{O}_6$ is quadratic in $\mathbf{q}$ while $\mathcal{O}_{13}$ is linear in $\mathbf{q}$ {\it and}  $\mathbf{v}^\perp_{\rm el}$, and at the same time scattering on argon and xenon occurs with comparable probabilities.

\begin{figure*}
\centering
\includegraphics[width=0.45\textwidth]{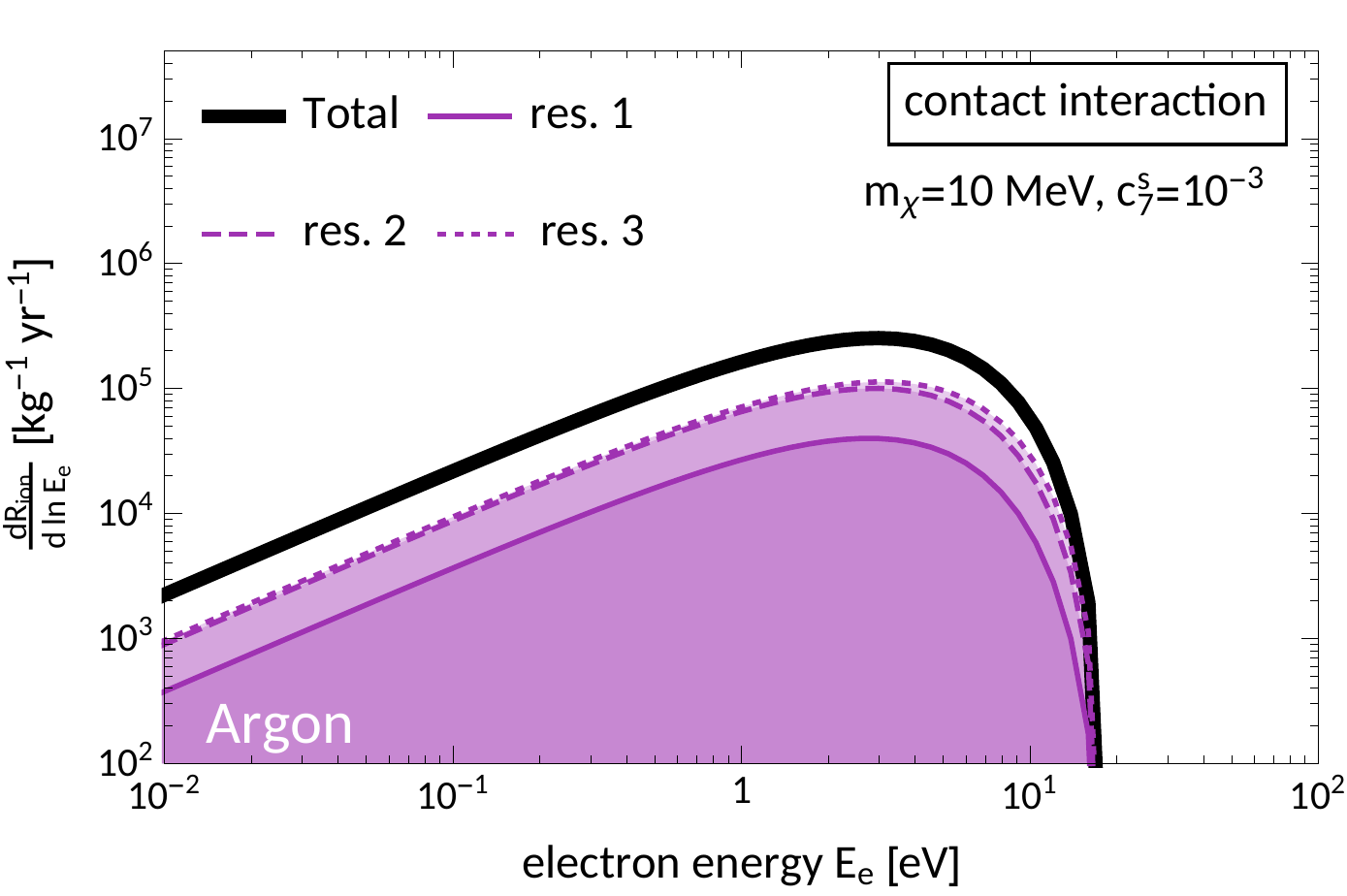}
\includegraphics[width=0.45\textwidth]{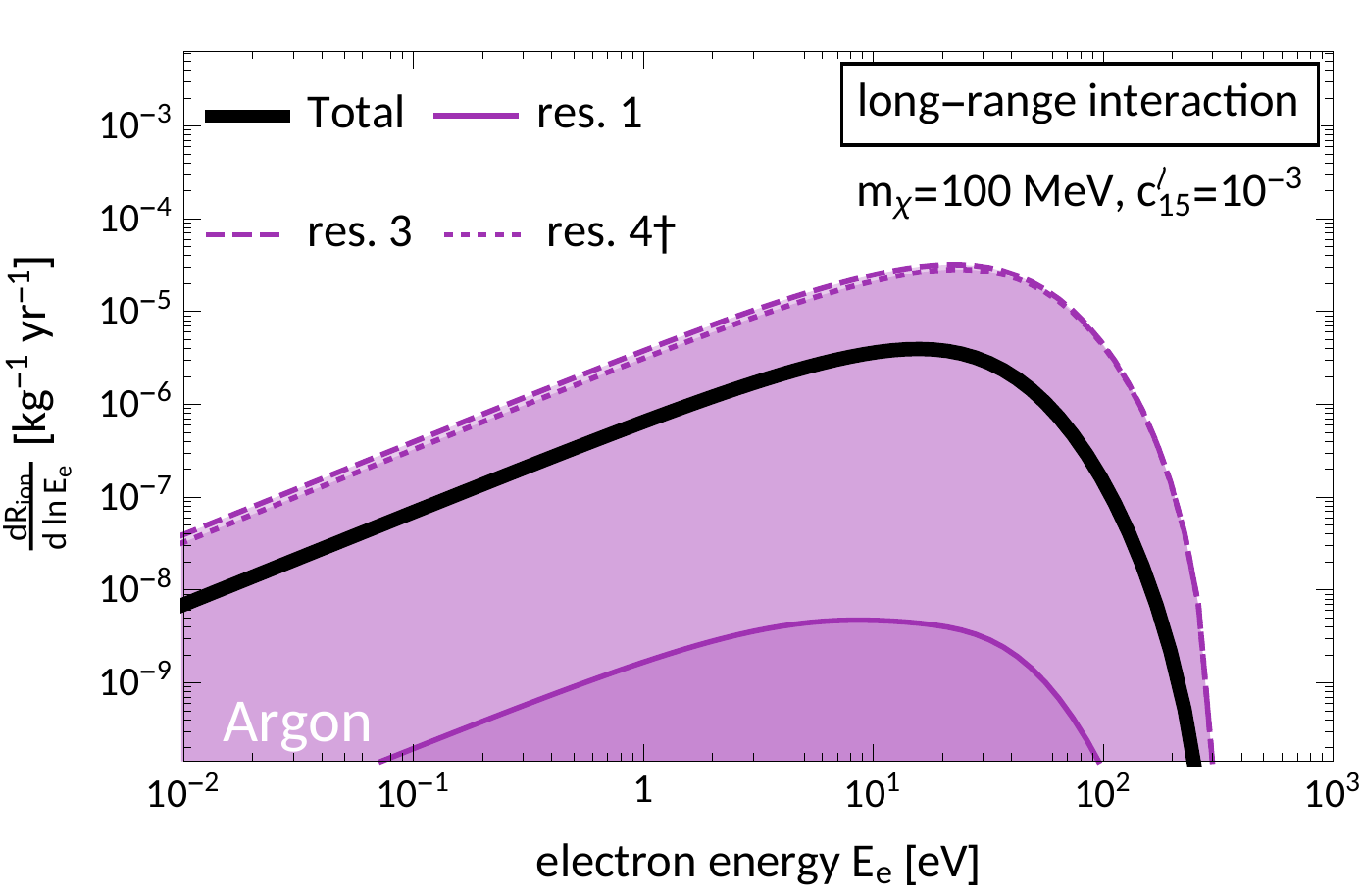}
\includegraphics[width=0.45\textwidth]{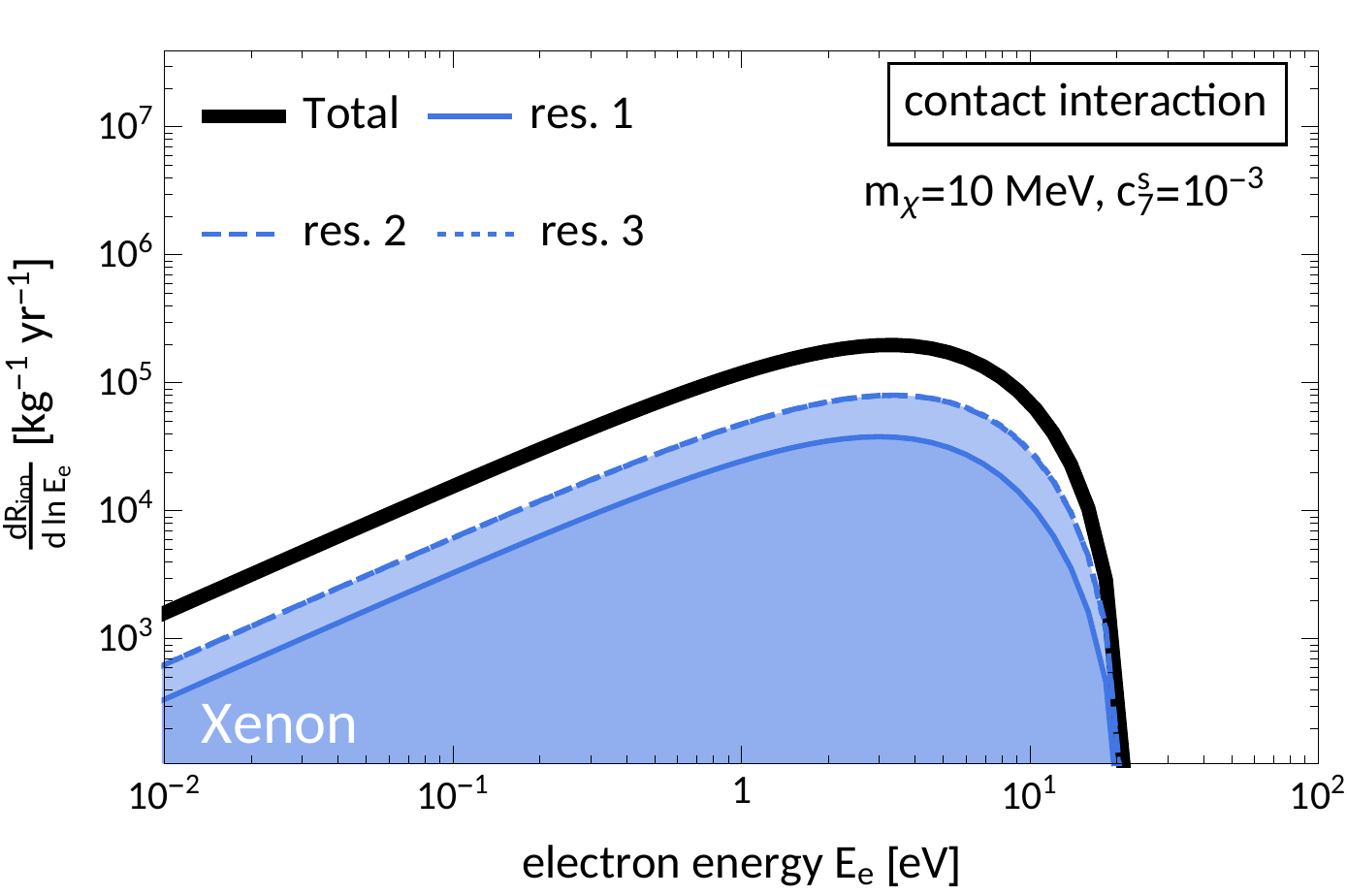}
\includegraphics[width=0.45\textwidth]{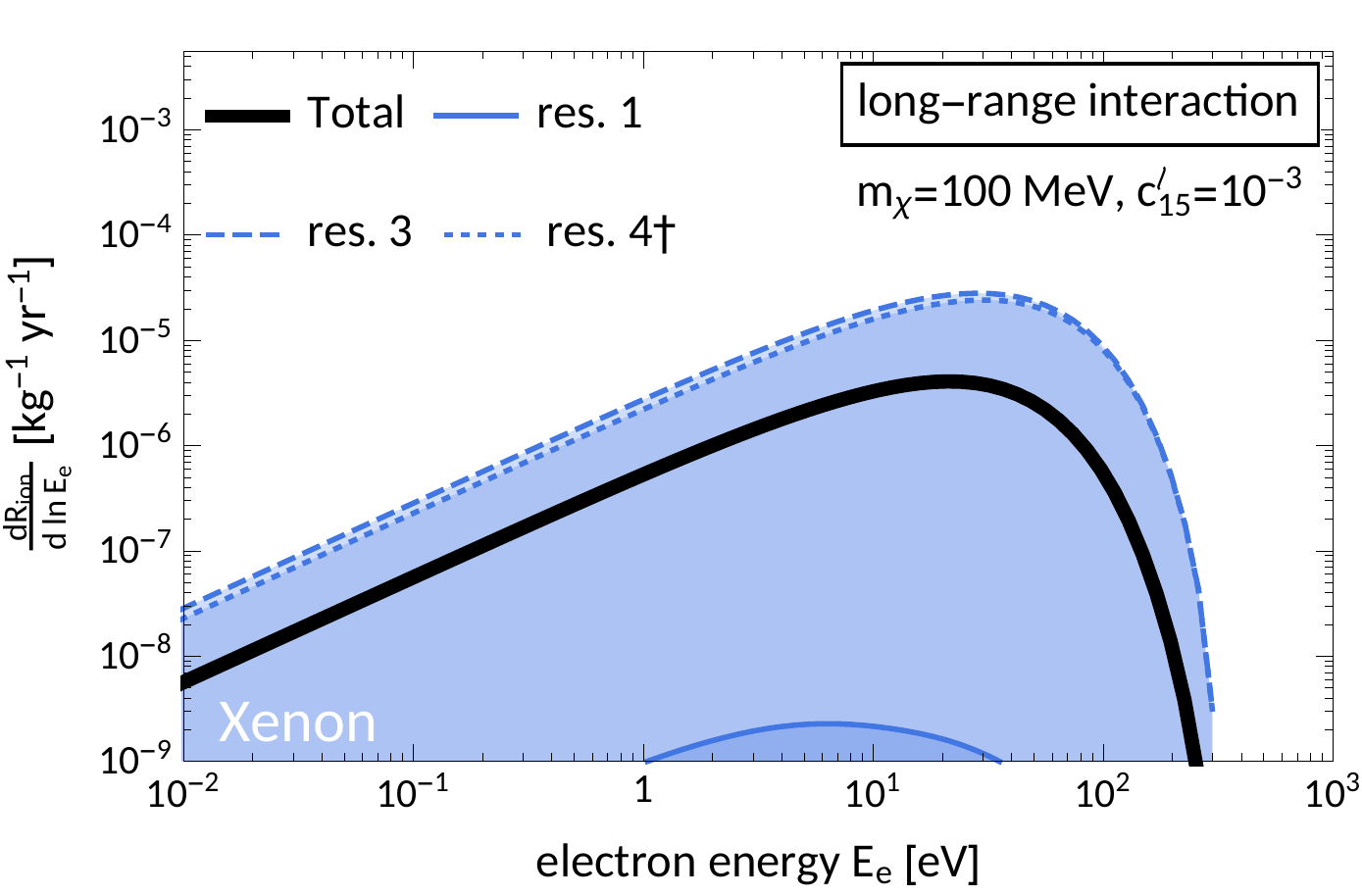}
\caption{The four $R_j^{n\ell}W_j^{n\ell}$, $j=1,\dots,4$ contributions to the ionization energy spectrum (here generically denoted as ``responses'') for the two example couplings $c^{s,\ell}_7$ (left) and $c^{s,\ell}_{15}$ (right). We present our results for argon (above) and xenon (below), as well as for contact interactions with $m_\chi=10$~MeV (left) and long-range interactions with $m_\chi=100$~MeV (right).~A dagger~$\dagger$ indicates negative contributions.}
\label{fig: spectra response contributions}
\end{figure*}

To make the last observation more precise and complete, we now extend the above comparison to the 14~operators in Table~\ref{tab:operators} describing spin 1/2 DM.~For each of these operators, we compute the associated ionization spectrum considering scattering on both argon and xenon targets. 
Figure~\ref{fig: operator comparison} shows the integrated ionization spectra in argon and xenon that we find by integrating $\dd R_{\rm ion}/\dd \ln E_e$ from zero up to the maximum kinematically allowed energy.~We present our results for three different DM~particle masses, namely 10, 100, and 1000~MeV/$c^2$.~By comparing the magnitude of the ionization spectra for a given operator, we find that the integrated ionization spectra in argon and xenon are always comparable, while the xenon ionization rate slightly exceeds the rate in argon in most but not all cases.

By comparing the ionization spectra of different operators for a given target material, we identify~$\mathcal{O}_1$ and~$\mathcal{O}_4$ as the ``leading-order''~(LO) operators.~This is expected, since the corresponding terms in Eq.~\eqref{eq: DM response functions} are not suppressed by powers of~$\mathbf{q}$ or~$\vPerpEl$, and, in addition, they generate the first atomic response function~$W_1^{n\ell}$, which is numerically found to be the one of largest magnitude.
The operators~$\mathcal{O}_{7}$ to~$\mathcal{O}_{12}$ can be regarded as ``next-to-leading-order''~(NLO) operators, in that they are linear in either~$\mathbf{q}$ or~$\vPerpEl$.
Similarly, the ``next-to-next-to-leading-order''~(NNLO) operators ($\mathcal{O}_{5}$, $\mathcal{O}_{6}$, $\mathcal{O}_{13}$, and $\mathcal{O}_{14}$) are quadratic in $\mathbf{q}$ or in a combination of $\mathbf{q}$ and $\vPerpEl$.~Compared to the LO and NLO cases, the total ionization spectrum of NNLO operators is further suppressed.
Lastly, the only~$\text{N}^{\text{3}}\text{LO}$ operator we include in our analysis is~$\mathcal{O}_{15}$, which is quadratic in~$\mathbf{q}$ and linear in~$\vPerpEl$.
The ionization spectrum due to this operator is the lowest and suppressed by up to ten orders of magnitude compared to~$\mathcal{O}_1$ for identical couplings.

This hierarchy of operators is essentially identical to that found when classifying DM-nucleus interactions within the effective theory of DM-nucleon interactions~\cite{Fan:2010gt,Fitzpatrick:2012ix}.~In that context, the equivalent of the $\mathcal{O}_1$ and $\mathcal{O}_4$ operators defined here correspond to the ``familiar'' spin-independent (SI) and spin-dependent~(SD) interactions -- a standard benchmark in the analysis of DM-nucleus scattering data.~Similarly to the SI/SD paradigm dominating the literature on direct DM~searches via nuclear recoils for a long time, the dark photon model underlies most studies of sub-GeV DM~searches nowadays.

\begin{figure*}
\centering
\includegraphics[width=0.45\textwidth]{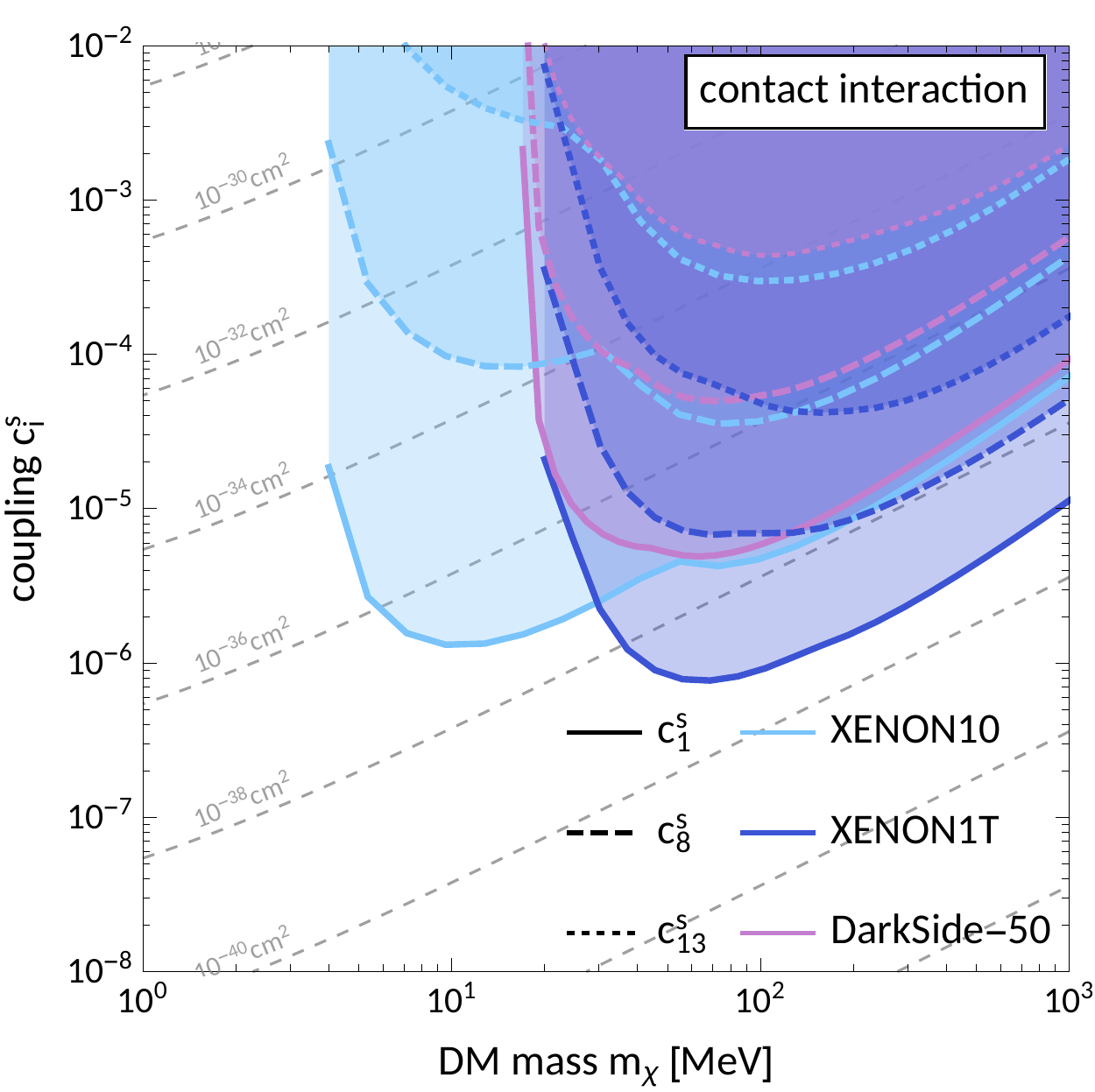}
\includegraphics[width=0.45\textwidth]{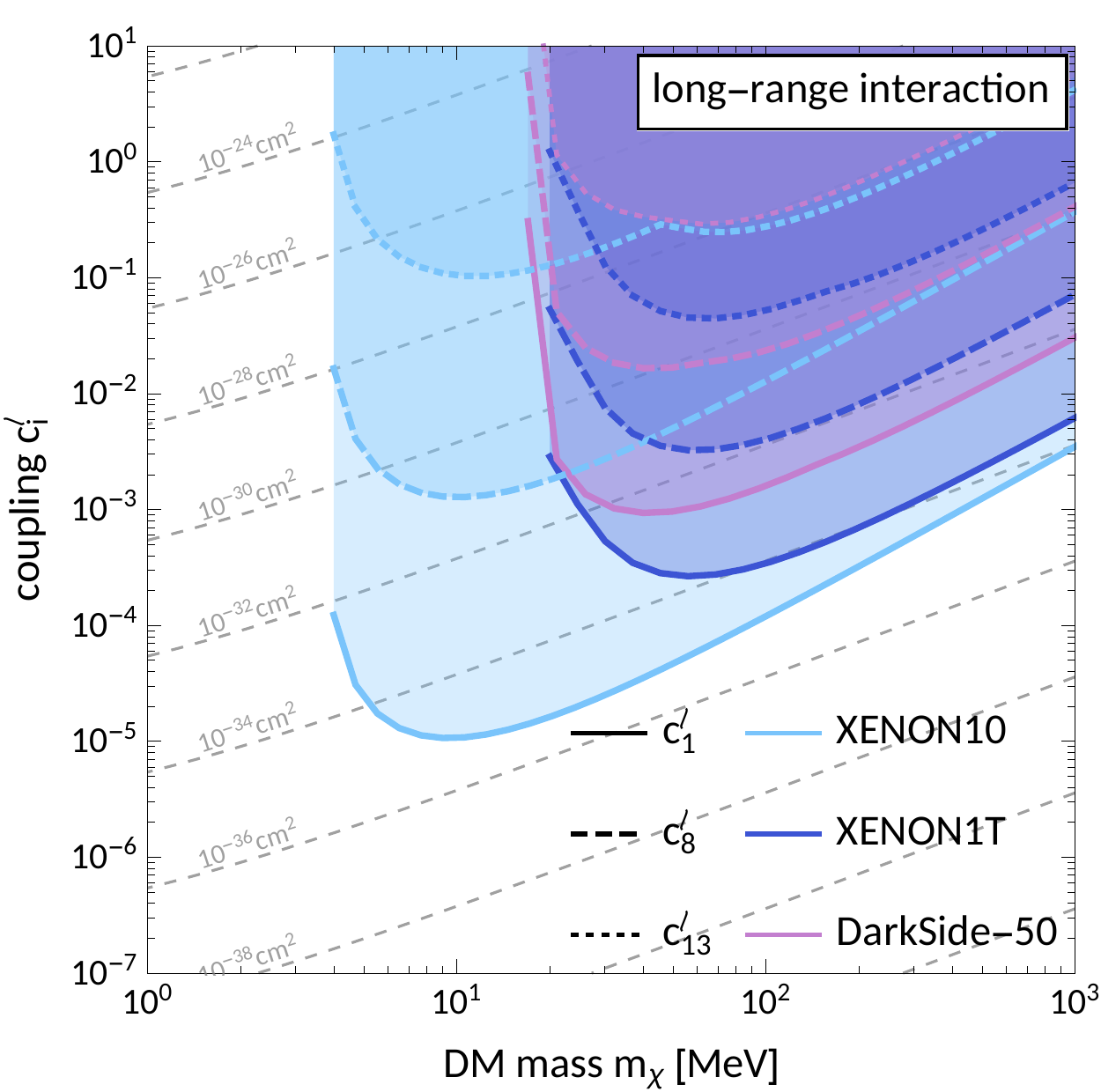}
\caption{Exclusion limits (90\% C.L.) from XENON10~\cite{Angle:2011th,Essig:2012yx}(blue), XENON1T~\cite{Aprile:2019xxb}(dark blue) and DarkSide-50~\cite{Agnes:2018oej}(purple) on a selection of individual effective couplings for both contact and long-range interactions. The dashed gray lines indicate constant reference cross sections $\overline{\sigma}_{e}$, see Eq.~\eqref{eq: reference cross section}. Note that the constraints on long-range interactions depend on a choice of the reference momentum transfer~$q_{\rm ref}$.}
\label{fig: constraint couplings}
\end{figure*}

Another interesting question in the context of isolated operators regards the relative contribution of the four different terms in Eq.~\eqref{eq:RW}.~Each of these terms is the product of a DM response function given in Eq.~(\ref{eq: DM response functions}) and an atomic response function given in Eq.~(\ref{eq: atomic responses}).
In order to answer the above question, in Fig.~\ref{fig: spectra response contributions} we show the ionization energy spectrum for two example operators, each of which generates multiple atomic responses.
Specifically, we show the total argon and xenon ionization spectra for $\mathcal{O}_7$, contact interaction and $m_\chi=10$~MeV/$c^2$ (left panels), and for $\mathcal{O}_{15}$, long-range interaction and $m_\chi=100$~MeV/$c^2$  (right panels).~As one can see from Fig.~\ref{fig: spectra response contributions}, we find that in the case of contact interactions of type $\mathcal{O}_7$ the four terms in Eq.~(\ref{eq: DM response functions}) give comparable contributions to the total ionization spectrum.~This is the result of a compensation between DM and atomic response functions.~Indeed, while the first atomic response function~$W_1^{n\ell}$ generally dominates over the second and third, which are in turn larger than~$W_4^{n\ell}$, in the case of contact interactions of type $\mathcal{O}_7$ we find an inverse hierarchy for the DM response functions $R^{n\ell}_j$, $j=1,2,3$, which scale as $(\vPerpEl)^2$, $(\vPerpEl\cdot\mathbf{q})m_e/q^2$ and $q^0$, respectively.~In this respect, the operator $\mathcal{O}_7$ is not unique (see for example $\mathcal{O}_8$, just to name one).~From this example, we conclude that the four terms in Eq.~(\ref{eq: atomic responses}) can in general contribute in a comparable manner and must therefore be taken into account in the interpretation of experimental data.

The contribution of an individual response may also be negative, an example of which is shown in the right panels of Fig.~\ref{fig: spectra response contributions} for the long-range ionization spectrum of~$\mathcal{O}_{15}$.\footnote{Negative contributions are marked with a dagger~($\dagger$).}
The effective operator~$\mathcal{O}_{15}$ generates the DM~responses~1,~3, and~4.
In this example, the contribution to the total ionization spectrum from the first atomic response function, $W_1^{n\ell}$, is suppressed by the first DM~response function, $R_1^{n\ell}$, which renders the $R_1^{n\ell} W_1^{n\ell}$ contribution to Eq.~(\ref{eq: atomic responses}) completely negligible.
In addition, the contribution of $R_4^{n\ell} W_4^{n\ell}$ is as anticipated negative, which originates from the sign of the $c_{15}$~term in~$R_4^{n\ell}$ in Eq.~\eqref{eq: DM response 4}.

We conclude this section with an analysis of present exclusion limits on the $c_i^{s,\ell}$ coupling constants from direct detection data.~Despite tremendous experimental efforts by a multitude of direct detection experiments (see Ref.~\cite{Undagoitia:2015gya} for a review), no conclusive DM~signal has been established, and DM~continues to evade direct searches to this day.
The largest of these experiments are dual-phase time-projection-chambers~(TPC) with xenon and argon targets, which is why we are focusing on these elements in this paper.~Here, we re-interpret the null results by the xenon target experiments XENON10~\cite{Angle:2011th} and XENON1T~\cite{Aprile:2019xxb}, as well as the argon experiment DarkSide-50~\cite{Agnes:2018oej} and set 90\% C.L. exclusion limits on the individual effective coupling constants. The details underlying the computation of the limits for these three experiments are summarized in Appendix~\ref{app: experiments}.

\begin{figure*}
    \centering
    \includegraphics[width=0.32\textwidth]{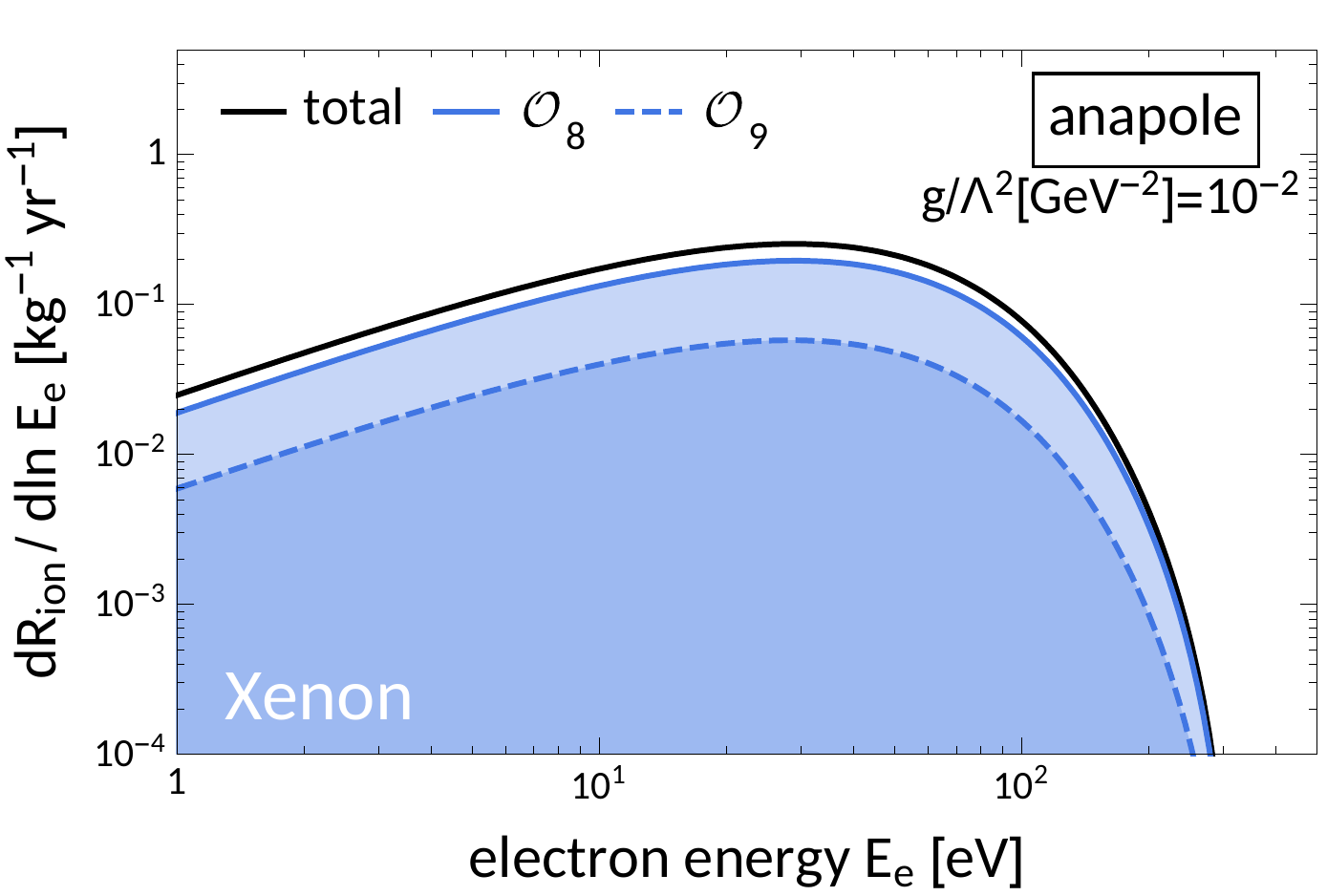}
    \includegraphics[width=0.32\textwidth]{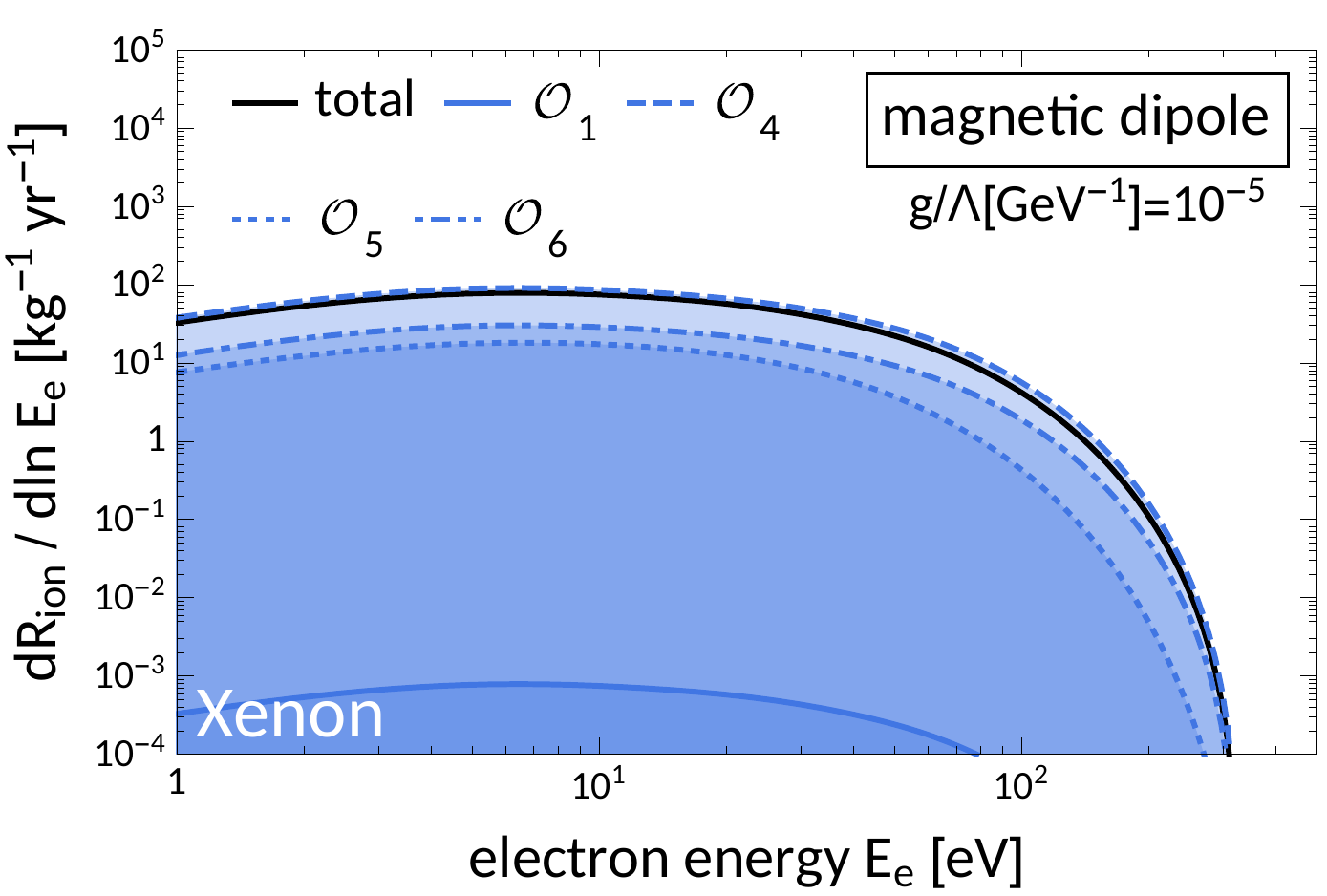}
    \includegraphics[width=0.32\textwidth]{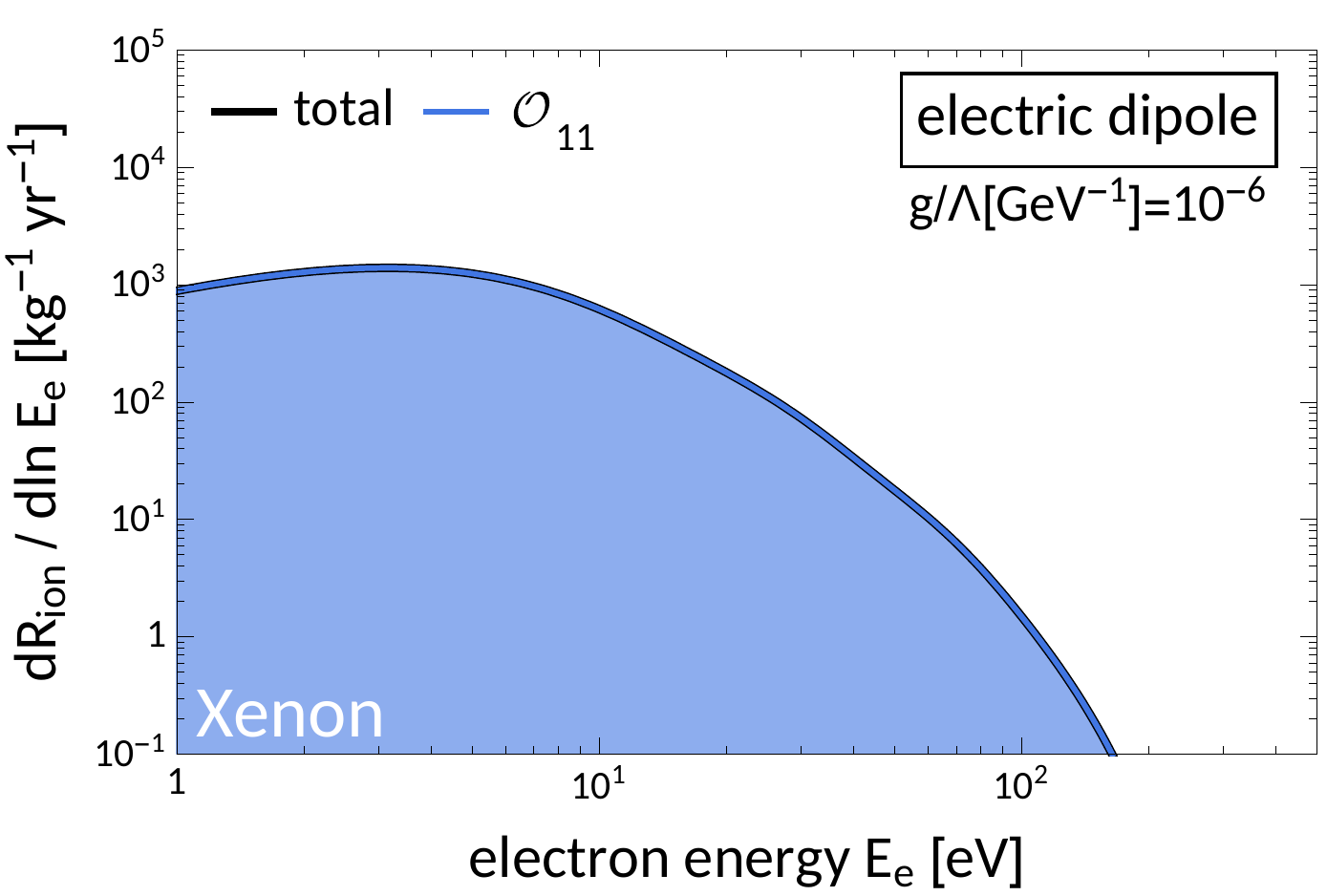}
    \caption{Ionization spectra for anapole, magnetic dipole, and electric dipole interactions shown in black. We set the DM~mass to 100~MeV and focus on xenon. The different blue lines show the contributions of the individual effective operator. Note that in the case of magnetic dipole interactions, the sum of the four individual contributions exceeds the total spectrum due to a negative interference between $\mathcal{O}_4$ and $\mathcal{O}_6$ in Eq.~\eqref{eq: DM response 1}.}
    \label{fig: spectra anapole}
\end{figure*}

Figure~\ref{fig: constraint couplings} shows our 90\% C.L. exclusion limits for three selected interaction operators, one LO~($\mathcal{O}_{1}$), one NLO~($\mathcal{O}_{8}$), and one NNLO~operator~($\mathcal{O}_{13}$), both for contact~(left) and long-range interactions~(right).\footnote{Due to the chosen parametrization of the scattering amplitude in Eq.~\eqref{eq:Mnr}, it is important to remember that the limits on the dimensionless long-range interaction coupling constants~$c_i^\ell$ depend on the choice for the reference momentum transfer~$q_{\rm ref}=\alpha m_e$.}
The dashed gray lines indicate constant values of a reference DM-electron scattering cross section that we define as follows,
\begin{align}
\label{eq: reference cross section}
    \overline{\sigma}_e\equiv\frac{\mu_{\chi e}^2 c_i^2}{16\pi m_\chi^2 m_e^2}\, . 
\end{align}
With this definition the constraints on~$c_1$ can be directly compared with previous works which described the DM-electron interaction using the dark photon model, see, e.g., Ref.~\cite{Essig:2012yx,Essig:2017kqs,Agnes:2018oej}.

\subsection{Anapole, magnetic dipole, and electric dipole interactions}
\label{sec:ame}
As an illustration of how different combinations of effective operators, $\mathcal{O}_i$, can arise from specific models for DM-electron interactions, we consider three examples:~the anapole, magnetic dipole, and electric dipole interaction. Electric and magnetic dipole interactions have previously been studied in the context of direct DM~searches via electron scatterings~\cite{Graham:2012su,Essig:2012yx,Chang:2019xva}. These studies used different approximations and did not exploit our new atomic response functions.

In Appendix~\ref{app:DMphoton}, we provide explicit Lagrangians for these types of interactions.~Therein we also present the associated nonrelativistic DM-electron scattering amplitudes, which we then match on to the effective theory expansion in Eq.~(\ref{eq:Mnr}).
In this way we can read off the connection between the Lagrangian parameters and our effective couplings.

Let us start with the anapole interaction.~From the amplitude in Eq.~\eqref{eq:M3}, we find that this model generates a linear combination of  the~$\mathcal{O}_8$ and~$\mathcal{O}_9$ operators in the nonrelativistic limit.~Only contact interactions are generated and the corresponding effective coupling constants are given by
\begin{subequations}
\label{eq: anapole effective couplings}
\begin{align}
    c_8^s &= 8 e m_e m_\chi\frac{g}{\Lambda^2}\, ,\\
    c_9^s &= -8 e m_e m_\chi\frac{g}{\Lambda^2}\, ,
\end{align}
\end{subequations}
where $g$ is a dimensionless coupling constant, and $\Lambda$ is the energy scale at which the anapole interaction term is generated.

For magnetic dipole interactions, we find that the scattering amplitude in Eq.~\eqref{eq: amplitude magnetic dipole} can be expressed as a linear combination of four interaction operators, namely~$\mathcal{O}_1$, $\mathcal{O}_4$, $\mathcal{O}_5$, and~$\mathcal{O}_6$.
Here, the effective coupling constants are given by
\begin{subequations}
\label{eq: magnetic dipole effective couplings}
\begin{align}
    c_1^s &= 4 e m_e \frac{g}{\Lambda}\, ,\\
    c_4^s &= 16 e m_\chi\frac{g}{\Lambda}\, ,\\
    c_5^\ell &= \frac{16em_e^2 m_\chi}{q_{\rm ref}^2} \frac{g}{\Lambda}\, ,\\
    c_6^\ell &= -\frac{16em_e^2 m_\chi}{q_{\rm ref}^2} \frac{g}{\Lambda}\, .
\end{align}
\end{subequations}

The situation is particularly simple for electric dipole interactions, with the scattering amplitude given in Eq.~\eqref{eq: amplitude electric dipole}.
It is linear in~$\mathcal{O}_{11}$ with the single effective coupling,
\begin{align}
    c_{11}^\ell &= \frac{16 e m_\chi m_e^2}{q_{\rm ref}^2}\frac{g}{\Lambda}\, .
\end{align}

Figure~\ref{fig: spectra anapole} shows the xenon ionization spectra for anapole, magnetic dipole and electric dipole interactions.~In the three cases we assume a DM~particle mass of 100~MeV$c^2$.
Different blue lines correspond to  contributions from the individual interaction operators to the total ionization spectrum (shown as a black solid line).
Most interesting is the spectrum of the magnetic dipole interaction.
While one might expect the interaction operator $\mathcal{O}_1$ to dominate the spectrum (since it is one of the two LO operators), its contribution is in fact negligible due to the small relative size of the four effective coupling constants.
This can be seen e.g. by the ratio $c_1^s/c_4^s\approx 10^{-3}$ for~$m_\chi=100$~MeV/$c^2$.

Another interesting aspect of the magnetic dipole interaction spectrum is the fact that the contribution of the single operator~$\mathcal{O}_4$ exceeds the total result.
This is explained by the interference term between~$\mathcal{O}_4$ and~$\mathcal{O}_6$ in the first DM~response function in Eq.~\eqref{eq: DM response 1}, whose overall contribution is negative due to the sign of~$c_6^\ell$ in Eq.~\eqref{eq: magnetic dipole effective couplings}.

\begin{figure*}
    \centering
    \includegraphics[width=0.32\textwidth]{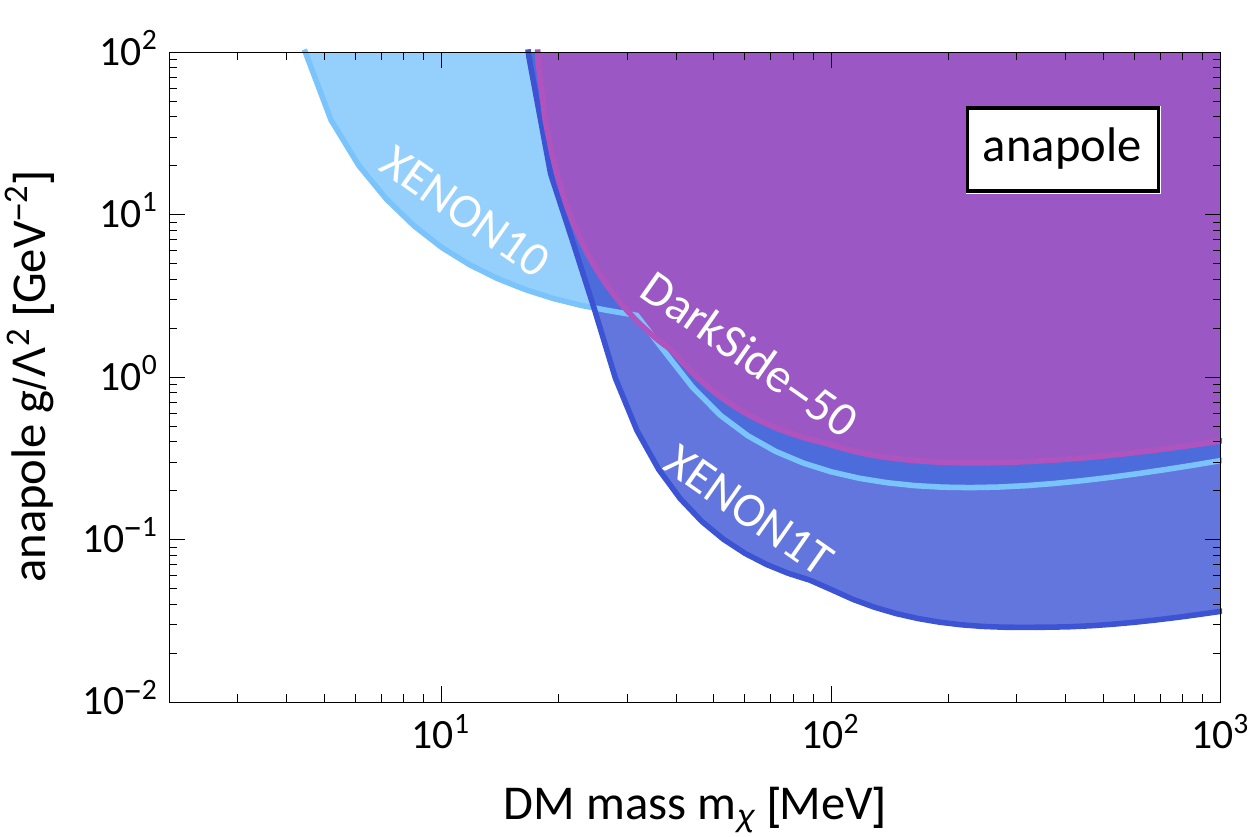}
    \includegraphics[width=0.32\textwidth]{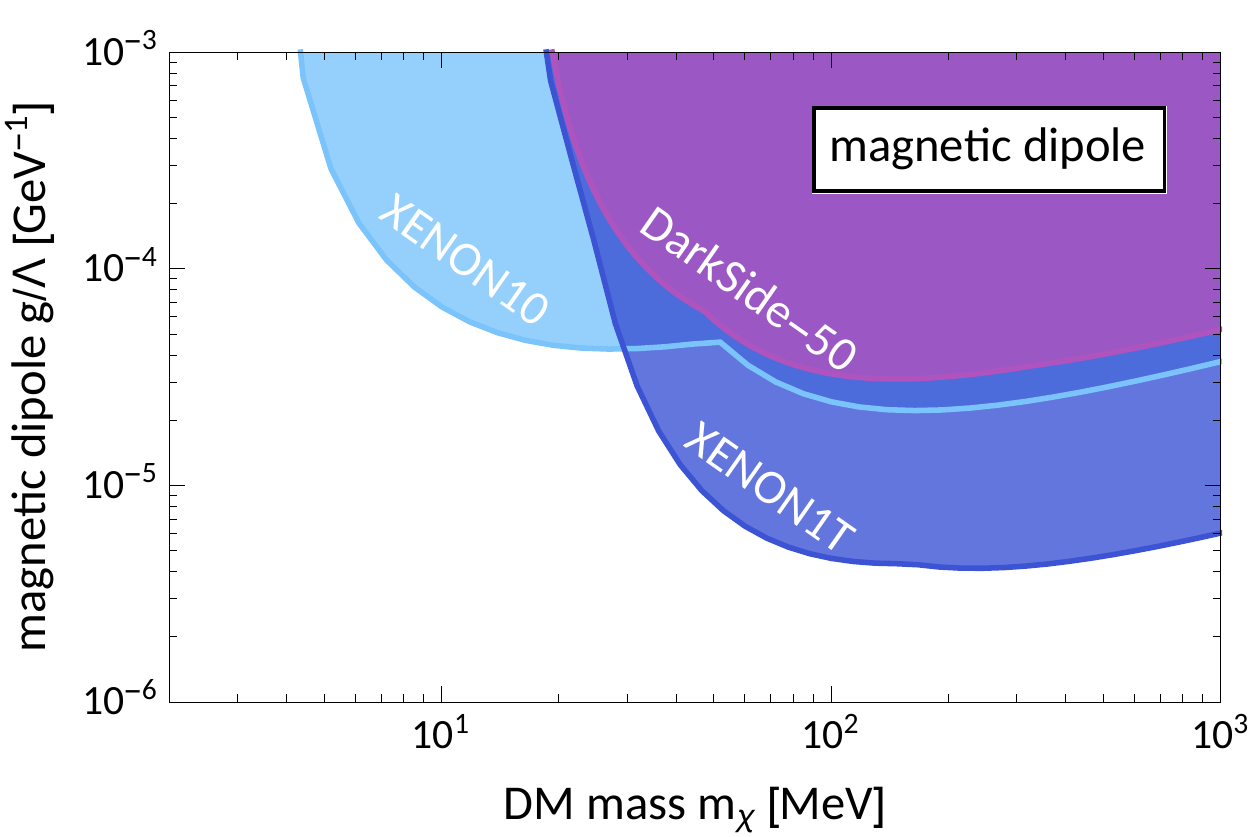}
    \includegraphics[width=0.32\textwidth]{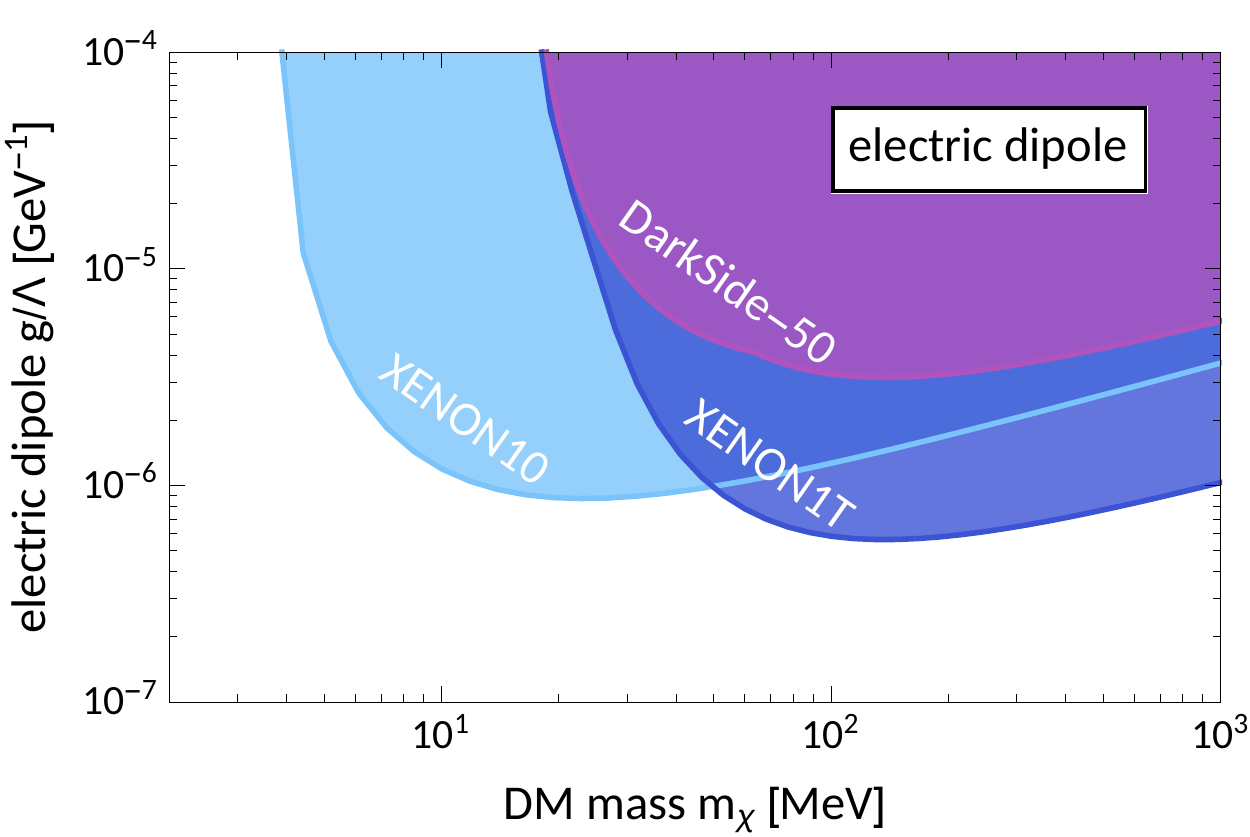}
    \caption{Constraints (90\% C.L.) on anapole, magnetic dipole, and electric dipole interactions (see Appendix~\ref{app:DMphoton}) from the null result reported by XENON10~\cite{Angle:2011th,Essig:2012yx} (light blue), XENON1T~\cite{Aprile:2019xxb}(dark blue), and DarkSide-50~\cite{Agnes:2018oej} (purple).}
    \label{fig: constraint anapole and dipole}
\end{figure*}

We conclude this chapter by presenting our 90\% C.L. exclusion limits on the ratio between dimensionless coupling constant $g$ and energy scale $\Lambda$ (or $\Lambda^2$) for the anapole, magnetic dipole, and electric dipole interaction.~Our exclusion limits are shown in Fig.~\ref{fig: constraint anapole and dipole} and based on a re-interpretation of the null results reported by the XENON10, XENON1T, and DarkSide-50 experiments.\footnote{The details of these experiments can be found in Appendix~\ref{app: experiments}.}~For these specific DM-electron interaction models, we find that exclusion limits from DarkSide-50 are generically less stringent, as compared to those arising from XENON10 and XENON1T data.~While XENON10 always sets the most stringent limits for DM masses below about 20-50~GeV/$c^2$ (depending on the assumed interaction), the minimum coupling constant value that XENON10 can exclude is comparable to the minimum coupling probed by XENON1T only for electric dipole interactions.

\section{Summary}
\label{sec:conclusions}
We computed the response of isolated xenon and argon atoms to general, nonrelativistic DM-electron interactions.~This is a key input to the interpretation of DM-electron scattering data.~For example, dual-phase argon and xenon targets are used in direct detection experiments searching for signals of nonrelativistic interactions between Milky Way DM~particles and electrons in the target materials.~We modelled the DM-electron scattering by formulating a nonrelativistic effective theory of DM-electron interactions which significantly extends the currently favoured framework, namely, the dark photon model.~Within our effective theory description of DM-electron interactions, the free amplitude for DM-electron scattering not only depends on the momentum transferred in the scattering (as in the dark photon model), it can also depend on a second, independent combination of particle momenta as well as on the DM~particle and electron spin operators.~Quantitatively, we defined the atomic response to an external probe in terms of the overlap between the initial and final state electron wave functions.

As a first application of the atomic responses found in this work, we computed the rate at which Milky Way DM~particles can ionize isolated argon and xenon atoms in target materials used in operating direct detection experiments.~We found that the final state electron ionization energy spectrum can in general be expressed as the sum of four independent terms.~Each of these terms is given by the product between an atomic response function, which depends on the initial and final state electron wave functions, and a DM response function, which only depends on kinematic variables, coupling constants, and the energy gap between initial and final electron states.~We investigated the structure and relative strength of the four terms contributing to the total ionization spectrum, describing under what circumstances they can be generated in nonrelativistic DM-electron scattering processes in argon and xenon targets.~The effective theory approach developed in this work has proven very useful in the classification and characterization of the individual atomic responses and contributions to the total ionization energy spectrum.~At the same time, the generality of the formalism developed here allows for straightforward applications to most specific DM interaction models.

We then used the argon and xenon atomic responses found here to compute 90\% C.L. exclusion limits on the strength of DM-electron interactions from the null result reported by the XENON10, XENON1T and DarkSide-50 direct detection experiments.~DarkSide-50 is the only direct detection experiment considered here that uses an argon target.~For contact DM-electron interactions, we found that XENON10 generically sets the most stringent constraints for DM~particle masses below $\mathcal{O}(10)$ MeV and that XENON1T is the leading experiment at larger masses.~The DM~particle mass above which XENON1T dominates depends on the specific DM-electron interaction model.~For long-range interactions, we found cases in which XENON10 sets the strongest exclusion limits on the entire mass window explored here.~Both for contact and long-range interactions, the DarkSide-50 experiment currently sets 90\% C.L. exclusion limits which are generically less stringent than those from XENON10 and XENON1T. 

Considering the plethora of proposed DM-electron scattering targets, the results presented in this work should be thought of as a first step within a far-reaching program which aims at exploring the response to general DM-electron interactions of condensed matter systems used in (or proposed for) DM direct detection experiments.~Indeed, the expressions derived in Sec.~\ref{sec:rate} and~\ref{sec:interactions} apply not just to isolated atoms, but also more generally to other target materials, such as semiconductor crystals and quantum materials.

Finally, we would like to stress that only one of the four atomic response functions computed here can arise from standard electromagnetic interactions between external electrons or photons and electrons bound to target materials.~The remaining three atomic responses can only be generated when DM is the external probe used to investigate the properties of condensed matter systems.~From this perspective, our findings show that the detection of Milky Way DM~particles has the potential to open up a new window on the exploration of materials' responses to external probes, revealing as of yet hidden properties of matter.

\acknowledgements
We would like to thank Rouven Essig, Sin\'ead Griffin, Felix Kahlhoefer, Yonathan Kahn, and Tien-Tien Yu for useful comments and valuable discussions. RC and TE were supported by the Knut and Alice Wallenberg Foundation (PI, Jan Conrad).~RC also acknowledges support from an individual research grant from the Swedish Research Council, dnr. 2018-05029.~NAS and WT were supported by the ETH Zurich, and by the European Research Council (ERC) under the European Union’s Horizon 2020 research and innovation programme grant agreement No 810451.~The research presented in this paper made use of the following software packages, libraries, and tools: hankel~\cite{hankel}, mpmath~\cite{mpmath}, NumPy~\cite{numpy}, SciPy~\cite{scipy2019}, WebPlotDigitizer~\cite{webplotdigitizer}, Wolfram Mathematica~\cite{Mathematica}. Some of the computations were performed on resources provided by the Swedish National Infrastructure for Computing (SNIC) at NSC.

\appendix

\section{Derivation of the atomic and DM~response functions}
\label{app: response derivation}
This appendix serves as a bridge between Eqs.~\eqref{eq:ionization amplitude 2} and~\eqref{eq:RW}. In particular, we identify and derive the atomic and DM~response functions given in Eqs.~\eqref{eq: DM response functions} and~\eqref{eq: atomic responses}. This involves the computation of the three terms in Eq.~\eqref{eq:ionization amplitude 2}, given more explicitly by
\begin{widetext}
\begin{align}
    \overline{| \mathcal{M}(\mathbf{q},\mathbf{v}_{\rm el}^\perp) |^2} &=\frac{1}{2j_\chi+1}\frac{1}{2j_e+1}\sum_{\rm spins}| \mathcal{M}(\mathbf{q},\mathbf{v}_{\rm el}^\perp) |^2\, ,\label{eq:M2}\\
     2m_e\overline{ \Re \left[ \mathcal{M} \nabla_{\mathbf{k}} \mathcal{M}^*\cdot f_{1\rightarrow 2}(\mathbf{q})\mathbf{f}^*_{1\rightarrow 2}(\mathbf{q})  \right]}&= \frac{1}{2j_\chi+1}\frac{1}{2j_e+1}\sum_{\rm spins}  2m_e\Re \left[ \mathcal{M} \nabla_{\mathbf{k}} \mathcal{M}^*\cdot f_{1\rightarrow 2}(\mathbf{q})\mathbf{f}^*_{1\rightarrow 2}(\mathbf{q}) \right]\, ,\label{eq:2MdM}\\
    m_e^2\overline{|\nabla_{\mathbf{k}} \mathcal{M}(\mathbf{q},\mathbf{v}_{\rm el}^\perp) \cdot \mathbf{f}_{1\rightarrow 2}(\mathbf{q}) |^2} &=\frac{1}{2j_\chi+1}\frac{1}{2j_e+1}\sum_{\rm spins}m_e^2|\nabla_{\mathbf{k}} \mathcal{M}(\mathbf{q},\mathbf{v}_{\rm el}^\perp) \cdot \mathbf{f}_{1\rightarrow 2}(\mathbf{q}) |^2\, ,\label{eq:dM2}
\end{align}
\end{widetext}
 where we average (sum) over initial (final) spins of the DM~particle and the electron. For the amplitude we substitute the general expression of Eq.~\eqref{eq:Mnr}, where we consider the first 14 operators listed in Table~\ref{tab:operators}.
 
When averaging (summing) the terms in Eqs.~\eqref{eq:M2} to~\eqref{eq:dM2} over the initial (final) spin of the DM~particle and the electron, the non-vanishing terms are of the form~\cite{Anand:2013yka},
\begin{widetext}
\begin{align}
   &\frac{1}{2j_i+1} \sum_{s,s^\prime} \bra{j_i s} \left\{
    \begin{array}{c} 
    \mathbf{S}_i \ket{j_is^\prime}\cdot\bra{j_i s^\prime}\mathbf{S}_i\\[0.15cm]
    \mathbf{A}\cdot\mathbf{S}_i \ket{j_is^\prime}\phantom{\cdot}\bra{j_i s^\prime}\mathbf{B}\cdot\mathbf{S}_i\\[0.15cm]
    \mathbf{A}\times\mathbf{S}_i \ket{j_is^\prime}\cdot\bra{j_i s^\prime}\mathbf{B}\times \mathbf{S}_i\\[0.15cm]
    \mathbf{A}\times\mathbf{S}_i \ket{j_is^\prime}\cdot\bra{j_i s^\prime}\mathbf{S}_i
    \end{array}
    \right\} \ket{j_i s}=\left\{\begin{array}{c}
         1  \\[0.15cm]
         \mathbf{A}\cdot\mathbf{B}/3\\[0.15cm]
         2\mathbf{A}\cdot\mathbf{B}/3\\[0.15cm]
         0
    \end{array}
    \right\}j_i(j_i+1)\, . \label{eq: spin sum identities}
\end{align}
\end{widetext}
Here, $s(s^\prime)$ is the initial (final) spin's index of either the electron or DM~particle, and~$\mathbf{A}$ and~$\mathbf{B}$ are two general vectors. All cross terms which are linear in~$\mathbf{S}_i$ vanish. It is important to keep in mind that the scalar product~$\mathbf{q}\cdot \vPerpEl$, which occurs frequently, does not vanish since we are not considering an elastic scattering process.
 
\subsection{Response 1}
\noindent We evaluate the spin sums in Eq.~\eqref{eq:M2} for the scattering amplitude including the first 14~effective operators. For the evaluation of the electron and DM spin sums, we make use of the identities of Eq.~\eqref{eq: spin sum identities}. We find
\begin{widetext}
\begin{align}
    &\overline{| \mathcal{M}(\mathbf{q},\mathbf{v}_{\rm el}^\perp) |^2} = |c_1|^2 + \frac{|c_3|^2}{4}\left( \frac{\mathbf{q}}{m_e}\times\vPerpEl \right)^2+ \frac{|c_7|^2}{4}\left(\vPerpEl\right)^2+ \frac{|c_{10}|^2}{4}\left(\frac{\mathbf{q}}{m_e}\right)^2 + \frac{\Im(c_7c_{10}^*)}{2}\left(\frac{\mathbf{q}}{m_e}\cdot \vPerpEl\right)\nonumber\\
    &\quad+\frac{j_\chi(j_\chi+1)}{12}\Bigg\{ 3 |c_4|^2+\left(4|c_5|^2-2\Re(c_{12}c_{15}^*)\right)\left( \frac{\mathbf{q}}{m_e}\times\vPerpEl \right)^2 + |c_6|^2 \left(\frac{\mathbf{q}}{m_e}\right)^4 + \left(4 |c_8|^2+2|c_{12}|^2\right) \left(\vPerpEl\right)^2  \nonumber\\
    &\qquad+ \left(2|c_{9}|^2+4|c_{11}|^2+2\Re(c_4^*c_6)\right)\left(\frac{\mathbf{q}}{m_e}\right)^2+\left(|c_{13}|^2+|c_{14}|^2\right)\left(\frac{\mathbf{q}}{m_e}\right)^2\left(\vPerpEl\right)^2 + |c_{15}|^2\left(\frac{\mathbf{q}}{m_e}\right)^2\left( \frac{\mathbf{q}}{m_e}\times\vPerpEl \right)^2 \nonumber\\
    &\qquad+2\bigg[\Im(c_4c_{13}^*)+\Im(c_4c_{14}^*)+4\Im(c_8c_{11}^*)+2\Im(c_9^*c_{12})+\left(\Im(c_6c_{13}^*)+\Im(c_6c_{14}^*)\right)\left(\frac{\mathbf{q}}{m_e}\right)^2\nonumber\\
    &\qquad\quad+\Re(c_{13}^*c_{14})\left(\frac{\mathbf{q}}{m_e}\cdot \vPerpEl\right)\bigg]\left(\frac{\mathbf{q}}{m_e}\cdot \vPerpEl\right)\Bigg\}\, .
\end{align}

For real couplings~$(c_i=c_i^*$ for all~$i$), the squared amplitude simplifies to
\begin{align}
    &\overline{| \mathcal{M}(\mathbf{q},\mathbf{v}_{\rm el}^\perp)|^2} = c_1^2 +\frac{c_3^2}{4}\left( \frac{\mathbf{q}}{m_e}\right)^2 (\vPerpEl)^2 -\frac{c_3^2}{4} \left(\frac{\mathbf{q}}{m_e}\cdot \vPerpEl\right)^2 +\frac{c_7^2}{4}(\vPerpEl)^2 + \frac{c_{10}^2}{4}\left(\frac{\mathbf{q}}{m_e}\right)^2   \nonumber\\
   &+ \frac{j_\chi(j_\chi+1)}{12}\Bigg\{ 3c_4^2 +c_6^2 \left(\frac{\mathbf{q}}{m_e}\right)^4 +(4c_8^2+2c_{12}^2)(\vPerpEl)^2 +(2c_9^2+4c_{11}^2+2c_4c_6)\left(\frac{\mathbf{q}}{m_e}\right)^2\nonumber \\
   &+\left(4c_5^2+c_{13}^2+c_{14}^2-2c_{12}c_{15}\right)\left(\frac{\mathbf{q}}{m_e}\right)^2(\vPerpEl)^2+c_{15}^2\left(\frac{\mathbf{q}}{m_e}\right)^4 \left(\vPerpEl\right)^2\nonumber\\
   &-c_{15}^2 \left(\frac{\mathbf{q}}{m_e}\right)^2 \left(\vPerpEl\cdot \frac{\mathbf{q}}{m_e}\right)^2+\left(-4c_5^2+2c_{13}c_{14}+2c_{12}c_{15}\right)\left(\vPerpEl\cdot\frac{\mathbf{q}}{m_e}\right)^2\Bigg\}\, .
\end{align}
\end{widetext}
The square of the amplitude can be identified as the first DM~response function,
\begin{align}
   R^{n\ell}_1\left(\vPerpEl,\frac{\mathbf{q}}{m_e}\right) = \overline{| \mathcal{M}(\mathbf{q},\mathbf{v}_{\rm el}^\perp)|^2}\, ,
\end{align}
which was given in Eq.~\eqref{eq: DM response 1} since the first atomic response function is of the form
\begin{align}
    W_1^{n\ell}(k^\prime,q)\propto \left| f_{1\rightarrow 2}(\mathbf{q})\right|^2\, ,
\end{align}
as we saw in Eq.~\eqref{eq: atomic response 1}.

\subsection{Response 2}
\noindent By using the spin-sum identities of Eqs.~\eqref{eq: spin sum identities}, the second term given by Eq.~\eqref{eq:2MdM} results in
\begin{widetext}
\begin{align}
    &2m_e\overline{ \Re \left[ \mathcal{M} \nabla_{\mathbf{k}} \mathcal{M}^*\cdot \mathbf{A}  \right]}= \left[\frac{|c_3|^2}{2}\left(\left(\frac{\mathbf{q}}{m_e}\cdot \vPerpEl\right) \frac{\mathbf{q}}{m_e}-\left(\frac{\mathbf{q}}{m_e}\right)^2\vPerpEl\right) - \frac{|c_7|}{2}\vPerpEl\right]\cdot\Re(\mathbf{A})+\frac{1}{2}\left( \frac{\mathbf{q}}{m_e}\times\vPerpEl \right)\cdot\Im((c_3c_7^*+c_3^*c_7)\mathbf{A})\nonumber\\
 &\quad+\frac{1}{2}\frac{\mathbf{q}}{m_e}\cdot\Im(c_7^*c_{10}\mathbf{A})+\frac{j_\chi(j_\chi+1)}{6}\Bigg\{ \bigg[\left(4|c_5|^2+|c_{15}|^2\left(\frac{\mathbf{q}}{m_e}\right)^2\right)\left(\left(\frac{\mathbf{q}}{m_e}\cdot \vPerpEl\right) \frac{\mathbf{q}}{m_e}-\left(\frac{\mathbf{q}}{m_e}\right)^2\vPerpEl\right)\nonumber\\
 &\qquad\qquad-\left(4|c_8|^2+2|c_{12}|^2+(|c_{13}|^2+|c_{14}|^2)\left(\frac{\mathbf{q}}{m_e}\right)^2\right)\vPerpEl\bigg]\cdot\Re(\mathbf{A})-\frac{\mathbf{q}}{m_e}\cdot\Im(c_4c_{13}^*\mathbf{A})-\frac{\mathbf{q}}{m_e}\cdot\Im(c_4c_{14}^*\mathbf{A})\nonumber\\
 &\qquad+4 \left( \frac{\mathbf{q}}{m_e}\times\vPerpEl \right)\cdot\Im((c_5c_8^*+c_5^*c_8)\mathbf{A}) - \left(\frac{\mathbf{q}}{m_e}\right)^2 \frac{\mathbf{q}}{m_e}\cdot\Im(c_6c_{13}^*\mathbf{A})- \left(\frac{\mathbf{q}}{m_e}\right)^2 \frac{\mathbf{q}}{m_e}\cdot\Im(c_6c_{14}^*\mathbf{A}) + 2\frac{\mathbf{q}}{m_e}\cdot\Im(c_9c_{12}^*\mathbf{A})\nonumber\\
 &\qquad+ 4\frac{\mathbf{q}}{m_e}\cdot\Im(c_{11}c_{8}^*\mathbf{A})-\left( \frac{\mathbf{q}}{m_e}\times\vPerpEl \right)\cdot\Im((c_{12}c_{13}^*+c_{12}^*c_{13})\mathbf{A})+\left( \frac{\mathbf{q}}{m_e}\times\vPerpEl \right)\cdot\Im((c_{12}c_{14}^*+c_{12}^*c_{14})\mathbf{A})\nonumber\\
 &\qquad-\left(\left(\frac{\mathbf{q}}{m_e}\cdot \vPerpEl\right) \frac{\mathbf{q}}{m_e}-\left(\frac{\mathbf{q}}{m_e}\right)^2\vPerpEl\right)\cdot\Re((c_{12}c_{15}^*+c_{12}^*c_{15})\mathbf{A})-\left(\frac{\mathbf{q}}{m_e}\cdot \vPerpEl\right)\frac{\mathbf{q}}{m_e}\cdot\Re((c_{13}c_{14}^*+c_{13}^*c_{14})\mathbf{A})\nonumber\\
 &\qquad-\left(\frac{\mathbf{q}}{m_e}\right)^2\left( \frac{\mathbf{q}}{m_e}\times\vPerpEl \right)\cdot\Im((c_{14}c_{15}^*+c_{14}^*c_{15})\mathbf{A})\Bigg\}\, .
\end{align}
To keep the expressions more manageable, we introduced the vector~$\mathbf{A}\equiv f_{1\rightarrow 2}(\mathbf{q})\mathbf{f}^*_{1\rightarrow 2}(\mathbf{q})$. For real couplings and a real vector~$\mathbf{A}$, this simplifies to
\begin{align}
    &2m_e\overline{ \Re \left[ \mathcal{M} \nabla_{\mathbf{k}} \mathcal{M}^*\cdot \mathbf{A}  \right]}=\nonumber\\
&=\frac{\mathbf{q}}{m_e}\cdot\mathbf{A}\;\left[\frac{c_3^2}{2}\left(\frac{\mathbf{q}}{m_e}\cdot \vPerpEl\right)+\frac{j_\chi(j_\chi+1)}{6}\left(\frac{\mathbf{q}}{m_e}\cdot \vPerpEl\right)\left\{ 4 c_5^2-2(c_{12}c_{15}+c_{13}c_{14})+c_{15}^2\left(\frac{\mathbf{q}}{m_e}\right)^2\right\} \right]\nonumber\\
&+\vPerpEl\cdot\mathbf{A}\;\Bigg[-\frac{c_3^2}{2}\left(\frac{\mathbf{q}}{m_e}\right)^2-\frac{c_7^2}{2}+\frac{j_\chi(j_\chi+1)}{6}\Bigg\{-4c_5^2\left(\frac{\mathbf{q}}{m_e}\right)^2-c_{15}^2\left(\frac{\mathbf{q}}{m_e}\right)^4 - 4c_8^2-2c_{12}^2\nonumber\\
&\hspace{1.75cm}-(c_{13}^2+c_{14}^2)\left(\frac{\mathbf{q}}{m_e}\right)^2+2c_{12}c_{15}\left(\frac{\mathbf{q}}{m_e}\right)^2\Bigg\} \Bigg]\, .\label{eq:2MdM 2}
\end{align}
\end{widetext}
The first line corresponds to the second response in Eq.~\eqref{eq:RW}, since we recognize the second atomic response function,
\begin{align}
     W_{2}^{n\ell}(k^\prime,\mathbf{q}) &\propto\sum_{m = -\ell}^\ell \sum_{\ell^\prime=0}^\infty \sum_{m^\prime = -\ell^\prime}^{\ell^\prime}\frac{\mathbf{q}}{m_e}\cdot \mathbf{A}\, .
\end{align}
Indeed, the vector~$\mathbf{A}^\prime\equiv\sum_{mm^\prime}\mathbf{A}$ is real which justifies our assumption above. 

Furthermore, it is not obvious that the second term in Eq.~\eqref{eq:2MdM 2}, which is proportional to~$(\vPerpEl\cdot\mathbf{A})$, becomes part of the second DM~response function as well. The reason for this is the fact that the vector~$\mathbf{A}^\prime$ is (anti-)parallel to~$\mathbf{q}$, which allows us to write
\begin{align}
    \vPerpEl\cdot\mathbf{A}^\prime &= \frac{\mathbf{q}}{m_e}\cdot \mathbf{A}^\prime \left(\frac{\mathbf{q}}{m_e}\right)^{-2}\frac{\mathbf{q}}{m_e}\cdot\vPerpEl\, .\label{eq: A q parallel}
\end{align}
Hence, we obtain the final expression for the second DM~response function,
\begin{align}
    &R^{n\ell}_2\left(\vPerpEl,\frac{\mathbf{q}}{m_e}\right) =\left(\frac{\mathbf{q}}{m_e}\cdot \vPerpEl\right)\Bigg[-\frac{c_7^2}{2}\left(\frac{\mathbf{q}}{m_e}\right)^{-2} \nonumber\\
    &- \frac{j_\chi(j_\chi+1)}{6}\left\{(4c_8^2+2c_{12}^2)\left(\frac{\mathbf{q}}{m_e}\right)^{-2}+(c_{13}+c_{14})^2\right\}\Bigg]\, .
\end{align}
This is the expression we presented in Eq.~\eqref{eq: DM response 2} of this paper. Note that the~$c_3$, $c_5$, and $c_{15}$ terms as well as the~$c_{12}c_{15}$ interference term cancel due to the use of Eq.~\eqref{eq: A q parallel}. 

\subsection{Response 3 and 4}
\noindent Lastly, we evaluate the third term in a similar fashion. Substituting the 14 first operators into Eq.~\eqref{eq:dM2}, we find
\begin{widetext}
\begin{align}
    &m_e^2\overline{|\nabla_{\mathbf{k}} \mathcal{M}(\mathbf{q},\mathbf{v}_{\rm el}^\perp) \cdot \mathbf{f}_{1\rightarrow 2}(\mathbf{q}) |^2}=\left(\frac{|c_3|^2}{4}\left(\frac{\mathbf{q}}{m_e}\right)^2+\frac{|c_7|^2}{4}\right)|\mathbf{f}_{1\rightarrow 2}(\mathbf{q})|^2 -\frac{|c_3|^2}{4}\left|\frac{\mathbf{q}}{m_e}\cdot \mathbf{f}_{1\rightarrow 2}(\mathbf{q})\right|^2\nonumber\\
&\qquad+\frac{j_\chi(j_\chi+1)}{12}\Bigg\{ |\mathbf{f}_{1\rightarrow 2}(\mathbf{q})|^2 \left[(4|c_5|^2+|c_{13}|^2+|c_{14}|^2-2\Re(c_{12}^*c_{15}))\left(\frac{\mathbf{q}}{m_e}\right)^2+4|c_8|^2+2|c_{12}|^2 +|c_{15}|^2\left(\frac{\mathbf{q}}{m_e}\right)^4\right]\nonumber\\
&\qquad\qquad+ \left|\frac{\mathbf{q}}{m_e}\cdot \mathbf{f}_{1\rightarrow 2}(\mathbf{q})\right|^2 \left[ -4|c_5|^2-|c_{15}|^2\left(\frac{\mathbf{q}}{m_e}\right)^2+2\Re(c_{12}^*c_{15})+2\Re(c_{13}^*c_{14})\right] \nonumber\\
&\qquad\qquad+ \frac{\mathbf{q}}{m_e}\cdot i\left( \mathbf{f}_{1\rightarrow 2}(\mathbf{q})\times\mathbf{f}^*_{1\rightarrow 2}(\mathbf{q})\right) \left[8\Re(c_3c_7^*)+8\Re(c_5c_8^*)-2\Re(c_{12}c_{13}^*)+2\Re(c_{12}^*c_{14})-8\Re(c_{14}^*c_{15})\left(\frac{\mathbf{q}}{m_e}\right)^2\right] \Bigg\}\, .
\end{align}
For real couplings, this becomes
\begin{align}
&=|\mathbf{f}_{1\rightarrow 2}(\mathbf{q})|^2\left[\frac{c_3^2}{4}\left(\frac{\mathbf{q}}{m_e}\right)^{2}+\frac{c_7^2}{4}+\frac{j_\chi(j_\chi+1)}{12}\left\{4c_8^2+2c_{12}^2+\left(4c_5^2+c_{13}^2+c_{14}^2-2c_{12}c_{15}\right)\left(\frac{\mathbf{q}}{m_e}\right)^{2}+c_{15}^2\left(\frac{\mathbf{q}}{m_e}\right)^{4}\right\}\right]\nonumber\\
&+\left|\frac{\mathbf{q}}{m_e}\cdot \mathbf{f}_{1\rightarrow 2}(\mathbf{q})\right|^2\left[-\frac{c_3^2}{4}+\frac{j_\chi(j_\chi+1)}{12}\left\{-4c_5^2-c_{15}^2\left(\frac{\mathbf{q}}{m_e}\right)^{2}+2c_{12}c_{15}+2c_{13}c_{14}\right\}\right]\nonumber\\
&+ \frac{\mathbf{q}}{m_e}\cdot i\left(\mathbf{f}_{1\rightarrow 2}(\mathbf{q})\times\mathbf{f}^*_{1\rightarrow 2}(\mathbf{q})\right)\left[\frac{j_\chi(j_\chi+1)}{12}\left\{8c_3c_7+8c_5c_8-2c_{12}c_{13}+2c_{12}c_{14}-8c_{14}c_{15}\left(\frac{\mathbf{q}}{m_e}\right)^2\right\}\right]\, .\label{eq: dM2}
\end{align}
\end{widetext}
From the definition of the third and fourth atomic response functions in Eqs.~\eqref{eq: atomic response 3} and~\eqref{eq: atomic response 4},
\begin{align}
     W_{3}^{n\ell}(k^\prime,\mathbf{q}) &\propto \sum_{m = -\ell}^\ell \sum_{\ell^\prime=0}^\infty \sum_{m^\prime = -\ell^\prime}^{\ell^\prime} |\mathbf{f}_{1\rightarrow 2}(\mathbf{q})|^2\, ,\\
    W_{4}^{n\ell}(k^\prime,\mathbf{q}) &\propto \sum_{m = -\ell}^\ell \sum_{\ell  ^\prime=0}^\infty \sum_{m^\prime = -\ell^\prime}^{\ell^\prime} \left|\frac{\mathbf{q}}{m_e}\cdot\mathbf{f}_{1\rightarrow 2}(\mathbf{q})\right|^2\, ,
\end{align}
we can directly assign the first and second line of Eq.~\eqref{eq: dM2} to the third and fourth DM~response functions,
\begin{align}
   &R^{n\ell}_3\left(\vPerpEl,\frac{\mathbf{q}}{m_e}\right) \equiv \frac{c_3^2}{4}\left(\frac{\mathbf{q}}{m_e}\right)^{2}+\frac{c_7^2}{4}+\frac{j_\chi(j_\chi+1)}{12}\nonumber\\
   &\times\Bigg\{4c_8^2+2c_{12}^2+\left(4c_5^2+c_{13}^2+c_{14}^2-2c_{12}c_{15}\right)\left(\frac{\mathbf{q}}{m_e}\right)^{2}\nonumber\\
   &+c_{15}^2\left(\frac{\mathbf{q}}{m_e}\right)^{4}\Bigg\}\, ,\\
   &R^{n\ell}_4\left(\vPerpEl,\frac{\mathbf{q}}{m_e}\right) \equiv -\frac{c_3^2}{4}+\frac{j_\chi(j_\chi+1)}{12}\nonumber\\
   &\times\Bigg\{-4c_5^2-c_{15}^2\left(\frac{\mathbf{q}}{m_e}\right)^{2}+2c_{12}c_{15}+2c_{13}c_{14}\Bigg\}\, .
\end{align}
These two response functions were presented in the paper as Eqs.~\eqref{eq: DM response 3} and~\eqref{eq: DM response 4}.

The third line of Eq.~\eqref{eq: dM2} does not contribute to ionization rates, since the vector $\sum_{m m^\prime}\left(\mathbf{f}_{1\rightarrow 2}(\mathbf{q})\times\mathbf{f}^*_{1\rightarrow 2}(\mathbf{q})\right)$ vanishes.\\[1em]

This completes the derivation of the atomic and DM~response functions.

\section{Evaluation of the atomic response functions}
\label{app: atomic responses}

\subsection{Useful identities}
\label{app: identities}
This chapter contains a number of mathematical identities necessary for the evaluation of the atomic response functions.

\paragraph{Spherical harmonic addition theorem}
\noindent The spherical harmonics, $Y_\ell^m(\theta,\phi)$ are given by
\begin{align}
	Y_\ell^m(\theta,\phi) = N_\ell^m e^{im\phi}P_\ell^m(\cos\theta)\, ,
\end{align}
where~$P_\ell^m(\cos\theta)$ are the associated Legendre polynomials, and $N_\ell^m$ is a normalization constant ensuring that $\int\dd\Omega\; (Y_{\ell^\prime}^{m^\prime })^*Y_\ell^m = \delta_{\ell\ell^\prime}\delta_{mm^\prime}$.
The normalization constant is given by
\begin{align}
	N_\ell^m&\equiv \sqrt{\frac{2\ell+1}{4\pi}\frac{(\ell-m)!}{(\ell+m)!}}\, .
\end{align}
The spherical harmonic addition theorem states that for two given unit vectors defined by the spherical angle coordinates~$(\theta_1,\phi_1)$ and $(\theta_2,\phi_2)$, the sum over all values of~$m$ yields~\cite{arfken2005}
\begin{subequations}
\label{eq: spherical harmonic addition theorem}
\begin{align}
	\sum_{m=-\ell}^\ell Y_\ell^m(\theta_1,\phi_1)Y_\ell^{m*}(\theta_2,\phi_2) = \frac{2\ell+1}{4\pi} P_\ell(\cos\omega)\, .
	\intertext{Here, $P_\ell(x)$ is a Legendre polynomial, and $\omega$ is the angle between the two vectors given by}
	\cos\omega= \cos\theta_1 \cos\theta_2 +\sin\theta_1\sin\theta_2\cos(\phi_1-\phi_2)\, . 
\end{align}
\end{subequations}

\paragraph{Plane-wave expansion}
\noindent A plane-wave $e^{i \mathbf{k}\cdot \mathbf{x}}$ can be expressed as a linear combination of spherical waves~\cite{edmonds1996},
\begin{subequations}
\label{eq: plane wave expansion}
\begin{align}
	&e^{i \mathbf{k}\cdot \mathbf{x}} = \sum_{\ell=0}^\infty i^\ell(2\ell+1) j_\ell(kr)P_\ell(\cos\omega)\, ,
	\intertext{where~$\omega$ is the angle between $\mathbf{k}$ and $\mathbf{x}$. Using the spherical harmonic addition theorem in Eq.~\eqref{eq: spherical harmonic addition theorem}, we can re-write this equation as}
	&= 4 \pi \sum_{\ell=0}^{\infty}i^\ell j_\ell(kr)\sum_{m=-\ell}^{+\ell}Y_\ell^{m*}(\theta_k,\phi_k)\,Y_\ell^m(\theta_x,\phi_x)\, . 
\end{align}
\end{subequations}

\paragraph{Identities with the Wigner 3j-symbol}
\noindent The angular integral of the product of three spherical harmonics can be expressed in terms of the Wigner 3j-symbol~\cite{arfken2005},
\begin{align}
	&\int\dd\Omega\; Y_{\ell_1}^{m_1}(\theta,\phi)Y_{\ell_2}^{m_2}(\theta,\phi)Y_{\ell_3}^{m_3}(\theta,\phi)=\nonumber\\
	& \qquad=\sqrt{\frac{(2\ell_1+1)(2\ell_2+1)(2\ell_3+1)}{4\pi}} \nonumber\\
	& \qquad\times \wignerj{\ell_1}{\ell_2}{\ell_3}{0}{0}{0}\wignerj{\ell_1}{\ell_2}{\ell_3}{m_1}{m_2}{m_3} \,.\label{eq:YYY}
\end{align}
The 3j-symbols satisfy the following orthogonality relation~\cite{edmonds1996},
\begin{align}
	&(2j+1)\sum_{m_1,m_2}\wignerj{j_1}{j_2}{j}{m_1}{m_2}{m} \wignerj{j_1}{j_2}{j^\prime}{m_1}{m_2}{m^\prime} =\nonumber\\
	&\qquad=\delta_{jj^\prime}\delta_{mm^\prime} \Delta(j_1,j_2,j)\, ,  \label{eq: 3j orthogonality}
\end{align}
where $\Delta(j_1,j_2,j)$ is equal to 1 if $(j_1,j_2,j)$ satisfy the triangular condition, i.e., $|j_1-j_2|\leq j_3\leq j_1+j_2$, and zero otherwise.

\subsection{Gradient of a wave function}
\label{app:gradient of wave function}
\noindent In order to compute the vectorial atomic form factor in Eq.~\eqref{eq: vectorial atomic form factor position space}, we need to evaluate the gradient of a wave function $\psi_{nlm}(\mathbf{x})$ as defined in spherical coordinates~$(r,\theta,\phi)$ by Eq.~\eqref{eq: wave function general} which is given by
\begin{align}
\label{eq:grad}
\boldsymbol{\nabla}\psi_{nlm}(\mathbf{x}) &= \frac{\dd R_{n\ell}}{\dd r} Y_\ell^m(\theta,\phi) \;\hat{\mathbf{r}}\;+ \nonumber\\
 &+\frac{R_{n\ell}(r)}{r} \left(\frac{\dd Y_\ell^m}{\dd\theta}\hat{\boldsymbol{\theta}}+\frac{1}{\sin\theta}\frac{\dd Y_\ell^m}{\dd\phi}\hat{\boldsymbol{\phi}}\right)\, , 
 \end{align}
 with the unit vectors
 \begin{align}
\hat{\mathbf{r}} = \begin{pmatrix} \sin\theta\cos\phi \\ \sin\theta\sin\phi\\\cos\theta\end{pmatrix}&,\,
\hat{\boldsymbol{\theta}}= \begin{pmatrix} \cos\theta\cos\phi \\ \cos\theta\sin\phi\\-\sin\theta\end{pmatrix},\,\nonumber\\ \hat{\boldsymbol{\phi}}= &\begin{pmatrix} -\sin\phi \\ \cos\phi\\0\end{pmatrix}\,.
\end{align}
This gradient can be expressed in terms of vector spherical harmonics~(VSH),
\begin{align}
    \boldsymbol{\nabla}\psi_{nlm}(\mathbf{x}) = \frac{\dd R_{n\ell}}{\dd r} \mathbf{Y}_\ell^m(\theta,\phi) + \frac{R_{n\ell}(r)}{r}\boldsymbol{\Psi}_\ell^m(\theta,\phi) \, ,
\end{align}
where the VSH are defined following the conventions of Barrera et al.~\cite{Barrera_1985},
\begin{subequations}
\label{eq: VSH}
\begin{align}
    \mathbf{Y}_{\ell}^{m}(\theta,\phi)&\equiv\hat{\mathbf{r}}\, Y_\ell^m(\theta,\phi) \, ,\\
    \boldsymbol{\Psi}_{\ell}^{m}(\theta,\phi)&\equiv r\, \boldsymbol{\nabla} Y_\ell^m(\theta,\phi)\, .
\end{align}
\end{subequations}
The Cartesian components of the VSHs can themselves be expanded in spherical harmonics with coefficients closely related to Clebsch-Gordan coefficients~\cite{edmonds1996},
\begin{align}
    \left(\mathbf{Y}_{\ell}^{m}(\theta,\phi)\right)_i= \sum_{\hat{\ell}=\ell-1}^{\ell+1}\sum_{\hat{m}=m-1}^{m+1} c_{\hat{\ell}\hat{m}}^{(i)} Y_{\hat{\ell}}^{\hat{m}}(\theta,\phi) \, ,\label{eq:VSH expansion Y}
\end{align}
with the non-vanishing coefficients
\begin{align*}
    c_{\ell-1\,m-1}^{(1)} &= -i c_{\ell-1\,m-1}^{(2)} = - \frac{A_{-1}^-}{2\sqrt{(2\ell-1)(2\ell+1)}}\, ,\nonumber\\
	c_{\ell+1\,m-1}^{(1)} &= -i c_{\ell+1\,m-1}^{(2)} =  \frac{A_{-1}^+}{2\sqrt{(2\ell+3)(2\ell+1)}}\, ,\nonumber\\
	c_{\ell-1\,m+1}^{(1)} &= i c_{\ell-1\,m+1}^{(2)} = \frac{A_1^-}{2\sqrt{(2\ell-1)(2\ell+1)}} \, ,\nonumber\\
	c_{\ell+1\,m+1}^{(1)} &= i c_{\ell+1\,m+1}^{(2)} = -\frac{A_{1}^+}{2\sqrt{(2\ell+3)(2\ell+1)}}\, ,\nonumber\\
	c_{\ell-1\,m}^{(3)} &=  \frac{A_{0}^-}{\sqrt{2(2\ell-1)(2\ell+1)}}\, ,\nonumber\\
	c_{\ell+1\,m}^{(3)} &= -\frac{A_{0}^+}{\sqrt{2(2\ell+3)(2\ell+1)}}\, .\nonumber
\end{align*}
Here, we used the notation of~\cite{edmonds1996},
\begin{subequations}
\begin{align}
    A_{1}^+ &= \sqrt{(\ell + m + 1)(\ell+m+2)} \, , \\
    A_{0}^+ &= -\sqrt{2(\ell + m + 1)(\ell-m+1)} \, , \\
    A_{-1}^+ &= \sqrt{(\ell-m+1)(\ell-m+2)} \, , \\
    A_{1}^- &= \sqrt{(\ell-m-1)(\ell-m)} \, , \\
    A_{0}^- &= \sqrt{2(\ell+m)(\ell-m)} \, , \\
    A_{-1}^- &= \sqrt{(\ell+m-1)(\ell+m)} \, .
\end{align}
\end{subequations}
Similarly, the second VSH can be expanded as
\begin{align}
    &\left(\boldsymbol{\Psi}_{\ell}^{m}(\theta,\phi)\right)_i  = \sum_{\hat{\ell}=\ell-1}^{\ell+1}\sum_{\hat{m}=m-1}^{m+1} d_{\hat{\ell}\hat{m}}^{(i)} Y_{\hat{\ell}}^{\hat{m}}(\theta,\phi) \, , \label{eq: VSH expansion psi}
\end{align}
with coefficients,
\begin{align*}
	d_{\ell-1\,m-1}^{(1)} &= -i d_{\ell-1\,m-1}^{(2)}=-\frac{(\ell+1)A_{-1}^{-}}{2\sqrt{(2\ell+1)(2\ell-1)}} \, ,\nonumber\\
	d_{\ell+1\,m-1}^{(1)} &= -i d_{\ell+1\,m-1}^{(2)} =-\frac{\ell\,A_{-1}^+}{2\sqrt{(2\ell+1)(2\ell+3)}}\, ,\nonumber\\
	d_{\ell-1\,m+1}^{(1)} &= i d_{\ell-1\,m+1}^{(2)} = \frac{(\ell+1)A_{1}^-}{2\sqrt{(2\ell+1)(2\ell-1)}}\, ,\nonumber\\
	d_{\ell+1\,m+1}^{(1)} &= i d_{\ell+1\,m+1}^{(2)} =\frac{\ell\, A_{1}^+}{2\sqrt{(2\ell+1)(2\ell+3)}}\, ,\nonumber\\
	d_{\ell-1\,m}^{(3)} &=\frac{(\ell+1)A_{0}^-}{\sqrt{2(2\ell+1)(2\ell-1)}} \, ,\nonumber\\
	d_{\ell+1\,m}^{(3)} &=\frac{\ell\,A_{0}^+}{\sqrt{2(2\ell+1)(2\ell+3)}}\, .\nonumber
\end{align*}

Eqs.~\eqref{eq:VSH expansion Y} and~\eqref{eq: VSH expansion psi} allow us to the express the gradient of the wave function $\psi_{lmn}$ in a compact expansion in spherical harmonics,
\begin{align}
    \boldsymbol{\nabla}\psi_{nlm}(\mathbf{x}) &= \sum_{i=1}^3\mathbf{e}_i\sum_{\hat{\ell}=\ell-1}^{\ell+1}\sum_{\hat{m}=m-1}^{m+1} \nonumber \\
    &\left(c_{\hat{\ell}\hat{m}}^{(i)} \frac{\dd R_{n\ell}}{\dd r} + d_{\hat{\ell}\hat{m}}^{(i)}\frac{R_{n\ell}(r)}{r}\right) Y_{\hat{\ell}}^{\hat{m}}(\theta,\phi)\label{eq:gradient of wavefunction}
\end{align}
where $\mathbf{e}_i$, ($i=1,2,3$) are the three Cartesian unit vectors.

\subsection{The atomic form factors}
\label{app: atomic form factors}
In Eqs.~\eqref{eq:standard atomic form factor} and~\eqref{eq:new atomic form factor}, we introduced the scalar and vectorial atomic form factors. In this section, we describe the details of evaluting these form factors.

First, we denote the initial~(`1') and final~(`2') state electron wave functions by their respective quantum numbers~$(n,\ell,m)$ and~$(k^\prime,\ell^\prime,m^\prime)$. Consequently, the atomic form factors take the form
\begin{align}
	f_{1\rightarrow 2}(\mathbf{q}) &= \int\frac{\dd^3 k}{(2\pi)^3}\psi^*_{k^\prime\ell^\prime m^\prime}(\mathbf{k}+\mathbf{q})\psi_{nlm}(\mathbf{k})\, ,\\
	\mathbf{f}_{1\rightarrow 2}(\mathbf{q}) &= \int\frac{\dd^3 k}{(2\pi)^3}\psi^*_{k^\prime\ell^\prime m^\prime}(\mathbf{k}+\mathbf{q})\;\frac{\mathbf{k}}{m_e}\;\psi_{nlm}(\mathbf{k})\, .
\end{align}
Moving to position space via a Fourier transformation, we find
\begin{align}
	f_{1\rightarrow 2}(\mathbf{q}) &= \int\dd^3 x \;\psi^*_{k^\prime \ell^\prime m^\prime}(\mathbf{x})e^{i \mathbf{x}\cdot\mathbf{q}}\psi_{n \ell m}(\mathbf{x})\, ,\\
	\mathbf{f}_{1\rightarrow 2}(\mathbf{q}) &= \int\dd^3 x \;\psi^*_{k^\prime \ell^\prime m^\prime}(\mathbf{x})e^{i \mathbf{x}\cdot\mathbf{q}} \frac{i\boldsymbol{\nabla}}{m_e}\psi_{n \ell m}(\mathbf{x})\, . \label{eq: vectorial atomic form factor position space}
\end{align}
In Eqs.~\eqref{eq: wave function general} and~\eqref{eq: wave function general final} we expressed the initial and final state electron wave functions in terms of spherical coordinates~$\mathbf{x}(r,\theta,\phi)$, where the angular dependence is given by the spherical harmonics,
\begin{align}
	\psi_{n\ell m}(\mathbf{x}) = R_{n\ell}(r)\,Y_{\ell}^m(\theta,\phi)\, .
\end{align}
We postpone the discussion of the radial components of the wave functions to Appendix~\ref{app: wave functions}. and start with the evaluation of the scalar atomic form factor~$f_{1\rightarrow 2}$. Afterwards, we discuss the vectorial atomic form factor~$\mathbf{f}_{1\rightarrow 2}$.

\paragraph{Scalar atomic form factor}
Focussing on the scalar atomic form factor for now, we replace the exponential~$e^{i\mathbf{x}\cdot\mathbf{q}}$ with the plane-wave expansion of Eq.~\eqref{eq: plane wave expansion}.
	\begin{align}
		f_{1\rightarrow 2}(\mathbf{q})&=\int \dd^3 x\; R^*_{k^\prime \ell^\prime}(r)Y_{\ell^\prime}^{m^\prime *}(\theta,\phi)R_{n\ell}(r)Y_{\ell}^{m}(\theta,\phi) \nonumber\\
		&\times 4 \pi \sum_{L=0}^{\infty}i^L j_L(qr)\sum_{M=-L}^{+L}Y_L^{M*}(\theta_q,\phi_q)\,Y_L^M(\theta,\phi) \nonumber \\
		&=4\pi \sum_{L=0}^\infty i^L \sum_{M=-L}^L I_1(q) Y_L^{M*}(\theta_q,\phi_q) \nonumber \\ 
		&\times \int\dd\Omega\; Y_{\ell^\prime}^{m^\prime *}(\theta,\phi)Y_{\ell}^{m}(\theta,\phi)Y_L^M(\theta,\phi)\,,
	\end{align} 
where we absorbed the radial integral into
\begin{align}
    I_1(q)\equiv \int \dd r\; r^2 R^*_{k^\prime \ell^\prime}(r) R_{n\ell}(r) j_L(q r)\,. \label{eq: radial integral 1}
\end{align}
This integral is the spherical Bessel transform of~$R^*_{k^\prime \ell^\prime}(r) R_{n\ell}(r)$, i.e. the product of the radial components of the final and initial wave functions. For its evaluation, we refer to Appendix~\ref{app: radial integrals}.

The integral over three spherical harmonics in the last expression can be re-written in terms of the Wigner 3j-symbols via Eq.~\eqref{eq:YYY},
\begin{align}
		f_{1\rightarrow 2}(\mathbf{q})&=\sqrt{4 \pi} \sum_{L=|\ell-\ell^\prime|}^{\ell+\ell^\prime}i^L I_1(q)\sum_{M=-L}^{+L} Y_L^{M*}(\theta_q,\phi_q) \nonumber\\
		&\times (-1)^{m^\prime}\sqrt{(2\ell+1)(2\ell^\prime+1)(2L+1)}\wignerj{\ell}{\ell^\prime}{L}{0}{0}{0}\nonumber\\
		&\times\wignerj{\ell}{\ell^\prime}{L}{m}{-m^\prime}{M}\, . \label{eq: scalar atomic form factor explicit}
\end{align}
At this point, we can already evaluate the first atomic response~$W_1^{n\ell}$ given by Eq.~\eqref{eq: atomic response 1}, which essentially given by the squared modulus of the scalar atomic form factor summed over~$m,m^\prime$,
\begin{align}
	\sum_{m=-\ell}^\ell & \sum_{m^\prime=-\ell^\prime}^{\ell^\prime}\left|f_{1\rightarrow 2}(\mathbf{q})\right|^2 = \sum_{mm^\prime}4\pi \sum_{L L^\prime} i^{L}(-i)^{L^\prime}\left|I_1(q)\right|^2\nonumber\\
	&\times \sum_{M,M^\prime} Y_L^{M*}(\theta_q,\phi_q) Y_{L^\prime}^{M^\prime}(\theta_q,\phi_q)\nonumber\\ 
	&\times (2\ell+1)(2\ell^\prime+1)\sqrt{(2L+1)(2L^\prime+1)}\nonumber\\
	&\times \wignerj{\ell}{\ell^\prime}{L}{0}{0}{0} \wignerj{\ell}{\ell^\prime}{L^\prime}{0}{0}{0}\nonumber\\
	&\times \wignerj{\ell}{\ell^\prime}{L}{m}{-m^\prime}{M} \wignerj{\ell}{\ell^\prime}{L^\prime}{m}{-m^\prime}{M^\prime}\, .
\end{align}
The orthogonality of the Wigner 3j-symbols, Eq.~\eqref{eq: 3j orthogonality}, allows us to sum over the $L^\prime$ and $M^\prime$ indexes,
\begin{align}
		\sum_{m=-\ell}^\ell\sum_{m^\prime=-\ell^\prime}^{\ell^\prime}\left|f_{1\rightarrow 2}(\mathbf{q})\right|^2&=4\pi \sum_{L} I_1(q)^2\sum_{M} Y_L^{M*}(\theta_q,\phi_q) \nonumber\\ &\times Y_{L}^{M}(\theta_q,\phi_q)(2\ell+1)(2\ell^\prime+1) \nonumber\\ 
		&\times \wignerj{\ell}{\ell^\prime}{L}{0}{0}{0}^2\, .
\end{align}
Lastly we apply the spherical harmonic addition theorem given in Eq.~\eqref{eq: spherical harmonic addition theorem},
\begin{align}
	\sum_{m=-\ell}^\ell\sum_{m^\prime=-\ell^\prime}^{\ell^\prime} \left|f_{1\rightarrow 2}(q)\right|^2 &=\sum_{L=|\ell-\ell^\prime|}^{\ell+\ell^\prime} (2\ell+1)(2\ell^\prime+1)\nonumber\\
	&\hspace{-3em}\times(2L+1) \left|I_1(q)\right|^2 \wignerj{\ell}{\ell^\prime}{L}{0}{0}{0}^2\, .
\end{align}
Note that the dependence on the momentum transfer's direction has disappeared. The explicit form of the first atomic response~$W_1^{n\ell}(q,k^\prime)$ for an atomic orbital~$(n,\ell)$ given by Eq.~\eqref{eq: atomic response 1} (or equivalently the `ionization form factor~$|f_{\rm ion}^{n,\ell}|^2$ defined in Eq.~\eqref{eq:standard ionization form factor}) is obtained by adding the remainder of the final state phase space\footnote{Regarding the numerical evaluation of the sum over~$\ell^\prime$, we sum up contributions up to~$\ell^\prime_{\rm max}=7$.}, see Eq.~\eqref{eq: final state phase space},
\begin{align}
    W_1^{n\ell}(q,k^\prime) &= V\frac{4k^{\prime 3}}{(2\pi)^3}\sum_{\ell^\prime=0}^\infty \sum_{L=|\ell-\ell^\prime|}^{\ell+\ell^\prime} (2\ell+1)(2\ell^\prime+1)\nonumber\\
    &\times (2L+1) 
    |I_1(q)|^2\wignerj{\ell}{\ell^\prime}{L}{0}{0}{0}^2\, . \label{eq: atomic response function 1 explicit}
\end{align}

In order to evaluate the three new atomic responses we need to follow similar steps for the vectorial atomic form factor as for the scalar one.

\paragraph{Vectorial atomic form factor}
We start from Eq.~\eqref{eq: vectorial atomic form factor position space} and substitute the gradient of the initial wave function using Eq.~\eqref{eq:gradient of wavefunction},
\begin{align}
  \mathbf{f}_{1\rightarrow 2}(\mathbf{q})&=\frac{i}{m_e} \sum_{i=1}^3\mathbf{e}_i\sum_{\hat{\ell}=\ell-1}^{\ell+1}\sum_{\hat{m}=m-1}^{m+1}\int\dd^3 x R^*_{k^\prime\ell^\prime}(r)Y^{m^\prime *}_{\ell^\prime}(\theta,\phi) \nonumber\\ &\times e^{i\mathbf{x}\cdot\mathbf{q}}\,\left(c_{\hat{\ell}\hat{m}}^{(i)} \frac{\dd R_{n\ell}}{\dd r} + d_{\hat{\ell}\hat{m}}^{(i)}\frac{R_{n\ell}(r)}{r}\right)Y_{\hat{\ell}}^{\hat{m}}(\theta,\phi) \,.
\end{align}
As for the scalar atomic form factor, the exponential in the above equation can be expressed in terms of spherical harmonics via Eq.~\eqref{eq: plane wave expansion},
\begin{align}
\mathbf{f}_{1\rightarrow 2}(\mathbf{q})&=\frac{4\pi i}{m_e}\sum_{i=1}^3 \hat{\mathbf{e}}_i \sum_{\hat{\ell}=\ell-1}^{\ell+1}\sum_{\hat{m}=m-1}^{m+1}\sum_{L=0}^{\infty}\sum_{M=-L}^{L} Y_{L}^{M *}(\theta_q,\phi_q)\nonumber\\
&\times i^L \int\dd\Omega \;
Y^{m^\prime *}_{\ell^\prime}(\theta,\phi)
Y^{M}_{L}(\theta,\phi)
Y^{\hat{m}}_{\hat{\ell}}(\theta,\phi) \nonumber\\ 
	&\times \left[ c_{\hat{\ell}\hat{m}}^{(i)} I_2(q)
	+d_{\hat{\ell}\hat{m}}^{(i)} I_3(q)	\right]\,,
\end{align}
where we defined two new radial integrals,
\begin{align}
I_2(q)&\equiv \int\dd r\,r^2 j_L(qr) R^*_{k^\prime\ell^\prime}(r) \frac{\dd R_{n\ell}}{\dd r} \,, \label{eq: radial integral 2}\\
I_3(q)&\equiv \int\dd r\,r j_L(qr) R^*_{k^\prime\ell^\prime}(r)R_{n\ell}(r) \,,\label{eq: radial integral 3}
\end{align}
which will require extra attention in Appendix~\ref{app: radial integrals}.

We again apply Eq.~\eqref{eq:YYY} to explicitly perform the angular integral,
\begin{align}
\mathbf{f}_{1\rightarrow 2}(\mathbf{q})&=\frac{i\sqrt{4\pi}}{m_e}\sum_{i=1}^3 \hat{\mathbf{e}}_i \sum_{\hat{\ell}=\ell-1}^{\ell+1}\sum_{\hat{m}=m-1}^{m+1}\sum_{L=0}^{\infty}\sum_{M=-L}^{L}Y_{L}^{M *}(\theta_q,\phi_q) \nonumber\\
&\times i^L (-1)^{m^\prime}\sqrt{(2\hat{\ell}+1)(2\ell^\prime+1)(2L+1)}\nonumber\\ 
	&\times \wignerj{\hat{\ell}}{\ell^\prime}{L}{0}{0}{0}\wignerj{\hat{\ell}}{\ell^\prime}{L}{\hat{m}} {-m^\prime}{M}\nonumber\\
	&\times \left[ c_{\hat{\ell}\hat{m}}^{(i)} I_2(q)+d_{\hat{\ell}\hat{m}}^{(i)} I_3(q) \right]\, .
\end{align}
It can be useful to evaluate the above expression in a particular coordinate frame, namely the one in which the $z$-axis points towards $\mathbf{q}$, such that $\theta_q = 0$. Then we can use $Y_{L}^{M}(\theta_q=0,\phi_q) = \sqrt{(2L+1)/(4\pi)}\delta_{M0}$ and sum over~$M$,
\begin{align}
	\mathbf{f}_{1\rightarrow 2}(\mathbf{q}) &= \frac{i}{m_e}\sum_{i=1}^3 \hat{\mathbf{e}}_i \sum_{\hat{\ell}=\ell-1}^{\ell+1}\sum_{\hat{m}=m-1}^{m+1}\sum_{L=|\hat{\ell}-\ell^\prime|}^{\hat{\ell}+\ell^\prime}i^L   \nonumber\\ 
	& (-1)^{m^\prime} (2L+1)\sqrt{(2\ell^\prime+1)(2\hat{\ell}+1)}
	\nonumber\\
	&\times \wignerj{\hat{\ell}}{\ell^\prime}{L}{0}{0}{0} \wignerj{\hat{\ell}}{\ell^\prime}{L}{\hat{m}} {-m^\prime}{0} \nonumber\\
	&\times\left[ c_{\hat{\ell}\hat{m}}^{(i)}I_2(q)  + d_{\hat{\ell}\hat{m}}^{(i)} I_3(q) \right]\, . \label{eq: vectorial atomic form factor explicit}
\end{align}
We explicitly checked that this choice of coordinate frame has no impact on the final values of the atomic response functions.

With the scalar and vectorial atomic form factors given in the forms of Eqs.~\eqref{eq: scalar atomic form factor explicit} and~\eqref{eq: vectorial atomic form factor explicit}, it is possible to evaluate the three new atomic response functions given by the Eqs.~\eqref{eq: atomic response 2} to~\eqref{eq: atomic response 4}. However, before doing so, we have to specify the radial components of the initial and final state electron wave functions without which we cannot evaluate the radial integrals defined in Eqs.~\eqref{eq: radial integral 1},~\eqref{eq: radial integral 2}, and~\eqref{eq: radial integral 3}.

\subsection{Initial and final state wave functions}
\label{app: wave functions}
For the radial part of the initial state wave function~$\psi_{n\ell m}(\mathbf{x})$, we assume Roothaan-Hartree-Fock~(RHF) ground state wave functions expressed as linear combinations of Slater-type orbitals,
\begin{align}
\label{eq:Rin}
R_{n \ell}(r)&= a_{0}^{-3 / 2}\sum_{j} C_{j \ell n} \frac{\left(2 Z_{j \ell}\right)^{n_{j \ell}^{\prime}+1 / 2}}{\sqrt{\left(2 n_{j \ell}^{\prime}\right) !}}\nonumber\\
&\times\left(\frac{r}{a_{0}}\right)^{n_{j \ell}^{\prime}-1} \exp\left(-Z_{j \ell} \frac{r}{a_{0}}\right)\, .
\end{align}
The coefficients~$C_{j\ell n}$, $Z_{j\ell}$, and $n^\prime_{j\ell}$ are tabulated for the atomic orbitals of argon and xenon in~\cite{Bunge:1993jsz} together with their respective binding energies~$E_B^{n\ell}$. Furthermore,~$a_0$ denotes the Bohr radius.

The final electron state is described by a positive energy continuum solution of the Schr\"odinger equation with a hydrogenic potential~$-Z_{\rm eff}/r$~\cite{bethe1957,Agnes:2018oej}.
\begin{align}
\label{eq:Rout}
	R_{k^\prime \ell^\prime}(r)&=\frac{(2 \pi)^{3 / 2}}{\sqrt{V}}(2 k^\prime r)^{\ell^\prime} \frac{\sqrt{\frac{2}{\pi}}\left|\Gamma\left(\ell^\prime+1-\frac{i Z_{\mathrm{eff}}}{k^\prime a_{0}}\right)\right| e^{\frac{\pi Z_{\mathrm{eff}}}{2 k^\prime a_{0}}}}{(2 \ell^\prime+1) !} \nonumber\\
	&\hspace{-2em}\times e^{-i k^\prime r} \Kummer{\ell^\prime+1+\frac{i Z_{\rm eff}}{k^\prime a_0}}{2 \ell^\prime+2}{2 i k^\prime r}\, .
\end{align}
Here, $\Kummer{a}{b}{z}$ is the so-called Kummer's function, or confluent hypergeometric function of the first kind. Note that both radial wave functions are purely real. It should also be noted that the factor of~$1/\sqrt{V}$ involving the volume always cancels due to the application of the integral operator in Eq.~\eqref{eq: final state phase space}, or equivalently with the factor of $V$ in Eqs.~\eqref{eq: atomic response 1} to~\eqref{eq: atomic response 4}.

The factor~$Z_{\rm eff}$ is determined by matching the binding energy~$E_B^{n\ell}$ of the ionized orbital~\cite{Essig:2012yx,Essig:2017kqs,Agnese:2018col},
\begin{align}
   E_B^{n\ell} &\stackrel{!}{=}  13.6\text{ eV}\frac{\left(Z_{\rm eff}^{n\ell}\right)^2}{n^2}\, ,\nonumber\\
    \Rightarrow Z_{\rm eff}^{n\ell} & = \sqrt{\frac{E_B^{n\ell}}{13.6\text{ eV}}}\;\times\;n\, . \label{eq: Z effective}
\end{align}

With the definition of the radial parts of the wave functions in place, we can continue with the evaluation of the radial integrals.

\subsection{Numerical evaluation of the radial integrals}
\label{app: radial integrals}
It is fair to say that the largest obstacle to the evaluation of the scalar and vectorial atomic form factors is the computation of the radial integrals\footnote{These types of integrals are known in the mathematical literature as spherical Bessel and Hankel transforms.}
\begin{align}
I_1(q) &=	\int_0^\infty\dd r\; r^2 j_{L}(qr) R^*_{k^\prime\ell^\prime}(r) R_{n\ell}(r)\, ,\\
I_2(q) &=	\int_0^\infty\dd r\; r^2 j_{L}(qr) R^*_{k^\prime\ell^\prime}(r) \frac{\dd R_{n\ell}}{\dd r}\, ,\\
I_3(q) &=	\int_0^\infty\dd r\; r j_{L}(qr) R^*_{k^\prime\ell^\prime}(r) R_{n\ell}(r)\,.
\end{align}
The highly oscillatory behavior of the spherical Bessel function can hinder the success of standard numerical integration methods, especially for large values of~$q$.

Below, we outline two of the methods we used to solve this integral. The first one is exact, but requires the numerical integration of an oscillating function, the second one is approximate, but fully analytical.

The python code we used for the evaluation of the radial integrals, the atomic form factors, and the atomic response functions is publicly available~\cite{Emken2019}.

\paragraph{Numerical solution}
One of the most straight-forward ways to evaluate the radial integrals is the use of specific numerical integration methods of numerical libraries like e.g. \textit{NumPy}~\cite{numpy}, or computer algebra systems such as \textit{Wolfram Mathematica}~\cite{Mathematica}. The numerical integration method ``DoubleExponential'' of the Wolfram language yielded reliable results. However, for large values of the momentum transfer, the method becomes critically slow.

It is usually beneficial for the performance of the numerical integration to split up the integration domain,
\begin{align}
    \int_0^\infty \dd r\; f(r) = \sum_{i=0}^\infty \int_{i \Delta r}^{(i+1)\Delta r}\dd r\; f(r)
\end{align}
with~$\Delta r = a_0$ and truncate the sum at a finite $i$ after the result has converged to a desired tolerance. Since the integrand in our case is highly oscillatory and this method does not ensure that we integrate from one root to the next, it is possible that an accidental cancellation of positive and negative contributions over one sub-interval of the domain can be misinterpreted as convergence. To avoid this subtle problem, we always verify the series' convergence on two consecutive terms.

\paragraph{Analytic solution}
The second method effectively replaces the integral with an infinite sum, which, if it converges sufficiently fast, yields accurate and prompt results.
This method worked best when applied to large values of~$q$, i.e., exactly where the numerical integration is problematic.

The Kummer function (or confluent hypergeometric function of the first kind), which appears in Eq.~\eqref{eq:Rout}, can be written as a power series,
\begin{align}
	\Kummer{a}{b}{z} &= \sum_{s=0}^{\infty} \frac{a^{(s)}}{b^{(s)}}\frac{z^{s}}{s !}=\sum_{s=0}^{\infty} \frac{\Gamma(s+a)\Gamma(b)}{\Gamma(a)\Gamma(s+b)} \frac{z^s}{s!}\, , \label{eq: Kummer sum}
\end{align}
with $x^{(s)}\equiv x(x+1)\ldots(x+s-1)$ being the rising factorial. In combination with the analytic form of the initial wave function in Eq.~\eqref{eq:Rin}, we find that the above radial integrals are infinite sums of simpler integrals of the form
\begin{align}
	I_L(q,\alpha,\beta)= \int \dd r\; r^\alpha \exp(-\beta r) j_L(q r)\, ,
\end{align}
where the values of the parameters~$\alpha$ and $\beta$ depend on whether we evaluate $I_1(q)$, $I_2(q)$ or $I_3(q)$.~Importantly, this integral can be solved analytically (provided that $\Re(L+\alpha)>-1$),
\begin{align}
 &I_L(q,\alpha,\beta)=
 \frac{\sqrt{\pi}q^L}{2^{L+1}\beta^{\alpha+L+1}}\frac{\Gamma(L+\alpha +1)}{\Gamma \left(L+\frac{3}{2}\right)}  \nonumber\\
&\times\Hypergeometric{\frac{(L+\alpha +1)}{2}}{\frac{(L+\alpha+2)}{2} }{L+\frac{3}{2}}{-\frac{q^2}{\beta ^2}}\,
\end{align}
where the hypergeometric function~$\Hypergeometric{a}{b}{c}{z}$ appears which is defined via
\begin{align}
    \Hypergeometric{a}{b}{c}{z}\equiv \sum_{s=0}^{\infty} \frac{a^{(s)}b^{(s)}}{c^{(s)}}\frac{z^s}{s!}\, .
\end{align}

Provided that the sum in Eq.~\eqref{eq: Kummer sum} converges after a finite number of terms, we can compute the radial integrals analytically. For large~$q$ the convergence occurs quickly. For lower values, we run into numerical precision problems, since we have to include very large numbers of terms of~\eqref{eq: Kummer sum}.~In this case, the numerical approach described in the previous section performs better.

As an example, by using the analytical method described here, the first radial integral, $I_1$(q), can now be written as follows:
\begin{align}
	I_1(q) &=  \sum_{s=0}^\infty \sum_j \frac{4\pi (2k^\prime)^{\ell^\prime}}{(2\ell^\prime+1)!}\left|\Gamma\left(l+1-\frac{i Z_{\mathrm{eff}}}{k a_{0}}\right)\right| e^{\frac{\pi Z_{\mathrm{eff}}}{2 k a_{0}}} \nonumber\\ &\times \frac{C_{jln}(2Z_{jl})^{(n^\prime_{jl}+1/2)}}{a_0^{(n^\prime_{jl}+1/2)}\sqrt{(2n^\prime_{jl})!}}\frac{\Gamma(s+a)\Gamma(b)}{\Gamma(s+b)\Gamma(a)}\frac{(2 i k^\prime)^s}{s!}\nonumber\\
	&\times\frac{\sqrt{\pi}q^L}{2^{L+1}\beta^{\alpha+L+1}}\frac{\gamma(L+\alpha+1)}{\Gamma(L+3/2)}\nonumber\\
	&\times \Hypergeometric{\frac{L+\alpha+1}{2}}{\frac{L+\alpha+2}{2}}{L+3/2}{-\frac{q^2}{\beta^2}}\, ,
\end{align}
where we defined
\begin{align}
	\alpha &= 1+\ell^\prime+n^\prime_{jl}+s\, ,\quad  \beta = \frac{Z_{jl}}{a_0} + i k^\prime\, ,\\
	a&=l^\prime +1 +\frac{iZ_{\rm eff}}{k^\prime a_0}\, , \quad b = 2\ell^\prime+1\, .
\end{align}
The sum can be truncated when the series converges to a given tolerance, provided that it converges sufficiently quickly.

\section{Scattering kinematics}
\label{app: kinematics}
In this appendix, we summarize a few basic kinematic relations for DM-electron scatterings.

The initial and final energies of the inelastic scattering process are given in Eqs.~\eqref{eq: energy initial} and~\eqref{eq: energy final}. By the conservation of energy one can show that
\begin{align}
    \mathbf{v}\cdot \mathbf{q} = \Delta E_{1\rightarrow 2}+\frac{q^2}{2m_\chi}\, . \label{eq: energy conservation v.q}
\end{align}
From this expression, it is clear that for a given momentum transfer~$q$, the minimum DM~speed necessary for an electron ionization with~$\Delta E_{1\rightarrow 2}=E_B+k^{\prime 2}/(2m_e)$ can be obtained by setting~$\cos\theta_{qv}=1$,
 \begin{align}
    v_{\rm min}(k^\prime,q) &= \frac{E_B+k^{\prime 2}/(2m_e)}{q} + \frac{q}{2m_\chi}\, .
\end{align}

Since the local DM~velocity distribution of the galactic halo is assumed to have a maximum speed~$v_{\rm esc}$, DM~particles are only able to ionize a bound electron with binding energy~$E_B$ if~$v_{\rm min}<v_{\rm max}=v_{\rm esc}+v_\oplus$. Thereby, we find the range of momentum transfers which can contribute to the ionization,
\begin{align}
     q_{\rm min} &= m_\chi v_{\rm max} - \sqrt{m_\chi^2 v_{\rm max}^2-2m_\chi E_B}\, ,\\
     &= \frac{E_B}{v_{\rm max}}\, ,\quad\text{for }m_\chi\rightarrow\infty\, ,\\
     q_{\rm max} &= m_\chi v_{\rm max} + \sqrt{m_\chi^2 v_{\rm max}^2-2m_\chi E_B}\, .
 \end{align}
 
 From the requirement that~$v_{\rm min}<v_{\rm max}=v_{\rm esc}+v_\oplus$, we can also derive the maximum final momentum of the electron for a given momentum transfer~$q$,
 \begin{align}
          k^\prime&< \sqrt{2m_e\left(v_{\rm max}q-\frac{q^2}{2m_\chi}+E_B \right)}\, ,
 \end{align}
which were shown as white lines in Figs.~\ref{fig: atomic responses argon} and~\ref{fig: atomic responses xenon}.

Finally, there are a few identities involving~$\vPerpEl$, which are necessary for the evaluation of the DM~response functions in the Eqs.~\eqref{eq: DM response functions}. We can write $\vPerpEl$ as
\begin{align}
    \vPerpEl &= \mathbf{v}-\frac{\mathbf{q}}{2\mu_e}-\frac{\mathbf{k}}{m_e}\, .
\end{align}
Hence the square of the~$\vPerpEl$ (with~$\mathbf{k}=\mathbf{0}$) is given by
\begin{align}    
    (\vPerpEl)^2|_{\mathbf{k}=\mathbf{0}} &= v^2 + \frac{q^2}{4\mu_e^2}\frac{m_\chi-m_e}{m_e +m_\chi} - \frac{\Delta E_{1\rightarrow 2}}{\mu_e}\, .
\end{align}
We will also need the scalar product with the momentum transfer~$\mathbf{q}$,
\begin{align}
 (\vPerpEl\cdot \mathbf{q})|_{\mathbf{k}=\mathbf{0}}&=\Delta E_{1\rightarrow 2} +\frac{q^2}{2m_\chi}-\frac{q^2}{2\mu_e}\nonumber\\
 &=\Delta E_{1\rightarrow 2}-\frac{q^2}{2m_e}\, .
\end{align}
For the last two relations, we used energy conservation via Eq.~\eqref{eq: energy conservation v.q}. The dependence on the binding energies via~$\Delta E_{1\rightarrow 2}$ also explains why the DM~response functions~$R_{i}^{n\ell}$ carry the atomic quantum numbers~$(n\ell)$ as indices.

\section{Anapole, magnetic and electric dipole dark matter}
\label{app:DMphoton}
The anapole, magnetic dipole and electric dipole DM models are described by the following interaction Lagrangians
\begin{align}
\mathscr{L}_{\rm anapole}&= \frac{g}{2\Lambda^2} \, \bar{\chi}\gamma^\mu\gamma^5\chi \, \partial^\nu F_{\mu\nu} \,, \\
\mathscr{L}_{\rm magnetic}&= \frac{g}{\Lambda} \, \bar{\psi}\sigma^{\mu\nu}\psi \, F_{\mu\nu}\,, \\
\mathscr{L}_{\rm electric}&= \frac{g}{\Lambda} \, i\bar{\psi} \sigma^{\mu\nu} \gamma^5 \psi \, F_{\mu\nu} \,,
\label{eq:L}
\end{align}
where $\chi$ ($\psi$) is a Majorana (Dirac) spinor describing the DM~particle, $g$ a dimensionless coupling constant and $\Lambda$ a mass scale.~For simplicity, we use the same notation for the coupling constants of the three interactions, but they can in principle be different.~The Lagrangians in Eq.~(\ref{eq:L}) induce a DM-electron coupling via the electromagnetic current, $J^\nu$, and Maxwell equations, $\partial_\mu F^{\mu\nu}=e J^\nu$, where $\psi_e$ is the four component electron spinor, $F_{\mu\nu} = \partial_\mu A_\nu - \partial_\nu A_\mu$ is the photon field strength tensor, and $A^\nu$ the photon field.

Taking the nonrelativistic limit of the free field solution of the equations of motion for $\psi$ (or $\chi$), and using Gordon's identity to express the matrix element of the current $J^\nu$ between single particle electron states in terms of electromagnetic form factors, at leading order we find the following amplitudes for DM scattering by free electrons
\begin{align}
\mathscr{M}_{\rm anapole}&=\frac{4 e g}{\Lambda^2} m_\chi m_e \Bigg\{
    2 \left(\mathbf{v}_{\rm el}^\perp \cdot \xi^{\dagger s'} \mathbf{S}_\chi \xi^s \right) \delta^{\lambda' \lambda}
    \nonumber\\
    &+  
    g_e \left( \xi^{\dagger s'} \mathbf{S}_\chi \xi^s \right) \cdot \left( i\frac{\mathbf{q}}{m_e} \times \xi^{\dagger \lambda'} \mathbf{S}_e \xi^\lambda  \right)
    \Bigg\} \,, \label{eq: amplitude anapole}\\ 
\mathscr{M}_{\rm magnetic}&=
\frac{e g}{\Lambda}  \Bigg\{
    4m_e\delta^{s's}\delta^{\lambda'\lambda} \nonumber\\
    &+\frac{16m_\chi m_e}{q^2}  i\mathbf{q} \cdot \left(\mathbf{v}_{\rm el}^\perp \times \xi^{\dagger s'} \mathbf{S}_\chi \xi^s \right)\delta^{\lambda'\lambda} 
    \nonumber\\
    &-  
    \frac{8 g_em_\chi}{q^2} \Bigg[\left( \mathbf{q} \cdot \xi^{\dagger s'} \mathbf{S}_\chi \xi^s \right)\left( \mathbf{q} \cdot \xi^{\dagger \lambda'} \mathbf{S}_e \xi^\lambda \right)
    \nonumber\\
    &- q^2  \left( \xi^{\dagger s'} \mathbf{S}_\chi \xi^s \right)\cdot \left( \xi^{\dagger \lambda'} \mathbf{S}_e \xi^\lambda \right)
    \Bigg] \Bigg\}\,,\label{eq: amplitude magnetic dipole}\\
\mathscr{M}_{\rm electric}&= \frac{e g}{\Lambda} \frac{16 m_\chi m_e}{q^2} i\mathbf{q} \cdot \left( \xi^{\dagger s'} \mathbf{S}_\chi \xi^s \right)\delta^{\lambda' \lambda}\label{eq: amplitude electric dipole} \,.
\end{align}
Here, the notation is the same one used in the main body of the paper.~Matching these expressions on to the effective theory expansion in Eq.~(\ref{eq:Mnr}), one can express the effective coupling constants given in Sec.~\ref{sec:ame} in terms of $g$ and $\Lambda$.~In the numerical calculations, we set $g_e=2$.

\section{Details of direct detection and exclusion limits}
\label{app: experiments}
The exclusion limits presented in this paper are based on a re-interpretation of published data and results by DarkSide-50~\cite{Agnes:2018oej}, XENON10~\cite{Angle:2011th}, and XENON1T~\cite{Aprile:2019xxb}.
In this section, we briefly summarize the necessary details to compute the exclusion limits.

An essential input for the prediction of ionization rates is the local density and velocity~distribution of the DM~particles of the galactic halo, which first enters Eq.~\eqref{eq:transition rate}. The local DM~number density is simply given by
\begin{align}
    n_\chi = \frac{\rho_\chi}{m_\chi}\, ,
\end{align}
where we set~$\rho_\chi=~0.4\text{ GeV/cm}^{3}$~\cite{Catena:2009mf}. For the velocity distribution, we use the standard halo model~(SHM) which approximates the distribution by a truncated Maxwell-Boltzmann~distribution that has been boosted into the Earth's rest frame,
\begin{align}
    f_\chi(\mathbf{v})&= \frac{1}{N_{\rm esc}\pi^{3/2}v_0^3}\exp\left[-\frac{(\mathbf{v}+\mathbf{v}_\oplus)^2}{v_0^2} \right]\nonumber\\
    &\times\Theta\left(v_{\rm esc}-|\mathbf{v}+\mathbf{v}_\oplus|\right)\, .
\end{align}
Here, $N_{\rm esc}\equiv \erf(v_{\rm esc}/v_0)-2 (v_{\rm esc}/v_0)\exp(-v_{\rm esc}^2/v_0^2)/\sqrt{\pi}$ is a normalization constant, and the standard choices for the other parameters are~$v_0=220\text{km sec}^{-1}$ for the Sun's circular velocity~\cite{Kerr:1986hz}, and $v_{\rm esc} = 544\text{km sec}^{-1}$ for the galactic escape velocity~\cite{Smith:2006ym}. For the speed of the Earth/the observer in the galactic rest frame, we choose~$v_\oplus\approx 244\text{km sec}^{-1}$.

In all cases, we have to translate the energy spectrum given in Eq.~\eqref{eq:ionization spectrum} into the spectrum as a function of the number of final electrons~$n_e$,
\begin{align}
    \frac{\dd\mathscr{R}_{\rm ion}^{n\ell}}{\dd n_e} = \int \dd E_e\; \mathrm{P}(n_e |E_e)\frac{\dd\mathscr{R}_{\rm ion}^{n\ell}}{\dd E_e}\, . \label{eq: dRdn}
\end{align}
In modeling the probability~$\mathrm{P}(n_e |E_e)$ for~$n_e$ electrons to reach the gas phase of the TPC given an initial electron energy of~$E_e$, we follow~\cite{Essig:2012yx,Essig:2017kqs}.

For the three experiments, we obtain the 90\%~C.L. exclusion limits by applying Poisson statistics to each event bin independently.

\subsection{XENON10 and XENON1T}
\label{app: XENON}

\begin{table}[ht!]
	\centering
	\begin{tabular}{|ll|c|ll|}
	\cline{1-2}\cline{4-5}
	\textbf{XENON10}			&			&\phantom{space}&\textbf{XENON1T}	&			\\
	bin~[S2]				&obs. events		&&bin~[S2]				&obs. events		\\
	\cline{1-2}\cline{4-5}
	$\text{[}$14,41)			&126			&&$\text{[}$150,200)		&8			\\
	$\text{[}$41,68)			&60			&&$\text{[}$200,250)		&7			\\
	$\text{[}$68,95)			&12			&&$\text{[}$250,300)	&2			\\
	$\text{[}$95,122)		&3			&&$\text{[}$300,350)	&1			\\
	$\text{[}$122,149)		&2			&&-	&-			\\
	$\text{[}$149,176)		&0			&&-	&-			\\
	$\text{[}$176,203)		&2			&&-	&-			\\
	\cline{1-2}\cline{4-5}
	\end{tabular}
	\caption{Events at XENON10~(left) and XENON1T~(right) observed in the given S2 bins.}
	\label{tab: observed events}
\end{table}

For XENON10 and XENON1T, the number of observed signals given in bins of the ionization signal~(S2) are available and listed in Table~\ref{tab: observed events}. It is necessary to have the spectrum in terms of S2, i.e. the number of photoelectrons~(PEs) in the photomultiplier tubes,
\begin{align}
    \frac{\dd\mathscr{R}_{\rm ion}^{n\ell}}{\dd \mathrm{S2}} = \epsilon(S2)\sum_{n_e=1}^\infty \mathrm{P}(S2|n_e)\frac{\dd\mathscr{R}_{\rm ion}^{n\ell}}{\dd n_e}\, .
\end{align}
For a given number of electrons~$n_e$, we assume that the resulting number of PEs follows a Gaussian distribution with mean~$n_e g_2$ and width~$\sqrt{n_e}\sigma_{S2}$~\cite{Aprile:2013blg,Essig:2017kqs,Aprile:2019dme},
\begin{align}
    \mathrm{P}(S2|n_e) &= \mathrm{Gauss}(S2|n_e g_2,\sqrt{n_e}\sigma_{S2})\, .
\end{align}
The used parameters for the distribution's mean and width are the secondary-scintillation gain factor~$g_2=27(33)$ and the associated width factor~$\sigma_{S2}=6.7(7)$ for XENON10~(XENON1T)\cite{Aprile:2015uzo,Essig:2017kqs,Aprile:2019xxb,Essig:2019xkx}.

For XENON10, the binned observed numbers of events are given on the left hand side of Table~\ref{tab: observed events}, which correspond to an exposure of 15~kg~days. The efficiency~$\epsilon(S2)$ is given by the product of a flat cut efficiency of~92\%~\cite{Angle:2011th} multiplied by the trigger efficiency given in Fig.~1 of~\cite{Essig:2012yx}.

In the case of XENON1T, we extracted the number of observed events passing all the cuts from Fig.~3 of~\cite{Aprile:2019xxb}.
They are listed on the right side of Table~\ref{tab: observed events}.
The efficiency~$\epsilon(S2)$ is the product of a flat efficiency of~93\% and the different efficiencies given in Fig.~2 of~\cite{Aprile:2019xxb}\footnote{Note that the `radius' efficiency needs to be squared.}. The exposure~$\mathcal{E}$ is given by
\begin{align}
    \mathcal{E} = R^2 \pi \times \Delta z\times \rho_{\rm Xe}\times \Delta t\, ,
\end{align}
where the target's radius is~$R=47.9$cm, the height of the search's volume is $\Delta z=20$cm (corresponding to $z\in [-30\mathrm{cm},-10\mathrm{cm}]$), the density of liquid xenon is taken to be $\rho_{\rm Xe}=3.1\mathrm{g\;cm}^{-3}$, and the time of the search data~$\Delta t =180.7$ days~\cite{Aprile:2019xxb}.
This yields an exposure of $\sim$~80755 kg~days.

Compared to the official XENON1T limits on the standard DM-electron scattering cross section (corresponding to our effective coupling~$c_1$), our constraints are weaker by a factor of a few and therefore conservative, which can potentially be explained by the additional background subtraction the XENON collaboration performs.

\subsection{DarkSide-50}
\label{app: darkside}
Just like XENON10 and XENON1T, DarkSide is a dual-phase TPC. However, instead of xenon, it uses argon as target~\cite{Agnese:2018col}.
The DarkSide-50 experiment had an exposure of 6786~kg~days, and a threshold of~$n_e=3$ electrons.
The number of observed events we base our limits on have been extracted from Fig.~3 of~\cite{Agnese:2018col}.
The official constraints on the standard DM-electron interaction (corresponding to the effective coupling~$c_1$) turn out to be stronger than our own limits by a factor of a few.

 \bibliography{ref}

\end{document}